\documentclass[a4paper,11pt]{article}
\pdfoutput=1 

\usepackage{jcappub} 

\usepackage[T1]{fontenc} 
\usepackage{amssymb}
\usepackage{amsmath}
\usepackage{graphics,epsfig}
\usepackage{epstopdf}
\usepackage{physics}
\usepackage{subfigure}
\usepackage{changes}
\usepackage[colorlinks=true,linkcolor=blue,urlcolor=blue,citecolor=blue]{hyperref}
\usepackage[toc,page]{appendix}
\usepackage[normalem]{ulem}
\usepackage{adjustbox}
\usepackage{latexsym}
\usepackage{amsfonts}
\usepackage{times}
\usepackage{graphicx}
\usepackage{dcolumn}
\usepackage{amsmath}
\usepackage{array}
\usepackage{wrapfig}
\usepackage{multirow}
\usepackage{tabularx}

\usepackage{todonotes}
\usepackage{graphicx,setspace}
\usepackage{dcolumn}
\usepackage{bm}
\usepackage[mathlines]{lineno}
\usepackage{changes}

\newcommand{\be}{\begin{equation}}
\newcommand{\ee}{\end{equation}}
\newcommand{\bea}{\setlength\arraycolsep{2pt} \begin{eqnarray}}
\newcommand{\eea}{\end{eqnarray}}

\graphicspath{{./images/} }

\newcolumntype{M}[1]{>{\centering\arraybackslash}m{#1}}
\newcolumntype{N}{@{}m{0pt}@{}}

\title{Nonlinear electrodynamics effects on the black hole shadow, deflection angle, quasinormal modes and greybody factors}

\author[a,1]{M. Okyay,\note{Corresponding author.}}
\author[b]{A. \"{O}vg\"{u}n}

\affiliation[a]{Department of Physics, University of Miami, Coral Gables, FL 33124, USA.}
\affiliation[b]{Physics Department, Eastern Mediterranean
University, Famagusta, 99628 North Cyprus via Mersin 10, Turkey.}

\emailAdd{mokyay@miami.edu} 
\emailAdd{ali.ovgun@emu.edu.tr}

\abstract{In this paper, we discuss the effects of nonlinear electrodynamics (NED) on non-rotating black holes, parametrized by the field coupling parameter $\beta$ and magnetic charge parameter $P$ in detail. Particularly, we survey a large range of observables and physical properties of the magnetically charged black hole, including the thermodynamic properties, observational appearance, quasinormal modes and absorption cross sections. Initially, we show that the NED black hole is always surrounded by an event horizon and any magnetic charge is permissible. We then show that the black hole gets colder with increasing charge. Investigating the heat capacity, we see that the black hole is thermally stable between points of phase transition. Introducing a generalized uncertainty principle (GUP) with a quantum gravity parameter $\lambda$ extends the range of the stable region, but the effect on temperature is negligible. Then we compute the deflection angle at the weak field limit, by the Gauss-Bonnet theorem and the geodesic equation, and find that even at the first order, the magnetic charge has a contribution due to the "field mass" term. Small changes of the charge contributes greatly to the paths of null geodesics due to the $P^2$ dependence of the horizon radius. Using a ray-tracing code, we simulate the observational appearance of a NED black hole under different emission profiles, thin disk and spherical accretion. We find that the parameter $P$ has a very strong effect on the observed shadow radius, in agreement with the deflection angle calculations. We finally consider quasinormal modes under massless scalar perturbations of the black hole and the greybody factor. We find that the charge introduces a slight difference in the fundamental frequency of the emitted waveform. We find that the greybody factor of the NED black hole is strongly steepened by the introduction of increasing charge. To present observational constrains, we show that the magnetic charge of the M87$^*$ black hole is between $ 0 \leq P \leq 0.024$ in units of M, in agreement with the idea that real astrophysical black holes are mostly neutral. We also find that LIGO/VIRGO and LISA could detect NED black hole perturbations from BHs with masses between $5 M_\odot$ and $8.0\cdot 10^8 M_\odot$. We finally show that for black holes with masses detected with LIGO so far, charged NED black holes would deviate from Schwarzschild by $5\sim10$ Hz in their fundamental frequencies.
}
\date{\today}

\begin{document}
\maketitle
\flushbottom

\section{\label{sec:level1}Introduction}
According to Einstein's theory of general relativity, massive objects warp the fabric of space-time \cite{Einstein:1916vd}. Black holes (BHs) are strange regions where gravity is strong enough to bend light, warp space and even produce so-called singularities in spacetime \cite{Penrose:1964wq}. The spacetimes formed by black holes can be perturbed by other black holes, especially by compact objects, such as neutron stars and other black holes. In 2016, Laser Interferometer Gravitational-Wave Observatory (LIGO) detected the first observation of gravitational waves, which had emanated from the coalescence two black holes, moreover, LIGO even realized the “ring down”: which is the last part of the waveform emitted after the merger of binary black holes, consisting of a few rapidly fading oscillations \cite{LIGOScientific:2016aoc,LIGOScientific:2017vwq}. Another big advance in black hole observations is the image of the shadow of the black hole which is surrounded by a “photon ring”. The Event Horizon Telescope (EHT) collaboration imaged the shadow of the black hole on the center of the elliptical galaxy Messier 87* in 2019 by reporting a bright ring of emission surrounding circular dark region. These experiments provide us a number of enlightening answers to probe general relativity, understanding the properties of black holes and also test the other modified theories \cite{EventHorizonTelescope:2019dse}. Recently EHT measured the polarization of M87*, a signature of magnetic fields, crucial to understanding the launching energetic jets from its core \cite{EventHorizonTelescope:2021bee}.

One of the biggest problem in general relativity is the singularities which lay at the beginning of the universe and also at center of a black hole. On the other hand, Maxwell's equations are known to exhibit singularities which cause the divergence problems in Maxwell's theory.  To solve the divergence of self-energy of a point charge in Maxwell's theory, Born and Infeld introduced the Born–Infeld electrodynamics in 1934 \cite{Born:1934dia}. Afterwards, in 1936, W. Heisenberg and H. Euler proposed the Euler–Heisenberg electrodynamics, in which the self-coupling of the electromagnetic field (EM) induced by virtual electron-positron pairs for energies below the electron mass is present, and this can be seen as an effective field theory which is the first picture of the vacuum polarization effect present in the Quantum Electrodynamics \cite{Heisenberg:1936nmg}. Such extensions of the Maxwell electrodynamics theories, often called "Nonlinead Electrodynamics" (NEDs), usually come as actions derived from string theories or other theoretical frameworks encompassing gravity and other fields \cite{Tseytlin:1986ti, Fradkin:1985qd}. The solutions of NED coupled with general relativity have been studied in various papers \cite{Guendelman:1995xy,Vasihoun:2014pha,Ayon-Beato:1998hmi,Bronnikov:2000vy,Gibbons:1995cv,Dymnikova:2004zc,Kuang:2018goo,Kruglov:2020aqm,Kruglov:2017mpj,Daghigh:2021psm,Fan:2016hvf,Sheykhi:2014gia,Sheykhi:2014ipa,Hendi:2015hoa,Hendi:2015oda,Olmo:2011ja,Guerrero:2020uhn,Novello:2003kh,Ovgun:2016oit,Ovgun:2017iwg,Benaoum:2021tec}.

Before the outstanding discovery of black hole shadow made by EHT \cite{EventHorizonTelescope:2019dse}, many scientists had tried to figure out how the observational appearance of a black hole surrounded by luminous	material would be \cite{Luminet:2019hfx}. In 1979 Luminet drew the luminous accretion disk around the black hole by hand \cite{Luminet:1979nyg}. Afterwards, Falcke	et	al.	created	a	ray-tracing	code to figure out the images of Sgr A*	using different value of spin and inclination angles \cite{Falcke:1999pj}, and Falcke termed it the “shadow of the black hole”. Basically the  hot, optically thin accretion flows create radiation around the black hole and the gases around the black hole behave like optically thin medium to its own radiation. 
The shadow of a black hole is caused by gravitational light deflection. The intensity of the deflected light rays leading to a dark interior and bright ring which can be observed from the distant observers. So far, there are numerous works in literature which study the shadows of the black holes \cite{Junior:2021dyw,Lima:2021las,Herdeiro:2021lwl,Cunha:2020azh,Cunha:2018gql,Wei:2018xks,Wei:2019pjf,Takahashi:2004xh,Takahashi:2005hy,Atamurotov:2013sca,Addazi:2021pty,Vagnozzi:2019apd,Vagnozzi:2020quf,Bambi:2019tjh,Allahyari:2019jqz,Khodadi:2020jij,Shaikh:2019hbm,Shaikh:2018lcc,Peng:2020wun,Konoplya:2021slg,Ovgun:2020gjz,Ovgun:2018tua,Ovgun:2021ttv,Konoplya:2019xmn,Kumar:2019pjp,Abdujabbarov:2016hnw,Atamurotov:2015nra,Toshmatov:2021fgm,Kumar:2019ohr,Afrin:2021imp,Qin:2020xzu,Cimdiker:2021cpz} as well as gravitational lensing of black holes \cite{Gibbons:2008rj,Werner:2012rc,Ovgun:2018fnk,Ovgun:2019wej,Javed:2019qyg,Jusufi:2017lsl,Belhaj:2020rdb,Jusufi:2017mav,Li:2020dln,Javed:2019rrg,Javed:2019ynm,Fu:2021akc,Li:2019mqw,Li:2019qyb,Li:2020wvn,Pantig:2021zqe,Pantig:2020odu,Ono:2018ybw,Arakida:2017hrm,Arakida:2020xil,Virbhadra:2007kw,Virbhadra:1999nm,Keeton:2005jd,Tsukamoto:2021fsz,Tsukamoto:2020bjm,Tsukamoto:2017fxq,Ono:2017pie,Ono:2018jrv,Ishihara:2016vdc,Bozza:2002zj,Ovgun:2020yuv}.

Gravitational waves radiated by perturbed black holes are dominated by quasinormal modes (QNMs), oscillations with complex frequencies. In addition, after perturbation, black holes experience three stages such as inspiral,
merger and ringdown phases. The ringdown phase equals to the quasinormal modes (QNMs) of the remnant BH. QNM frequencies depend only on the parameters of the black hole and also the parameters of the corresponding fields \cite{Daghigh:2011ty,Daghigh:2020mog,Daghigh:2008jz,Zhidenko:2003wq,Zhidenko:2005mv,Konoplya:2011qq,Chabab:2017knz,Lepe:2004kv,Gonzalez:2017shu,Lin:2016sch}. To check the deviations from general relativity, it is crucial to investigate QNM frequencies of black holes in alternative theories of gravity \cite{Berti:2009kk,Berti:2005ys,Cardoso:2016rao,Konoplya:2003ii,Baker:2002qf}. On the other hand, studying null geodesics is related to the classical scattering problem for the rays coming from infinity
with a critical impact parameter. Moreover, in the seminal paper by S. Hawking, it was shown that black holes are indeed gray because of emitted quantum radiation, known as Hawking radiation \cite{Hawking:1975vcx}. At the event horizon, the black hole radiates similarly with black body radiation, however  generated initial radiation varies during the traveling through the spacetime geometry \cite{Akhmedov:2006pg}, hence a distant observer at infinity detects a different spectrum than the original one, which is called greybody factor, a frequency- and geometry-dependent quantity. This is also thought as a filter of the initial Hawking radiation \cite{Visser:1998ke,Boonserm:2008zg}.

Near-horizon regions of black holes are expected to exhibit strong electrodynamic and gravitational fields. Linear Maxwell electrodynamics break down in strong field regions and nonlinear effects become relevant \cite{Heisenberg:1936nmg}. NED models coupled to Einstein gravity give a venue for studying such effects, of which we chose to investigate the recently introduced ‘double-logarithmic’ nonlinear electrodynamics (DL-NED hereafter) by Gullu et al. \cite{Gullu:2020ant}. Such NED theories are expected to obey particular criteria regarding limiting cases, such as \cite{Gullu:2020ant} (i) recovery of the Maxwell theory in the weak-field limit, (ii) providing finite and closed form expressions for the fields and self-energy of point particles, (iii) charged black hole solutions that reduce to the Reissner-Nordström case when nonlinearity parameters are turned off. These are fully satisfied by the DL-NED theory. Furthermore, astrophysical black holes carrying electric charge are expected to quickly neutralize \cite{Ghosh:2020tdu}. Symmetric formulations of Maxwell theory and extended NED models include the hypothetical magnetic monopole and hence have magnetically charged black hole solutions. Such solutions, on the other hand, sustain the magnetic charge and quickly tend toward extremality by Hawking radiation \cite{Maldacena:2020skw}. The DL-NED model is advantageous as the magnetic solution has the desirable property that the black hole is always non-extremal. This makes the DL-NED model of particular interest as cosmic censorship is not violated, with no modifications to  Einstein gravity.


The main aim of the paper is to probe physical properties the DL-NED black hole using the null geodesics, shadow and quasinormal modes. As NED models couple gravity to electrodynamics, the curvature generated by charge terms are expected to have significant contributions to paths of null geodesics. This should, in turn, affect the morphology of the observed shadow. In particular, it will be more interesting to investigate the corresponding light intensity of the shadow, especially spherical accretions, which can be classified into static and infalling. Moreover, to test the effects of the NED on the black hole, perturbative effects can also be investigated with varying charge. The charge term contributes to the Regge-Wheeler potential, directly affecting the ringdown phase of a black hole merger signal. We wish to determine if and how the signal would be affected by the presence of a magnetic charge, as any difference would allow for probing of the theory with the gravitational wave detections and simulations of mergers.


This paper is composed as follows: in section \ref{sec:level2}, we briefly review the black holes in the Double Logarithmic Nonlinear Electrodynamics theory. In section \ref{sec:level3}, we review and thermodynamics of the DL-NLED-BH and introduce corrections with a generalized uncertainty principle. In section \ref{sec:level4}, we compute the weak deflection angle of our spacetime in two distinct methods: using the Gauss-Bonnet theorem and the geodesic equation with a power series. In section \ref{sec:level5}, we investigate the visual appearances of an accretion disk around the DL-NED spacetime. We present the appearance of an optically and geometrically thin accretion disk, along with a spherically infalling accretion and investigate the effects of the theory parameters ($P, \beta$). In section \ref{sec:level6}, we compute the quasinormal modes of our black hole under a perturbation by a massless scalar field using the WKB method. To demonstrate the signal to be observed, we numerically solve the Regge-Wheeler equation and provide plots in the time-domain. We extract the quasi-normal modes from this signal using the Prony method and compare the results. We provide the ranges of masses of DL-NED black hole LISA and LIGO could detect. Finally, in \ref{sec:level7}, we compute the rigorous lower bounds on the greybody factor and compute the absorption cross section. We present how charge effects the size of the cross section for high frequencies. We conclude the paper by reviewing our findings and discuss the possible observational implications. 

\section{\label{sec:level2}Brief Review of Black Holes in Double-Logarithmic Nonlinear Electrodynamics (BH-DL-NED)}

    \subsection{\label{sec:sublevel2}Field equations for DL-NED Theory}
    The Born-Infeld-like Double-Logarithmic Nonlinear Electrodynamics Lagrangian \cite{Gullu:2020ant,Gullu:2020qni} is
    given by
    
    \begin{align}
    \mathcal{L}_{e} & =\frac{1}{2\beta}\left[\left(1-\sqrt{-2\beta\mathcal{F}}\right)\ln\left(1-\sqrt{-2\beta\mathcal{F}}\right)+\left(1+\sqrt{-2\beta\mathcal{F}}\right)\ln\left(1+\sqrt{-2\beta\mathcal{F}}\right)\right],\label{Lagrangian}
    \end{align}
    
    in which $\mathcal{F}=F_{\mu\nu}F^{\mu\nu}$ is the full contraction of the electromagnetic field tensor, which is the exterior derivative of the gauge potential $A$, given as $\text{F} = \text{d}A$, with components in a chart are given as $F_{\mu\nu}=\partial_{\mu}A_{\nu}-\partial_{\nu}A_{\mu}$ and the electric field and magnetic field are static and depend
    only on the radial coordinate. Taking the convectional Einstein-Hilbert Lagrangian $\mathcal{L}_{g}=R-2\Lambda,\label{Lagrangian_G}$ with $\Lambda$ as the cosmological constant and assuming minimal coupling, we construct the action as 
    
    \begin{equation}
    I=\int d^{4}x\sqrt{-g}\left(\frac{1}{2\kappa}\mathcal{L}_{g}+\mathcal{L}_{e}\right),\label{Action}
    \end{equation}
    
    where $\kappa=8\pi G$ and $G$ is Newton's constant in four dimensional spacetime. The usual 'variation of this action' recipe yields the field equations for the theory, detailed in \cite{Gullu:2020qni}:
    
    \begin{equation}
        G_\mu^\nu - \Lambda \delta_\mu^\nu = \kappa T_\mu^\nu, \label{eq:field}
    \end{equation}

    where $G$ is the usual Einstein tensor $G_{\mu}^{\nu}=R_{\mu}^{\nu}-\frac{1}{2}\delta_{\mu}^{\nu}R$ and the energy-momentum tensor is defined to be

    \begin{equation}
        T_{\mu}^{\nu}\equiv\mathcal{L}\delta_{\mu}^{\nu}-4\mathcal{L}_{\mathcal{F}}F_{\mu\lambda}F^{\nu\lambda}.\label{eq:EMT}
    \end{equation}
    
    \subsection{\label{sec:sublevel2.1}Solving the Magnetic Part}

    With the assumption $E_{r}=0$, Bianchi identity implies 
    $ B_{r}=\frac{P}{r^{2}}$ and thus $\mathcal{F}=\frac{2P^{2}}{r^{4}}$, in which $P$ is the magnetic monopole charge. Solving the field equations (\ref{eq:field}) for the spherically symmetric and static ansatz 

    \begin{align}
    ds^{2}= & -f\left(r\right)dt^{2}+\frac{1}{f\left(r\right)}dr^{2}+r^{2}(d\theta^{2}+\sin^{2}\theta d\phi^{2}),\label{spacetime_metric}
    \end{align}
     yields  \cite{Gullu:2020qni} 
     
    \begin{align*}
    	f(r) &= 1 - \frac{2GM}{r}+\frac{\Lambda}{3} r^{2} - \frac{4\kappa P^2}{r_0^2}\arctan\bigg(\frac{r_0^2}{r^2}\bigg) + \frac{2 \kappa     P^2}{3r_0^4}\ln\bigg(\frac{r^4 + r_0^4}{r^4}\bigg) \\
             &+\frac{4\sqrt{2}\kappa P^2}{3r_0r}\Bigg[\arctan\bigg(1-\frac{\sqrt{2}r}{r_0}\bigg)-\arctan\bigg(1+\frac{\sqrt{2}r}{r_0}\bigg)\Bigg]\\
             &+\frac{2\sqrt{2}\kappa P^2}{3r_0r}\ln\bigg(\frac{r^2+\sqrt{2}r_0r+r^2}{r^2-\sqrt{2}r_0r+r^2}\bigg),
    \end{align*}

    where $r_0^2 := 2P\sqrt{\beta}$.
    At the $r \rightarrow  \infty$ asymptotic limit, the metric function takes the form
    \begin{equation}
    f(r) = 1 - \frac{2GM}{r}+ \frac{2\kappa P^2}{r^2} - \frac{4\sqrt{2}\kappa \pi P^2}{3r_0r} +\frac{\Lambda}{3} r^{2},
    \label{metric_func}
    \end{equation}
    
    which is essentially a RN-dS like solution with extra term of order $-1$ in $r$. It can be absorbed into the Schwarzschild-like term by defining it as a contribution to the ADM mass, as $M_{\text{ADM}} = M + \frac{2\sqrt{2}\kappa \pi P^2}{3Gr_0}$, which can be interpreted as a "field mass" \cite{Gullu:2020qni}. For the rest of the paper, we assume $\Lambda = 0$, in this asymptotically flat setting, we obtain two horizons: the Cauchy horizon $r_c$ ("inner horizon") and the event horizon $r_h$ ("outer horizon"):
    
    \begin{align}
        r_c &= GM - \sqrt{G^2M_{\text{ADM}}^2 - 2 \kappa P ^2 }, \\
        r_h &= GM + \sqrt{G^2M_{\text{ADM}}^2 - 2 \kappa P ^2 }.
        \label{eq:horizons}
    \end{align}
    
Note that the charged black holes become extremal when the square root term becomes imaginary: we get a naked singularity. This happens when the charge P satisfies the following equation:
    
    \begin{equation}
        \frac{\sqrt{2 \kappa}P}{G} - \frac{2 \sqrt{2} P^2 \pi \kappa}{3 G r_0} - M = 0,
    \end{equation}
    
    which never has real roots for reasonable values of {M, $\beta$} pairs, so that the BH never becomes extremal, distinguishing our magnetic solution from the extended Maxwell magnetic black hole.
    Particularly, small magnetically charged Reissner-Nordstrom black holes rapidly evaporate to extremality \cite{Ghosh:2020tdu}. Such small black holes could have formed in the early universe \cite{Bai:2019zcd}. If so, possible naked singularities or at least extremal black holes could be astrophysically feasible. As shown, our magnetic solution \ref{metric_func} would not allow naked singularities, even in such extreme cases. This respects the well-known weak cosmic censorship conjecture \cite{Penrose:1969pc} which makes this model (\ref{Lagrangian}) interesting to study.

    Another remark must be made, which will come in handy later on. When we study the DL-NED metric in Eq. \ref{spacetime_metric} under the normalization that $r_H =1$, we see that the metric becomes more and more Schwarzschild-like with increasing charge.
      
\section{\label{sec:level3}Thermodynamics of a  BH-DL-NED}

    \subsection{\label{sec:level3.1}Semiclassial thermodynamics of a  BH-DL-NED}

\begin{figure}[ht]
            \centering
            \begin{minipage}[b]{0.4\linewidth}
            \includegraphics[width = 8 cm]{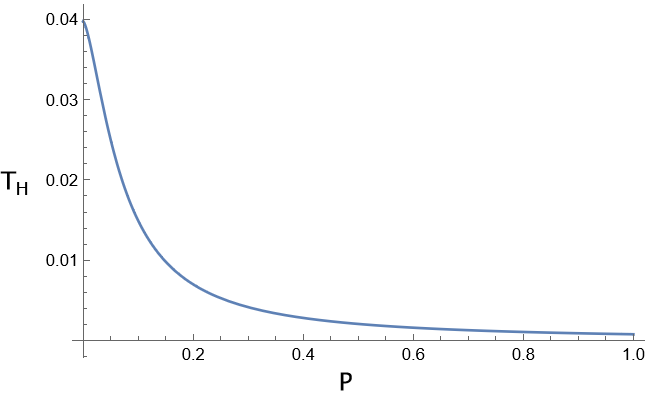}
            \caption{The Hawking temperature versus the magnetic charge P, with M = 1, $\beta$ = 1.}
            \label{fig:temperature}
            \end{minipage}
            \quad
            \begin{minipage}[b]{0.4\linewidth}
            \includegraphics[width = 8 cm]{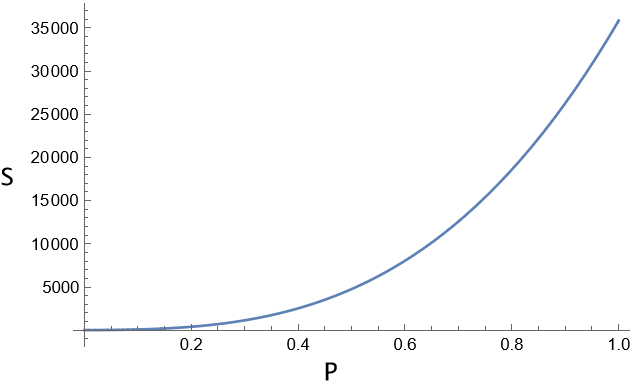}
            \caption{The Entropy versus the magnetic charge P, with M = 1, $\beta$ = 1.}
            \label{fig:entropy}
            \end{minipage}
        \end{figure}

        The thermodynamics of the magnetic solution of the DL-NED theory is considered in \cite{Gullu:2020qni} by use of Smarr relation. Here we consider the usual approach using the metric function. 
        The Hawking temperature of a black hole with a spherically symmetric metric, akin to (\ref{spacetime_metric}), is given by \cite{Hawking:1975vcx} $ T_H = \frac{f'(r)}{2} \bigg|_{r = r_H}$, which is simply
        
        \begin{equation}
            T_H = -\frac{\kappa P^2}{\pi r_H^3} + \frac{2GM + \frac{4\sqrt{2}\kappa \pi P^2}{3r_0r_H}}{4 \pi r_H^2},
        \end{equation}
        
        where $r_H$ itself goes like $~P^2$. We see that with increasing charge, the black hole gets colder, as shown in Fig. \ref{fig:temperature}.
        
\begin{figure}[ht]
            \centering
            \begin{minipage}[b]{0.4\linewidth}
            \includegraphics[width = 8 cm]{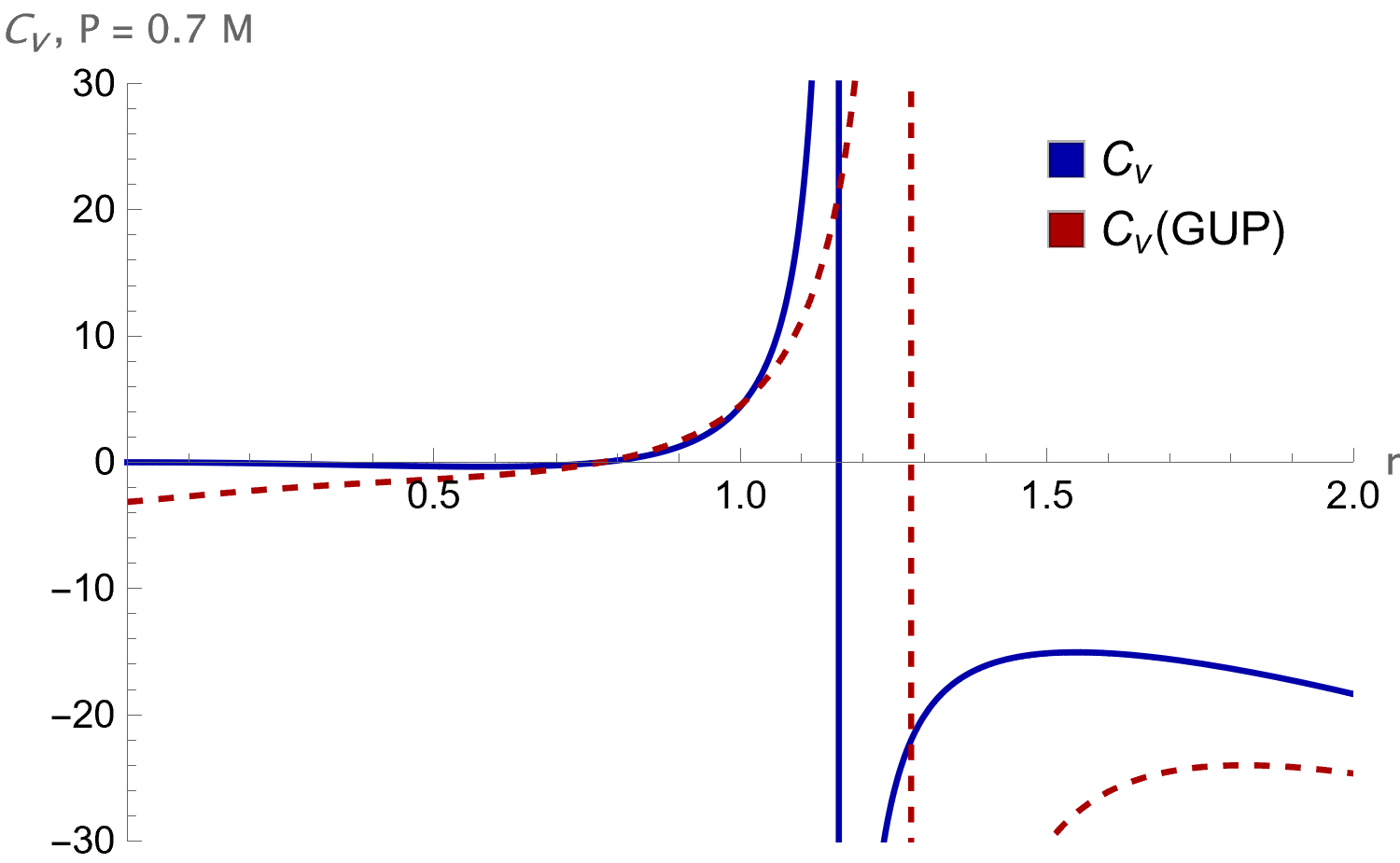}
            \caption{The heat capacity versus the radius of the event horizon for the usual and the GUP cases, given at fixed P = 0.7. The heat capacity turns negative around r = 0.8. As shown, the introduction of the GUP corrections greatly influences the size of the stable region, albeit strongly dependent on the choice of parameter $\lambda$.}
            	\label{fig:heatcap}
            \end{minipage}
            \quad
            \begin{minipage}[b]{0.4\linewidth}
            \includegraphics[width =8 cm]{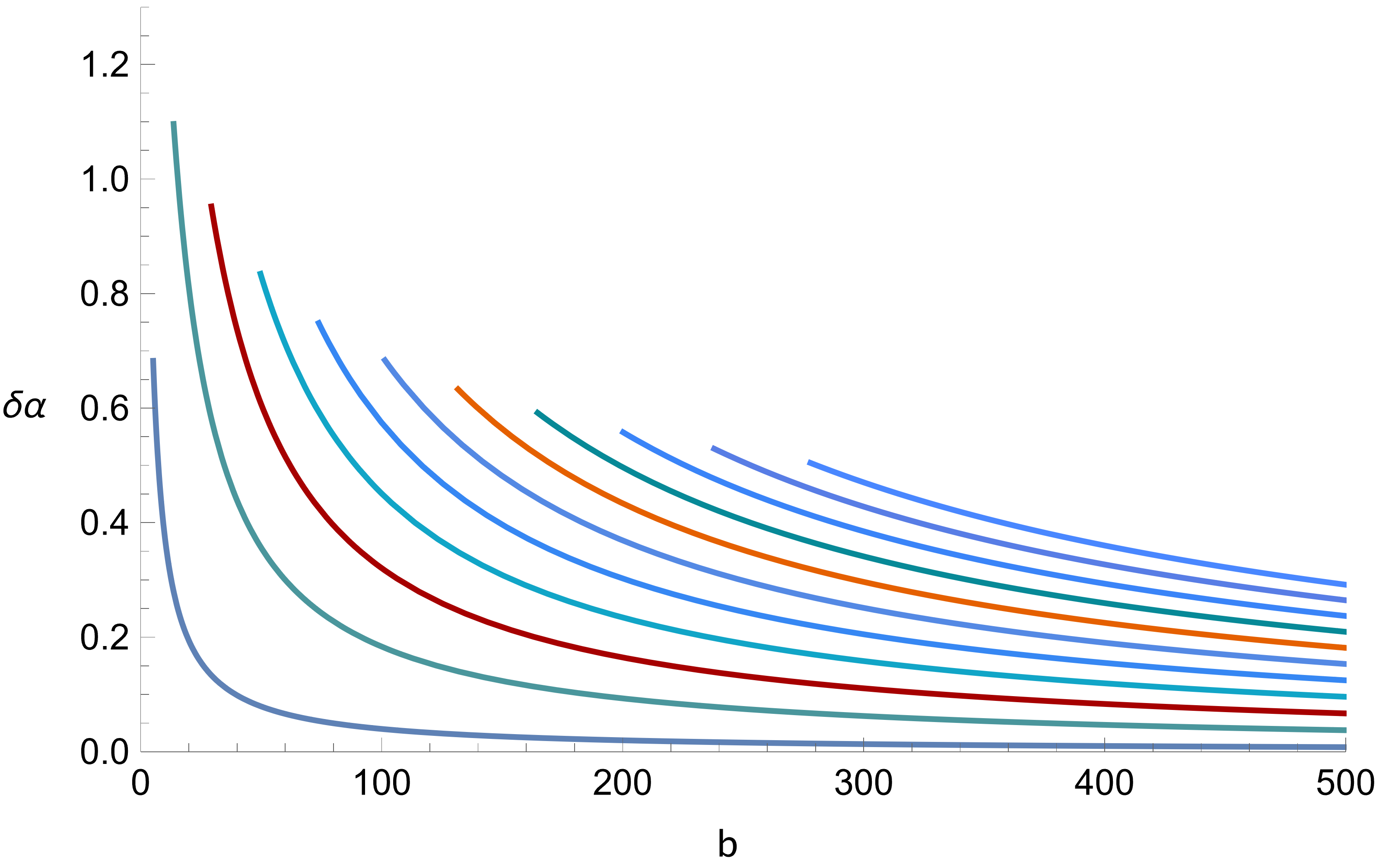}
            \caption{The variation of deflection angle $\delta\alpha$ with respect to the impact parameter $b$, for increasing charge $P$ in increments of $0.1$, from left to right. The curves begin at the critical impact parameter $b_c$ associated with their charges. $P = 0$ Schwarzschild case is given as the leftmost one. We correctly observe that the deflection angle vanishes with increasing impact parameter.}
            	\label{fig:deflectionangles}
            \end{minipage}
        \end{figure}

        This correctly approaches the Schwarzschild temperature at P = 0, $T_{Sch} = \frac{1}{4 \pi r_{Sch}}$
        The Hawking-Bekenstein entropy of a spherically symmetric spacetime is given by the area law
        
        \begin{equation}
            S = \pi r_H^2,
        \end{equation}
        
        which should be increasing with increasing $P$, as shown in Fig. \ref{fig:entropy}. The interesting part of black hole thermodynamics is the consideration of the stability of the black hole by computing its heat capacity. The heat capacity is given by
        
        \begin{equation}
            C_p = T_H  \pdv{S}{T_H} = T_H \bigg(\pdv{S}{r}\bigg)_{r = r_H}\bigg(\pdv{T_H}{r}\bigg)_{r = r_H}^{-1},
        \end{equation}
        
        where the subscript $P$ indicates that the charge is fixed.
        The analytical expression for the heat capacity is not very instructive, so we refer the reader to the graphical analysis. Here we consider the value of the heat capacity with fixed parameters of the theory (P, $\beta$) and varying the event horizon $r_H$. The following graphs \ref{fig:heatcap} show that we have a positive region of heat capacity between the type-1 transition point ($C_p = 0$) and the type-2 point ( $C_p \rightarrow \infty$). This analysis agrees fully with \cite{Gullu:2020qni}, which verifies the accuracy of the classic approach. 
    
    \subsection{Thermodynamics of a  BH-DL-NED  with a generalized uncertainty principle (GUP)}
    
        In this section we analyze the effects of considering a quantum correction to the usual black hole thermodynamics. These types of corrections are considered previously for various of black holes \cite{Kempf:1994su,Anacleto:2015mma,Ovgun:2017hje,Ovgun:2015jna,Bosso:2018ckz}. We consider an uncertainty principle of the form 
        
        \begin{equation}
            \Delta x \Delta p \geq \hbar \Bigg( 1 + \frac{\lambda^2l_p^2}{\hbar^2}\Delta p ^2 \Bigg),
        \end{equation}
        
        where $l_p$ is the Planck length and $\lambda$ is a dimensionless parameter, associated with a quantum gravity theory. Following the authors in \cite{Maluf:2018lyu}, we obtain a GUP-corrected temperature of the form 
        
        \begin{equation}
            T_{GUP} = T_H \Bigg(1+ \frac{\lambda^2l_p^2}{2 r_H^2} \Bigg)^{-1}.
        \end{equation}
        
        This essentially damps the temperature at higher values of the magnetic charge parameter, but is vastly overshadowed by the small magnitude of the Planck length $|l_p| \sim 10^{-35}$. The corrections to the entropy are computed by integrating of the first law of black hole thermodynamics $dM = T_H \, dS$:
        
        \begin{equation}
            S_{GUP} = \int{\frac{1}{T_{GUP}}dM} = \pi r_H^2+\lambda^2l_p^2\log{r_H},
        \end{equation}
        
        where we also obtain the so-called logarithmic term. Finally, the heat capacity is considered:
        
         \begin{equation}
            C_{GUP} = T_{GUP}  \pdv{S}{T_{GUP}} = T_{GUP} \bigg(\pdv{S}{r}\bigg)_{r = r_H}\bigg(\pdv{T_{GUP}}{r}\bigg)_{r = r_H}^{-1},
        \end{equation}
        
        which is also lengthy and does not provide intuitions about the behaviour of the heat capacity. From numerical solutions we observed that the stable region is slightly widened by the quantum gravity parameter $\lambda$. To verify the validity of our calculations, we found that the $C_{GUP}$ correctly reduces to its semiclassical part with the limit $\lambda \rightarrow 0$ and to the Schwarzschild case $C_{Sch} = -2\pi r_H^2$ with P = 0. 
        
\section{\label{sec:level4}Weak Deflection Angle}

    In this section we present the calculation of the deflection angle of a null geodesic in the DL-NED spacetime, and compare two approaches to the matter: the Gauss-Bonnet theorem with the optical metric method, provided by Gibbons and Werner \cite{Gibbons:2008rj}, and the null geodesics method, using the Lagrangian approach. 

    \subsection{\label{sec:level4.1}Calculation of weak deflection angle using Gauss-bonnet theorem}
    
        For null geodesics $ds^2 = 0$, and so the metric (\ref{spacetime_metric}) can be written as 
        
        \begin{equation}
            dt^2 = \gamma_{ij}dx^idx^j = \frac{1}{f^2}dr^2+\frac{r^2}{f}(d\theta^{2}+\sin^{2}\theta d\phi^{2}),
        \end{equation}
        
        where $i,j \in \{1,2,3\}$ and $\gamma_{ij}$ is the optical metric. The coordinate $\theta$ can be chosen to be $\frac{\pi}{2}$ and all derivatives of $\theta$ vanish by the virtue of the Euler Lagrange equations of the associated Lagrangian of (\ref{metric_func}). Then the motion is constrained to a plane and it suffices to consider motion in (r,$\phi$) plane. We calculate the Gaussian curvature $K$ of the optical metric in this plane by calculating the Ricci scalar and dividing by 2: 
        
        \begin{equation}
\begin{aligned} K=& \frac{2 G M+\frac{4 \sqrt{2} P^{2} \pi \kappa}{3 R_{0}^{2}}}{r^{3}}+\frac{-3 G^{2} M^{2}-6 P^{2} \kappa-\frac{8 P^{4} \pi^{2} \kappa^{2}}{3 R_{0}^{4}}-\frac{4 \sqrt{2} G M P^{2} \pi \kappa}{R_{0}^{2}}}{r^{4}} \\ &+\frac{12 G M P^{2} \kappa+\frac{8 \sqrt{2} P^{4} \pi \kappa^{2}}{R_{0}^{2}}}{r^{5}}-\frac{8 P^{4} \kappa^{2}}{r^{6}} \end{aligned}
\end{equation}
        
        Now for the Gauss-Bonnet Theorem \cite{Gibbons:2008rj}, we consider a collection of objects ($D, \chi, g$) to be a subset of a compact, oriented surface, with Euler characteristic $\chi$ and a Riemannian metric $g$ giving rise to a Gaussian curvature $K$. Define $\partial D: \mathbb{R} \rightarrow D$ be the piecewise smooth boundary of $D$ with geodesic curvature $\kappa$ and let $\alpha_i$ be the $\text{i}^\text{th}$ exterior angle. Then the Gauss Bonnet theorem states that 
        \begin{equation}
             \iint_D K\,dS + \int_{\partial D} \kappa\,dt + \sum_{i} \alpha_i = 2\pi \chi(D).\label{eq:GBT}
        \end{equation}
        
  Let S be a source and let observer be stationed at O. Let D to be bounded by a geodesic $C$, and another boundary $C1$ to be perpendicular to $C$ at S and O. Then the $\sum_{i} \alpha_i = \alpha_S + \alpha_O = \pi$, and $\chi(D) = 1$ as D is isomorphic to a 2-circle.

        The geodesic curvature $\kappa$ along a path $\gamma$ is defined as $k(\gamma) = |\nabla_{\dot{\gamma}} \dot{\gamma}|$ This vanishes on $C$ by definition, and so all that remains to consider is $\kappa(C1)$. The radial component of this is $\kappa(C1)^r = \dot{\gamma}^\phi \partial_\phi \dot{\gamma}^r + \Gamma^r_{\phi\phi}\dot{\gamma}^\phi\dot{\gamma}^\phi$. Over cosmological distances, $\gamma \sim r = \text{constant}$ so the first term vanishes, and the second term evaluates to $\frac{1}{r}.$ By a change of variables $dt \rightarrow r\,d\phi$ the boundaries of the integral over $\partial D$ becomes the sum of the exterior angles plus the deflection angle $\delta \alpha$.
        
        Then (\ref{eq:GBT}) becomes 
        \begin{equation}
            \delta \alpha = - \iint_D K \, dS.
        \end{equation}
        
        By using the weak deflection limit, we can set integration bounds $r:\; b/\sin\phi < r < \infty$ and $\phi:\; 0<\phi<\pi$, where $b$ is the impact parameter. The computation of the integral gives 
        
        \begin{equation}
            \delta \alpha = \frac{4 G M+\frac{8 \sqrt{2} P^2 \pi  \kappa }{3 R_0^2}}{b}-\frac{\frac{3}{4} G^2 M^2 \pi +\frac{3}{2} P^2 \pi  \kappa +\frac{2
            P^4 \pi ^3 \kappa ^2}{3 R_0^4}+\frac{\sqrt{2} G M P^2 \pi ^2 \kappa }{R_0^2}}{b^2}
             \label{eq:GBTangle}
        \end{equation}
        
        where we can recover the well-known Schwarzschild deflection angle $\frac{4GM}{b}$ at the $P\rightarrow0$ limit.

    \subsection{\label{sec:level4.2}Calculation of weak deflection angle using geodesics method}

        For the geodesic method, we consider the Lagrangian associated with (\ref{metric_func}):
        \begin{equation}
            2\mathcal{L} = -f(r)\dot{t}^2+ \frac{1}{f(r)}\dot{r}^2 + r^2 (\dot{\theta}^2 + \sin^2\theta\dot{\phi}^2),
            \label{eq:lagrangian}
        \end{equation}
        
        where $\dot{x} := \dv{x}{\lambda}$ and $\lambda$ is an affine parameter along curves $\gamma$. There are two cyclic coordinates ($t,\; \phi$) and hence two conserved quantities $L := \;r^2\dot{\phi}$ and $ E := -2f(r)\dot{t}$, where we employed the freedom to choose $\theta = \frac{\pi}{2}$. The equation for the radial coordinate of the geodesic can be converted to an equation of motion of type $u(\phi)$, where $u = r^{-1}$, by use of the conserved quantity L and the chain rule, to give 
        
        \begin{equation}
            \frac{d^2u}{d\phi^2} + \frac{u}{f(u)} + \frac{1}{2}\dv{}{u}\frac{1}{f(u)} = 0,
            \label{eq:radial}
        \end{equation}
        
        where the last term vanishes as $g_{tt} = \frac{1}{g_{rr}}$ for our metric. To solve this analytically, we propose a power series method in powers of $\frac{M}{b}$, where $M$ is the total ADM mass with the field term, and $b := \frac{L}{E}$ is the impact factor. We consider terms up to order 2:
        
        \begin{equation}
            u(\phi) = \frac{M}{b}\cos\phi + \Bigg(\frac{M}{b}\Bigg)^2u_1(\phi) + \Bigg(\frac{M}{b}\Bigg)^3u_2(\phi),
            \label{eq:powerseries}
        \end{equation}
        
        where the first term is the straight path of the null geodesic, and the higher order terms are the effects of the black hole on the path. By plugging Eq. \ref{eq:powerseries} in the equation Eq. \ref{eq:radial} and collecting terms of same order in $\frac{M}{b}$, we get equations for $u_1$ and $u_2$:
        
        \begin{align}
            \dv[2]{u_1}{\phi}&  + u_1(\phi) -3 G \cos^2{\phi} = 0\\
            \dv[2]{u_2}{\phi}&  + u_2(\phi) -6 G \cos{\phi}\;  u_1(\phi) + 4 \kappa P^2 \cos^3{\phi} = 0,
        \end{align}
        
        which has solutions ($u_1, u_2$):
        
        \begin{align}
            u_1(\phi) &= \frac{3-\cos\phi}{2}\\
            u_2(\phi) &= \frac{3}{16} G \cos(3 \phi )+\frac{P^2 \kappa  \cos(3 \phi )}{8 M^2}+\frac{3}{4} G \phi  \sin(\phi)-\frac{3 P^2
    \kappa  \phi  \sin(\phi )}{2 M^2}.
        \end{align}

        We wish the consider the weak lensing approximation, so we take a Taylor expansion of $u(\phi)$ for $r \rightarrow \infty$, which corresponds to $ u \rightarrow 0$. Assuming the zero of the coordinate $\phi$ is aligned with the point of closest approach, the relation between $\phi$ and the deviation angle $\delta \alpha$ from the straight path is given by $\phi = \frac{\pi}{2} + \frac{\delta \alpha}{2}$ at the $r \rightarrow \infty $ limit, where the path is assumed to be straight again. Taylor expanding $u(\phi)$ at this value for the angle gives us the full value for the deflection angle:
        
        \begin{equation}
            \delta \alpha = \frac{4 G M+\frac{8 \sqrt{2} P^2 \pi  \kappa }{3 R_0^2}}{b}-\frac{\frac{3}{4} G^2 M^2 \pi +\frac{3}{2} P^2 \pi  \kappa +\frac{2
            P^4 \pi ^3 \kappa ^2}{3 R_0^4}+\frac{\sqrt{2} G M P^2 \pi ^2 \kappa }{R_0^2}}{b^2},
        \end{equation}
        
        which confirms the Gauss-Bonnet result (\ref{eq:GBTangle}) up to second order as seen in Fig. \ref{fig:deflectionangles}.
        
\section{\label{sec:level5}Null Geodesics and Shadows Cast}
    
    For purposes of understanding the behaviours of null geodesics qualitatively, we can take a closer look on the radial E-L equation of \ref{eq:lagrangian}. It is:
    \begin{equation}
        \frac{1}{f(r)}\dot{r}^2 + r^2\dot{\theta}^2 - \frac{E^2}{f(r)} + \frac{L^2}{r^2\sin^2{\theta}} = 0,
    \end{equation}
    
    with the choices of the previous section \label{ref:geodesics}, we can write the equation as 
    \begin{equation}
        \dot{r}^2 + V_{\text{eff}}(r) =0,\;\;\; V_{\text{eff}}(r) = \frac{L^2}{r^2}f(r) - E^2.
    \end{equation}
    
    The turning point of the photon path is given by $\dot{r}|_{r_p} = 0, \;\; V_{\text{eff}}(r_p) = 0$. If at the turning point of a photon we have a local extrema of the potential,  where $\dv{V_{\text{eff}}}{r}\Big|_{r_p} = 0$, the radial coordinate of the turning point corresponds to a circular orbit, known as the photon sphere: 
    \begin{equation}
        r_{ph}=\frac{\sqrt{\left(6 G M r_0+4 \pi  \sqrt{2} \kappa  P^2\right)^2-64 \kappa  P^2 r_0^2}}{4 r_0}+\frac{3 G M}{2}+\frac{\pi  \sqrt{2} \kappa  P^2}{r_0}.
    \end{equation}
    The impact parameter of a geodesic is related to its turning point by 
    \begin{equation}
        b =\frac{r_{p}}{\sqrt{f(r_{p})}}.
    \end{equation}
    For the case of $r_p = r_{ph}$, we have the critical impact parameter $b_c$, for which all geodesics with $b < b_c$ fall into the event horizon. For this reason it is referred to as the radius of the shadow of the black hole, which is given by \cite{Perlick:2015vta}
    \begin{equation}
        R_{s}=r_{p} \sqrt{\frac{f\left(x_{0}\right)}{f\left(r_{p}\right)}},
    \end{equation}
    where we place the distant and static observer at position $x_0$. For large distant observer ($f(x_{0})=1$), it reduces to 
    \begin{equation*} 
        R_{s}=\frac{r_p }{\sqrt{f(r_p)}}
        \end{equation*}
        
           \begin{equation}
=\frac{2\left(\frac{1}{2} \sqrt{\left(6 G M r_{0}+4 \pi \sqrt{2} \kappa P^{2}\right)^{2}-64 \kappa P^{2} r_{0}^{2}}+3 G M r_{0}+2 \pi \sqrt{2} \kappa P^{2}\right)}{r_{0} \sqrt{\frac{{-\frac{3 G^{2} M^{2}}{\kappa P^{2}}+\frac{G M \sqrt{\left(6 G M r_{0}+4 \pi \sqrt{2} \kappa P^{2}\right)^{2}-64 \kappa P^{2} r_{0}^{2}}{2 \kappa P^{2} r_{0}}}{\sqrt{\pi \sqrt{2} \sqrt{\left(6 G M r_{0}+4 \pi \sqrt{2} \kappa P^{2}\right)^{2}-64 \kappa P^{2} r_{0}^{2}}}}-\frac{4 \sqrt{2 \pi G M}}{r_{0}}-\frac{8 \pi^{2} \kappa P^{2}}{3 r_{0}^{2}}}+8}{}}}
        \label{eq:shadow}
    \end{equation}
    The angular size of the M87$^*$ galactic center black hole shadow is $\theta_s = (42 \pm 3)\mu as$, as  reported by EHT Collaboration. Considering that the distance to M87$^*$ is $D = 16.8 $ Mpc and the mass of M87$^*$ central object $M = 6.5 \times 10^9$ M\textsubscript{\(\odot\)}. Ignoring rotation, this gives us the diameter of the shadow in units of mass $d_{M87^*}$ \cite{Falcke:1999pj,EventHorizonTelescope:2019dse}:
    
    \begin{eqnarray}
    d_{M87^*}=\frac{D \,\theta_s}{M_{87^*}}=11.0 \pm 1.5.
    \label{eq:M87d}
    \end{eqnarray}
    
    Because of the remarkable closeness of the M87$^*$ black hole shadow with a Schwarzschild one, we expect that if it had any DL-NED magnetic charge P, it would be very close to zero. With undoing the substitutions $r_0^2 = 2 P \sqrt{\beta}$ and $\kappa = 8 \pi G$, and choosing a value $\beta = 1$, we fit (\ref{eq:shadow}) to bounds of (\ref{eq:M87d}) and determine the possible values of magnetic charge for M87. We obtain that the magnetic charge should be in the range
    
    \begin{equation}
        0 \leq P \leq 0.024.
    \end{equation}
    
    with $ P = 0.01 $ for $d_{M87^*}= 11.0$. Moreover, we find that with very large values of $\beta$, the upper end of the charge constraint asymptotically approaches $P = 0.1$. With the understanding that the charge should be low, this suggests that the coupling parameter should not assume large values.
    \\ To obtain visual representations of these abstract calculations, we numerically solve the set of Euler-Lagrange equations of (\ref{eq:lagrangian}). For this purpose we created a simulation and visualisation scheme in \textit{Mathematica}. In this visualization, we assume that in the real universe, most black holes have accreting materials around it, which should influence the observational appearance of the Black Hole. To study the observational appearance, we consider two simplified models: an optically and geometrically thin disk, and a spherically infalling matter disk.

    \subsection{Classification of rays and Rings}

    The so-called "shadow" of a black hole is generated by the light coming from the background, which we treat as uniform in this paper. To classify the incoming light rays, we closely follow Gralla et al \cite{Gralla:2019xty}, we define the number of orbits $n(\gamma) = \phi/2\pi$, where $\phi$ is the final angular coordinate of the geodesic $\gamma$ once it is essentially free of the effect of the gravitational lens. The number of orbits measures the number of times any incoming geodesic crosses the accretion plane, which is the equatorial plane perpendicular to the plane generated by the observer and the position of the black hole \cite{Qin:2020xzu}.

    \begin{enumerate}
        \item $n < \frac{3}{4}$: Direct emission, intersects the accretion plane once. 
        \item $\frac{3}{4} < n < \frac{5}{4}$: Lensing ring, intersects the accretion plane twice
        \item $n > \frac{5}{4}$: Photon ring, intersects the accretion plane at least three times.
    \end{enumerate}

    We want to investigate the appearance of the shadow with different parameters of the theory. For this reason we first show the number of orbits n versus the impact parameter b associated with the geodesic $\gamma$.

    \begin{figure}[h!]
    	\centering
    	\includegraphics[width = 9.5 cm]{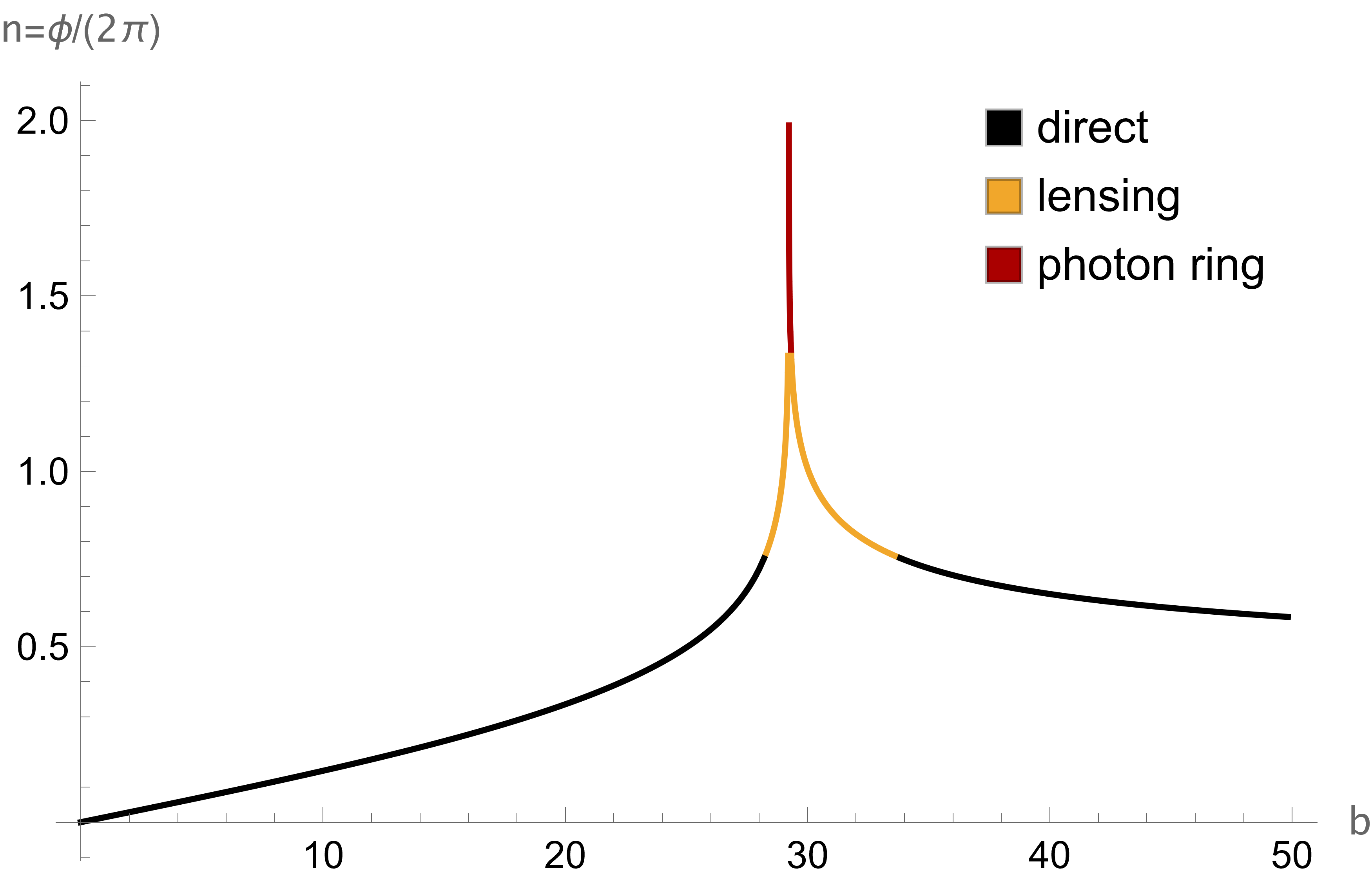}
    	\includegraphics[width = 8 cm]{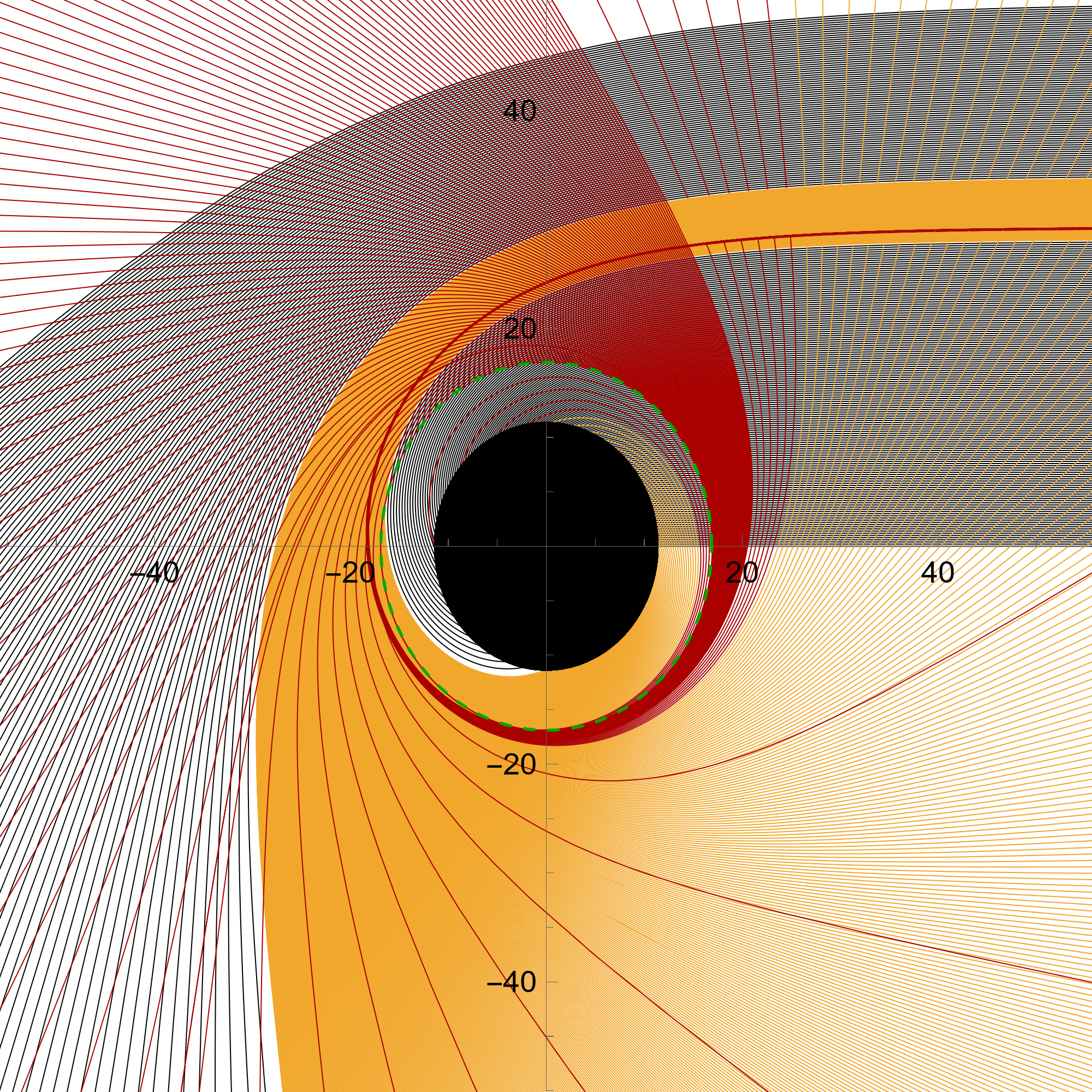}
    	\caption{Given are the number of orbits and behaviour of null geodesics, in the DL-NED spacetime,  with different values of the impact parameter. On the left, we show the
        fractional number of orbits, $n = \phi/(2\pi)$, where $\phi$ is the final azimuthal angle outside the horizon at $r\rightarrow \infty$ limit. The colors correspond to n $<$ 0.75
        (black), 0.75 $<$ n $<$ 1.25 (gold), and n $>$ 1.25 (red), defined as the direct, lensed, and photon ring trajectories, respectively.
        On the right is the spacetime traced out by null geodesics. We only show a selected portion, where the spacing between impact parameters are 1/5, 1/100, and 1/1000 in the direct, lensed, and photon ring bands, respectively. The photon ring is shown as green dashed line and the black circle is the event horizon.}
        
    	\label{fig:N}
    \end{figure}
    \begin{figure}
    	\centering
    	\includegraphics[width = 10 cm]{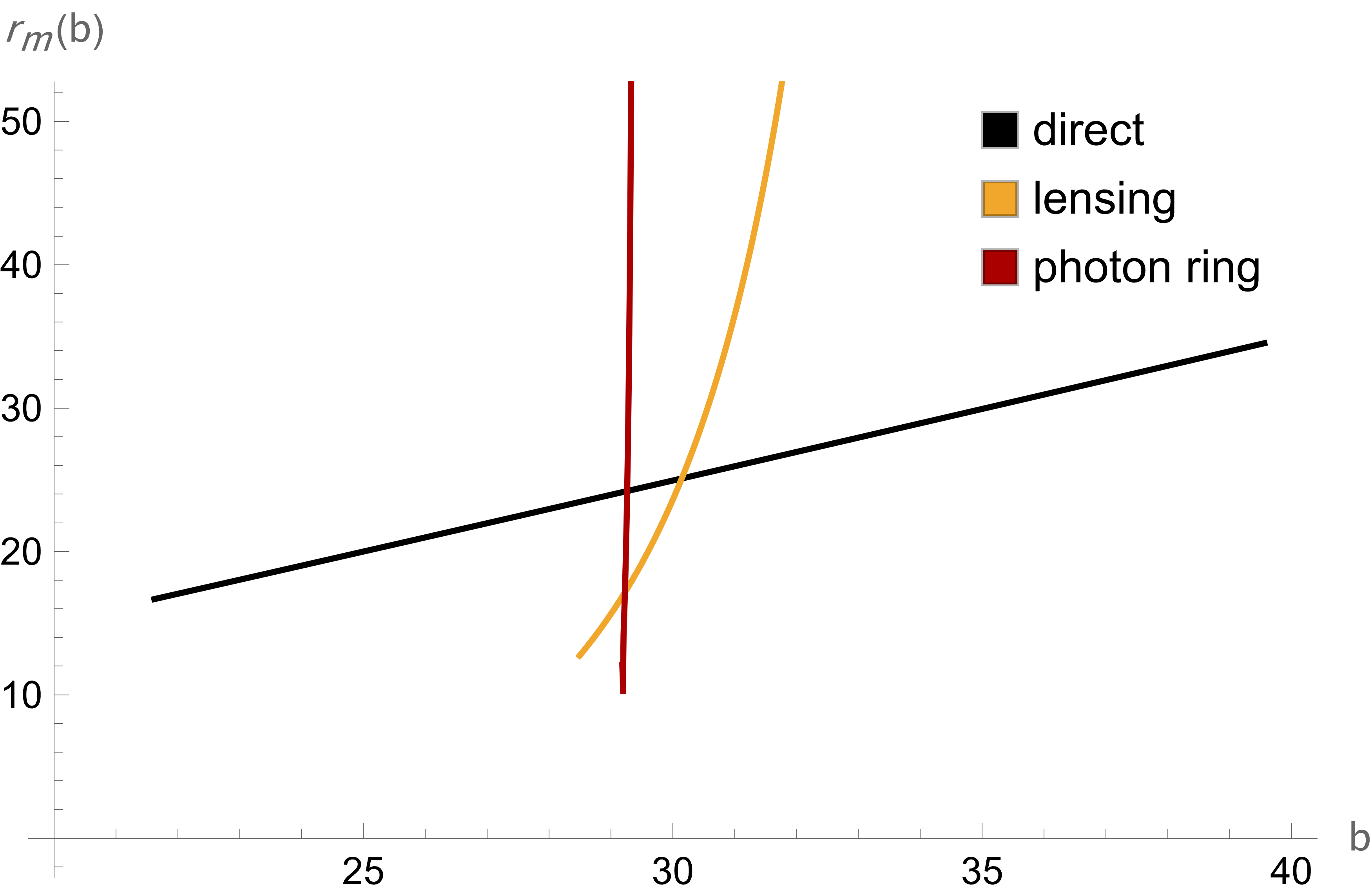}
    	\caption{The first three transfer functions $r_m(b)$ for a face-on
            thin disk in the DL-NED spacetime. Tracing a photon
            back from the detector, these represent the radial coordinate
            of the first (black), second (gold), and third (red) intersections
            with a face-on thin disk outside the horizon.}
    	\label{fig:transferFunc}
    \end{figure}

    Assuming that the only visible portion is due to the accretion disk, the specific intensity and frequency of the emission will be denoted by $I^{em}_\nu$. The observed intensity$I^{obs}_{\nu'}$ at some frequency $\nu'$ is given by

    \begin{equation}
        I^{obs}_{\nu'} = g^3I^{em}_\nu, \;\;\;g = \sqrt{f(r)},
        \label{eq:intensity}
    \end{equation}
    and integrating over all frequencies, we observe that $I^{obs} = g^4I_{EM}$ as $\nu' = gd\nu$ \cite{Gralla:2019xty} and $I_{EM} = \int I^{em}_\nu\,d\nu$. By our model of brightness by contact with the accretion disk, the total intensity received by the observer should be given by the sum of all crossings with the disk:

    \begin{equation}
        I(r) = \sum_n I^{obs}(r)|_{r = r_m(b)},
        \label{eq:totalI}
    \end{equation}

    where $r_m(b)$ is the radial coordinate of the mth intersection with the disk plane outside the horizon, which we will call the transfer function. 
    
    \subsection{Accretion Models and Shadows}
    
        \subsubsection{\label{sec:level5B1}Thin Disk Accretion}

\begin{figure}
        	\centering
        	\includegraphics[width = 6.5 cm]{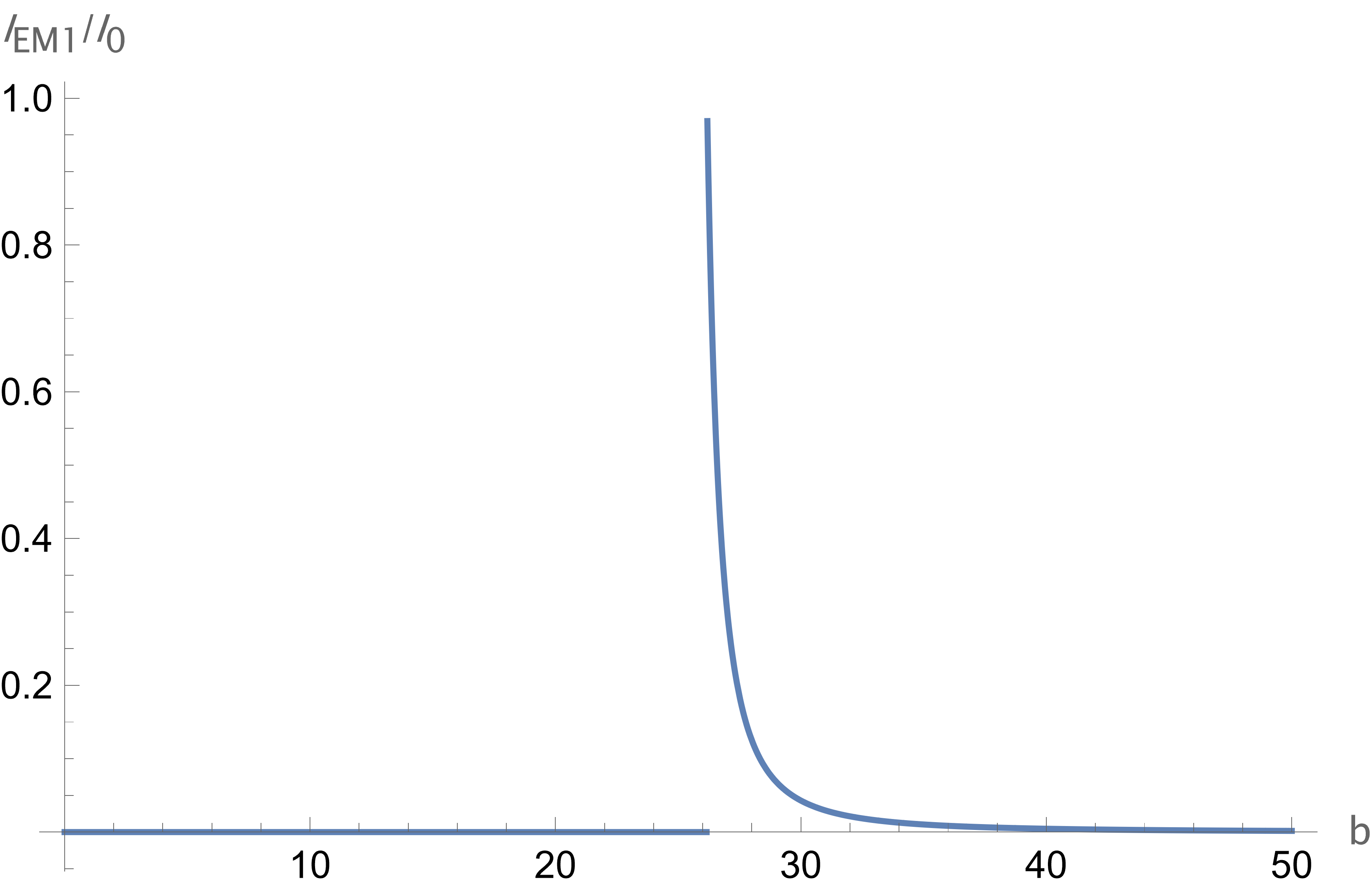}
        	\includegraphics[width = 6 cm]{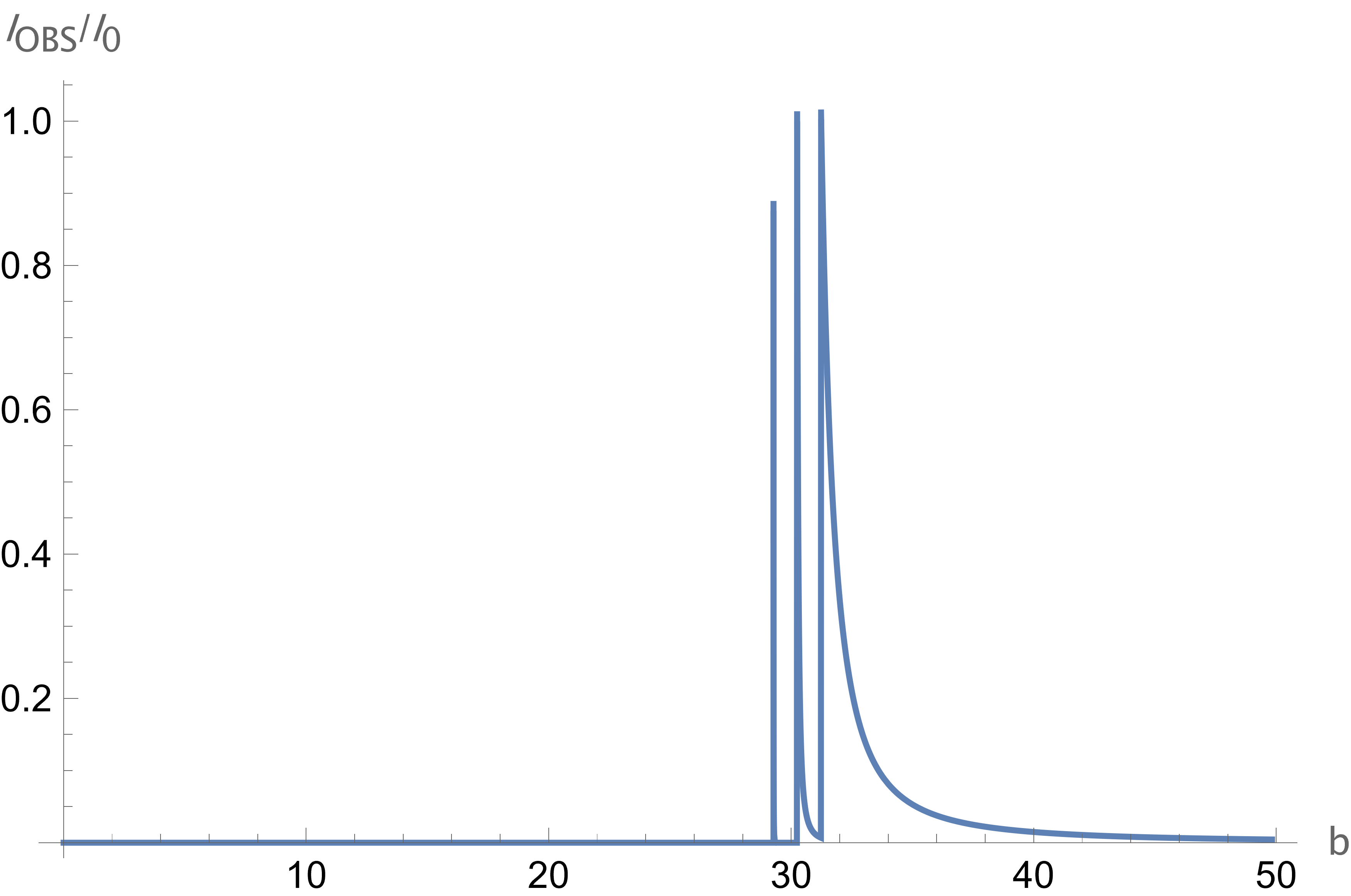}
        	\includegraphics[width = 5 cm]{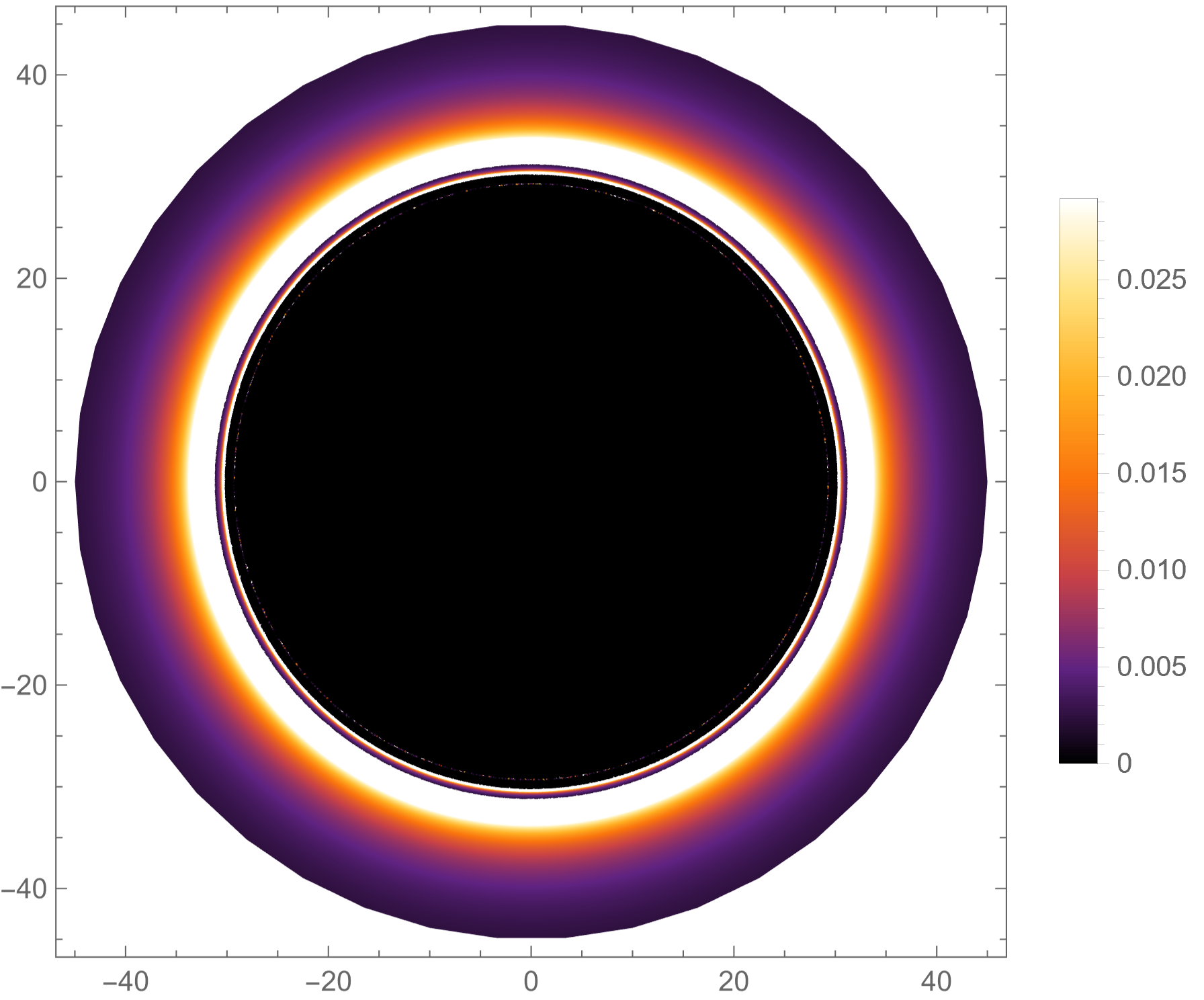}
        	\includegraphics[width = 6.5 cm]{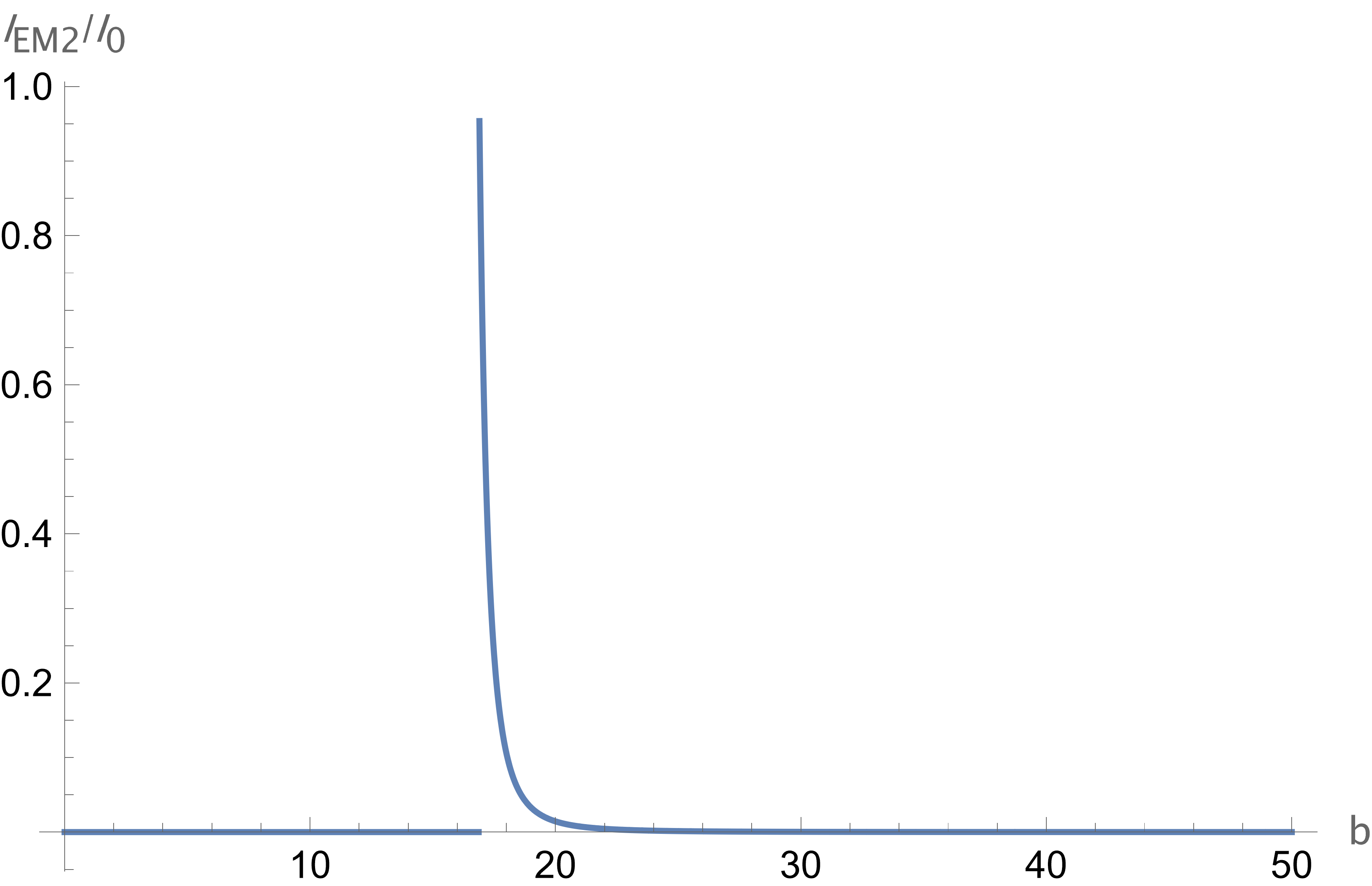}
        	\includegraphics[width = 6 cm]{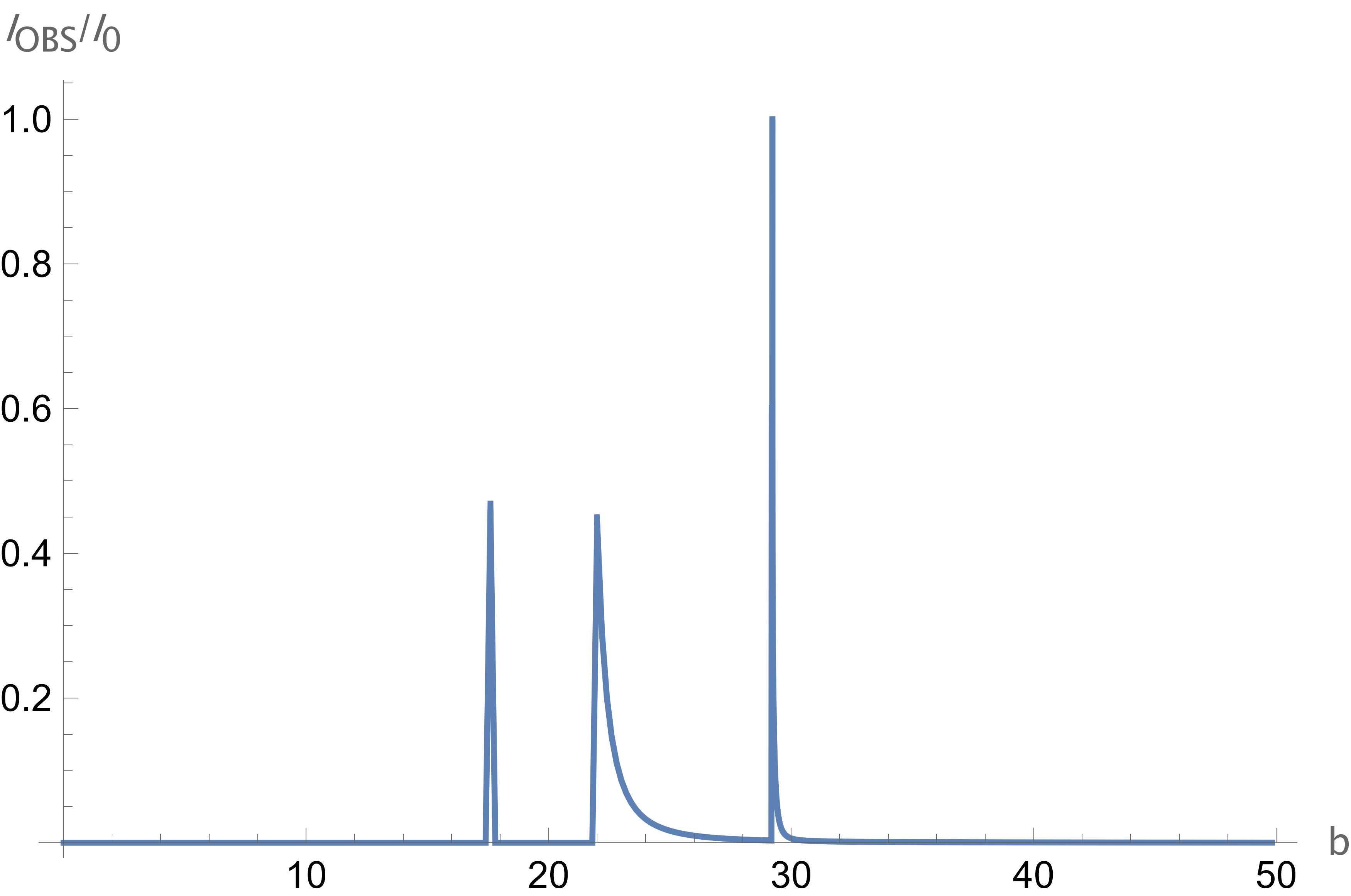}
        	\includegraphics[width = 5 cm]{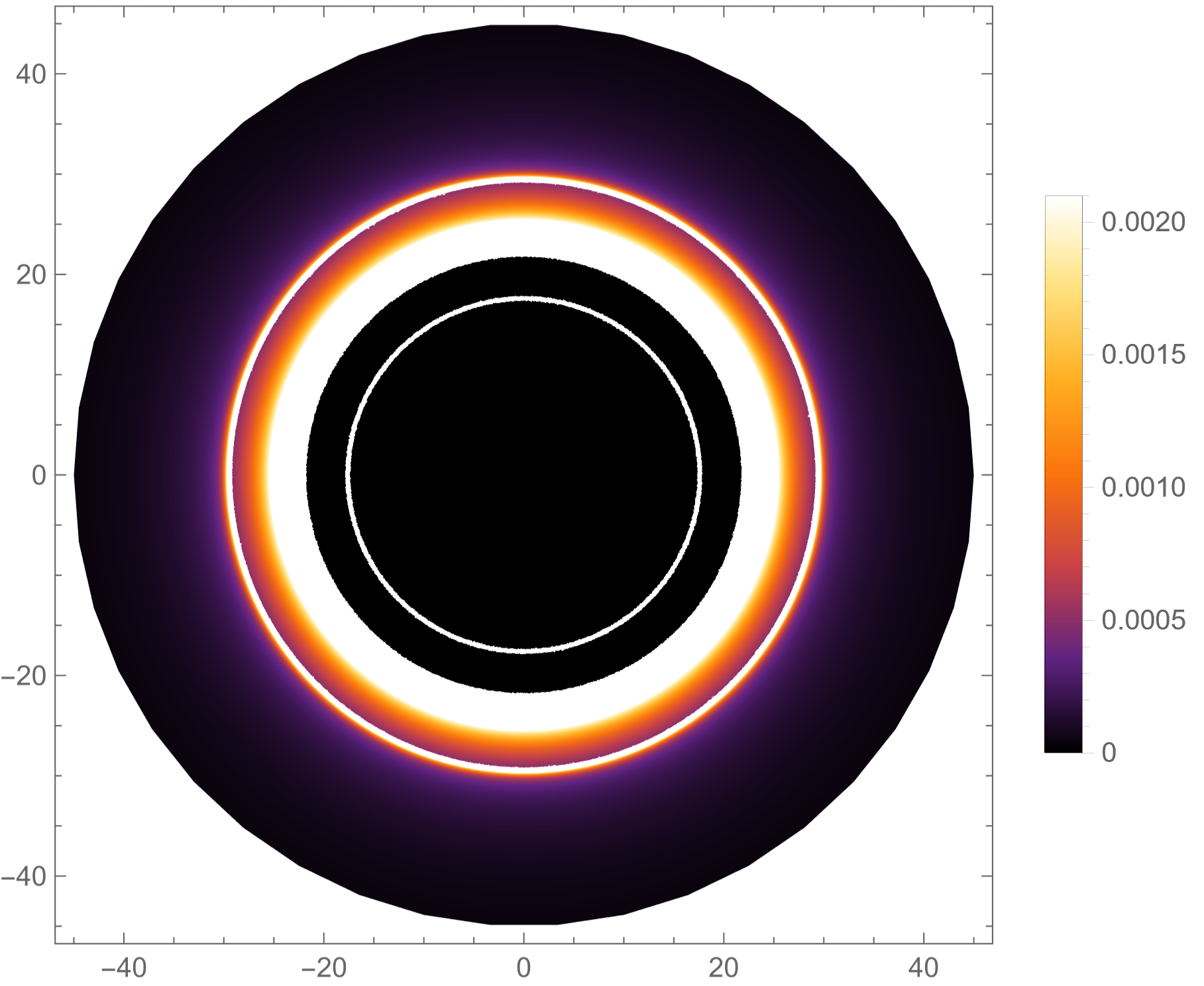}
        	\includegraphics[width = 6.5 cm]{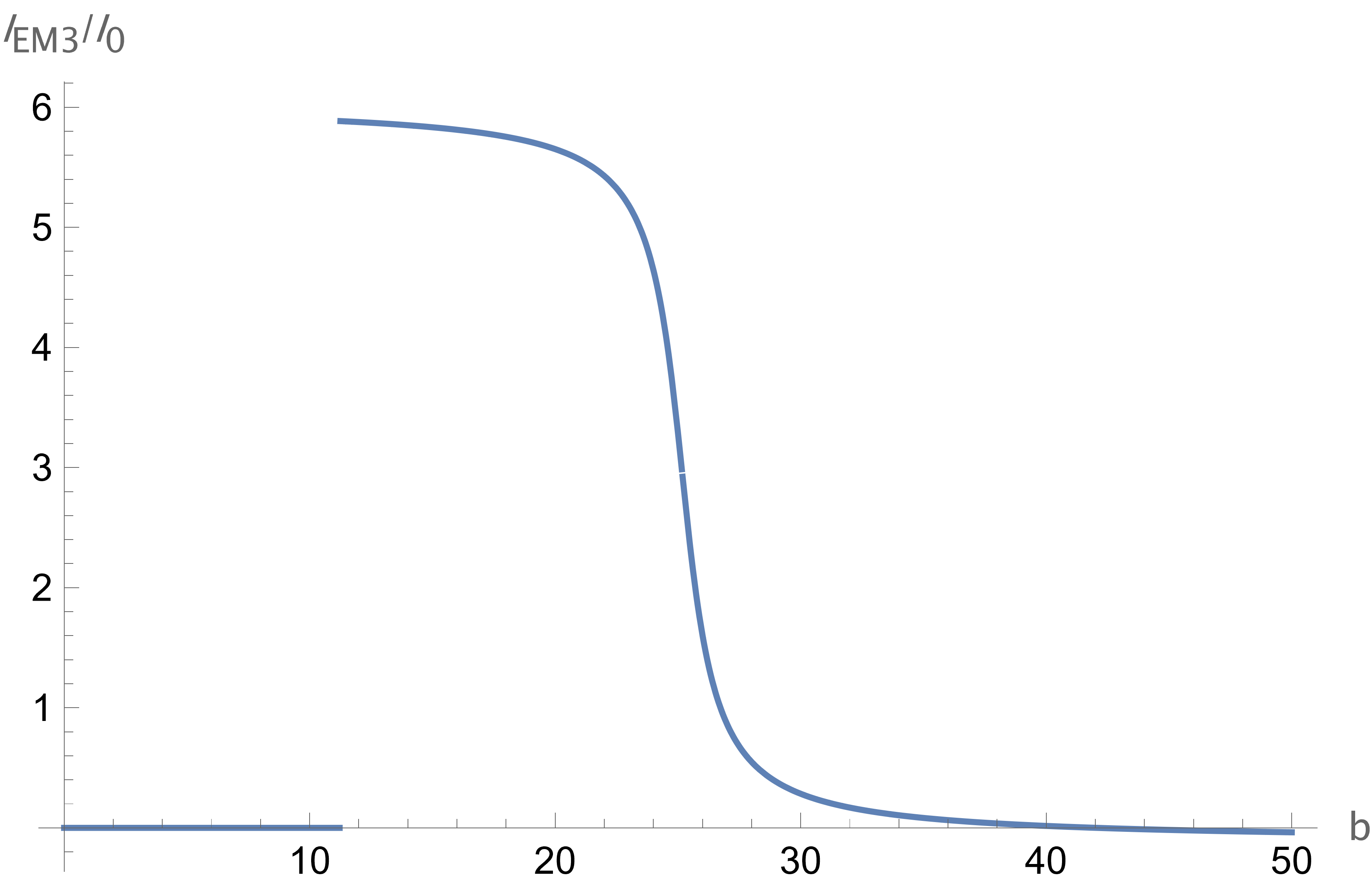}
        	\includegraphics[width = 6 cm]{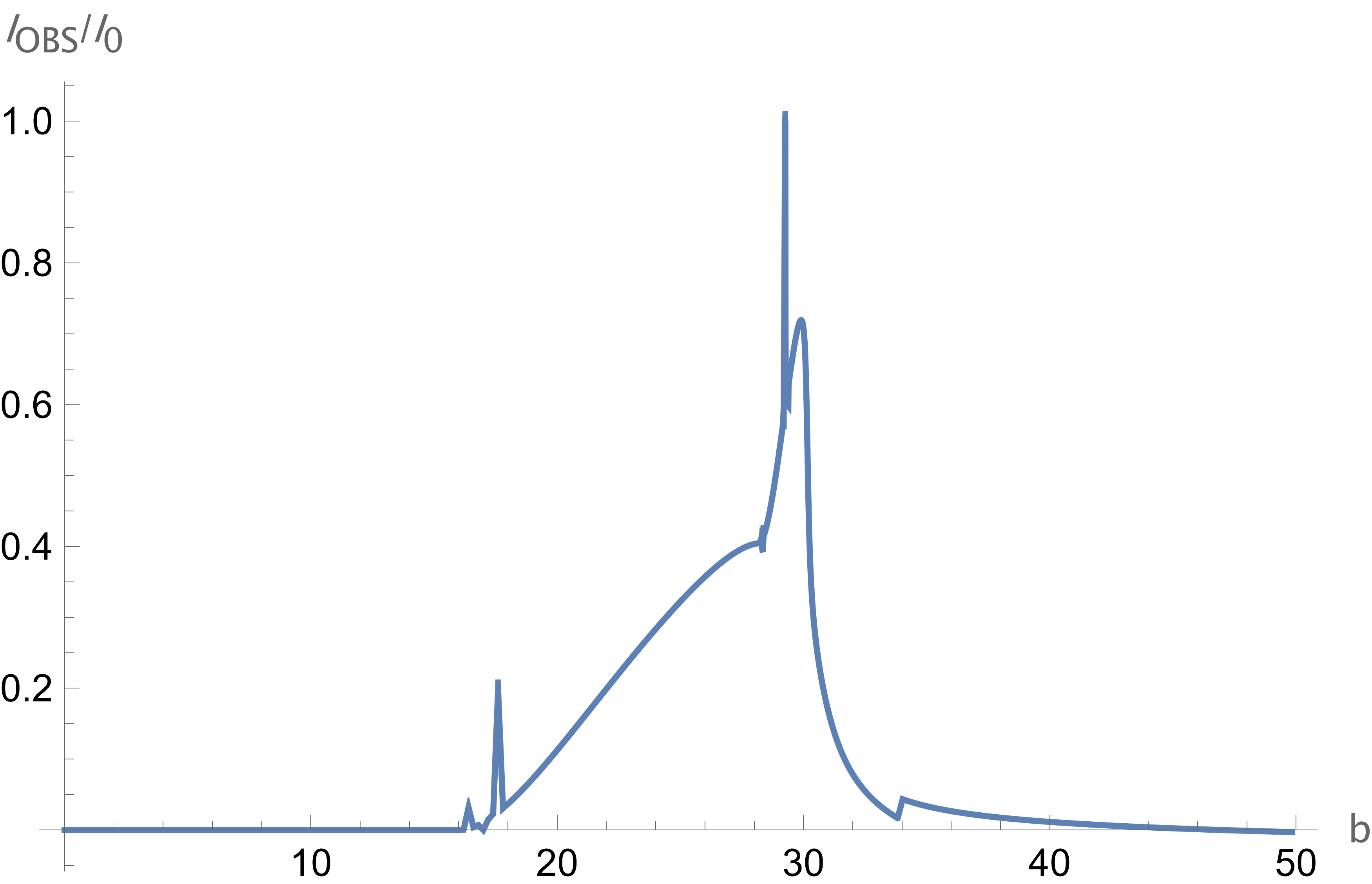}
        	\includegraphics[width = 5 cm]{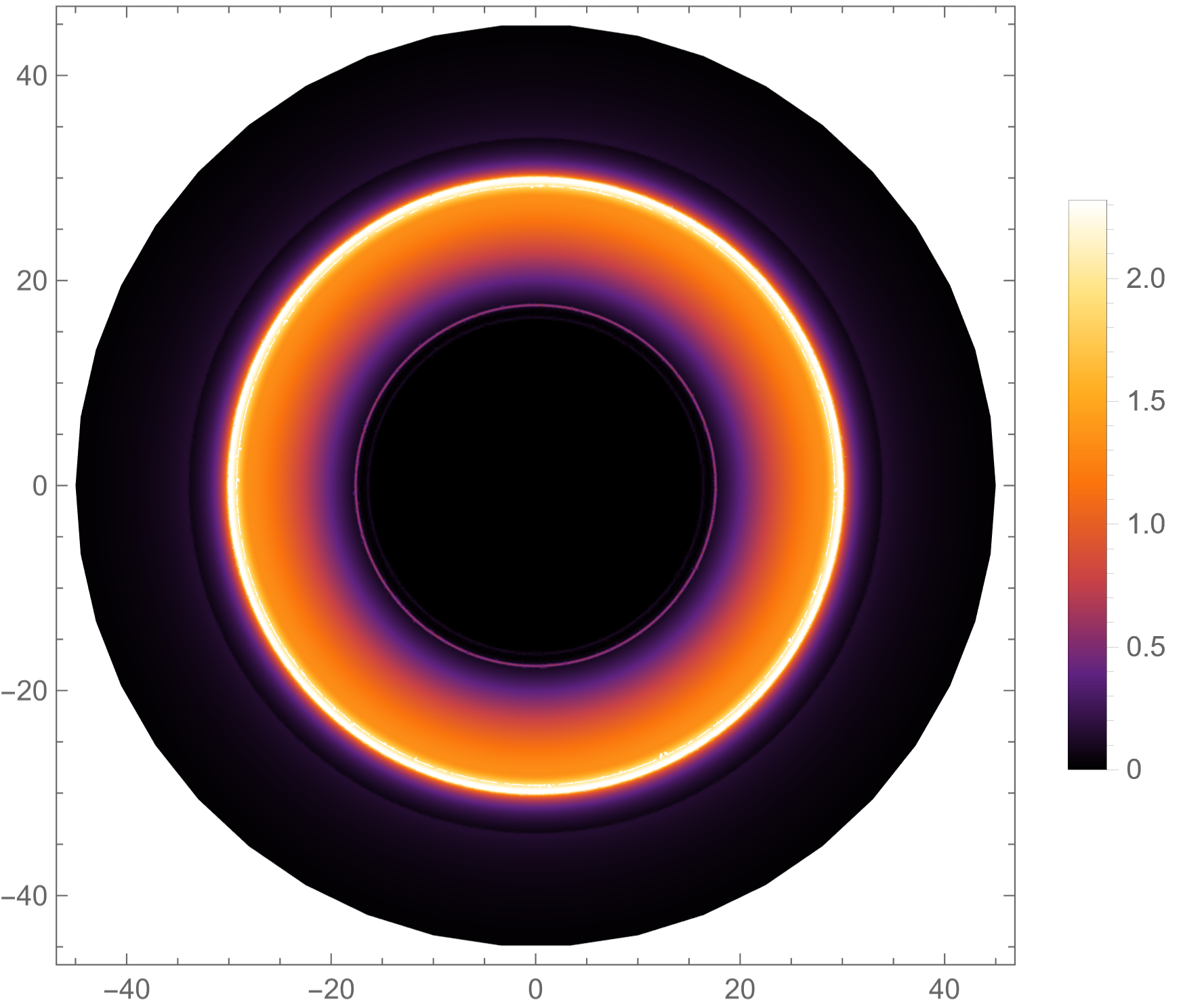}
                \caption{Observational appearance of an optically thin disk of emission near a DL-NED black hole with charge $P = 0.1$, $\beta = 1$, of all three emission profiles, viewed from a face-on orientation. The emitted and observed intensities $I_{em}$ and $I_{obs}$ are plotted as normalized to the maximum value. We observe the lensing ring around 20M, while the photon ring at 16.9M is barely visible, for only the transfer functions $m \leq 3$. When the emission model begins outside the event horizon, stops at some inner edge (e.g models I and II), the radius of the dark hole is the apparent position of the edge. When the emission extends to the horizon, the radius of the main dark hole becomes the observed radius of the event horizon (here $b \sim 11M$). The analytically calculated shadow radius (or the critical impact parameter) does not correspond to the observed radius of the shadow in these emission models.
                }
        	\label{fig:accmodels}
        \end{figure}

        Now that we obtained an expression for the observed intensity, the only missing ingredient is the $I_{EM}$. We consider three toy models, previously investigated by \cite{Zeng:2021dlj}, and visualize the results in Fig.s \ref{fig:N},\ref{fig:transferFunc},\ref{fig:accmodels}.

        \begin{itemize}
            \item Model I: The emission starts from the innermost stable circular orbit (ISCO), and the emission decays with the second power:
            \begin{equation}
                I^1_{EM}(r) = \begin{cases} 
                                \Big(\frac{1}{r - (r_{isco} - 1)}\Big)^2 & r\geq r_{isco}\\
                                0 & r\leq r_{isco}.
                            \end{cases}
            \end{equation}

            \item Model II: The emission peaks at the photon sphere but the rest of the emission has the similar center and asymptotic characteristics as the first model. But the rate of attenuation with the impact parameter is assumed to be slightly larger, which decays with the third power:
            
            \begin{equation}
                I^2_{EM}(r) = \begin{cases} 
                                \Big(\frac{1}{r - (r_{p} - 1)}\Big)^3 & r\geq r_{p}\\
                                0 & r\leq r_{p}.
                            \end{cases}
            \end{equation}

            \item Model III: we consider that the emission starts right off the event horizon, peaks at the photon ring, and decays much slower than the first two models:
            
            \begin{equation}
                I^3_{EM}(r) = \begin{cases} 
                                \frac{1 - \arctan(r-(r_{risco}-1))}{1-\arctan(r_p)} & r \geq r_H\\
                                0 & r\leq r_{H}.
                            \end{cases}
            \end{equation}
        \end{itemize}

        \subsubsection{\label{sec:level5B2}Spherically infalling accretion}

        In this section we study spherically free-falling accretion, following the technique of \cite{Bambi:2013nla}. The accretion is spherical and dynamic, contrary to the previous section, which was a static disc. For this dynamic model we again employ the number of orbits formalism of \cite{Gralla:2019xty}, but now the crossings are not with the equatorial plane but all throughout the spherical accretion, the sum in Eq. \ref{eq:totalI} becomes an integral over the null geodesic $\gamma$:

        \begin{equation}
            I(\nu_{obs},b_\gamma) = \int_\gamma g^3 j(\nu_e) dl_{prop},
            \label{eq:bambiI}
        \end{equation}

        where we compute the intensity observed at some specific frequency $\nu_{obs}$ from the null geodesic of impact parameter $b_\gamma$, $j$ is the emissivity per unit volume, and $dl_{prop}$ is the infinitesimal proper length and we must alter the redshift factor $g$:
        
        \begin{equation}
            g = \frac{k_\mu u^\mu_o}{k_\mu u^\mu_e}, \; k^\mu = \dot{x}_\mu,
        \end{equation}
        
        where $k^\mu$ is the 4-velocity of the photon, $u^\mu_o$ is the 4-velocity of the static observer at infinity. The $u^\mu_e$ is the 4-velocity of the infalling accretion:
        
        \begin{equation}
            u^\mu_e = \Big(\frac{1}{f(r)}, -\sqrt{1-f(r)}, 0, 0\Big),
        \end{equation}
        
        and the 4-velocity of the photon is 
        
        \begin{equation}
            k_t = \frac{1}{b}, \; k_r = \pm \frac{1}{bf(r)}\sqrt{1 - f(r)\frac{b^2}{r^2}},
        \end{equation}
        
        which allows us to write
        
        \begin{equation}
            g = \Big( u_e^t + \frac{k_r}{k_t}u_e^r \Big)^{-1}.
        \end{equation}
        
        The proper distance along a null geodesic $\gamma$ must be parametrized by an affine parameter other than the proper time, which is given by
        
        \begin{equation}
            dl_\gamma = k_\mu u^\mu_e d\lambda = \frac{k^t}{g |k_r|}dr.
        \end{equation}

        For the simplicity of the model we assume a monochromatic emission with rest-frame frequency $\nu_*$ and a $\frac{1}{r^2}$ radial profile:
        
        \begin{equation}
            j(\nu_e) \propto \frac{\delta(\nu_e - \nu_*)}{r^2},
        \end{equation}
        
        where $\delta$ is the delta function. Then integrating Eq. \ref{eq:bambiI} over all frequencies yields the total observed flux 
        
        \begin{equation}
            F(b_\gamma) \propto \int_\gamma \frac{g^3}{r^2} \frac{k_e^t}{k_e^r} dr,
        \end{equation}
        
\begin{figure}
        	\centering
        	\includegraphics[width = 8 cm]{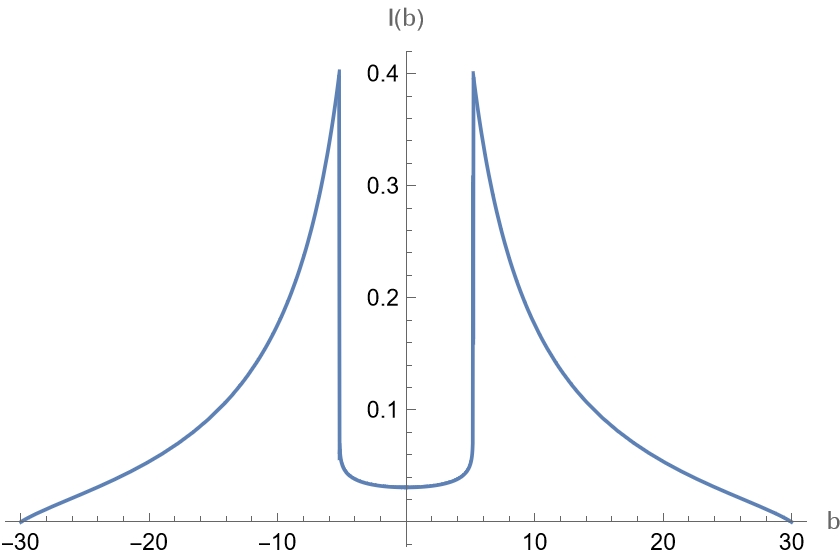}
        	\includegraphics[width = 6 cm]{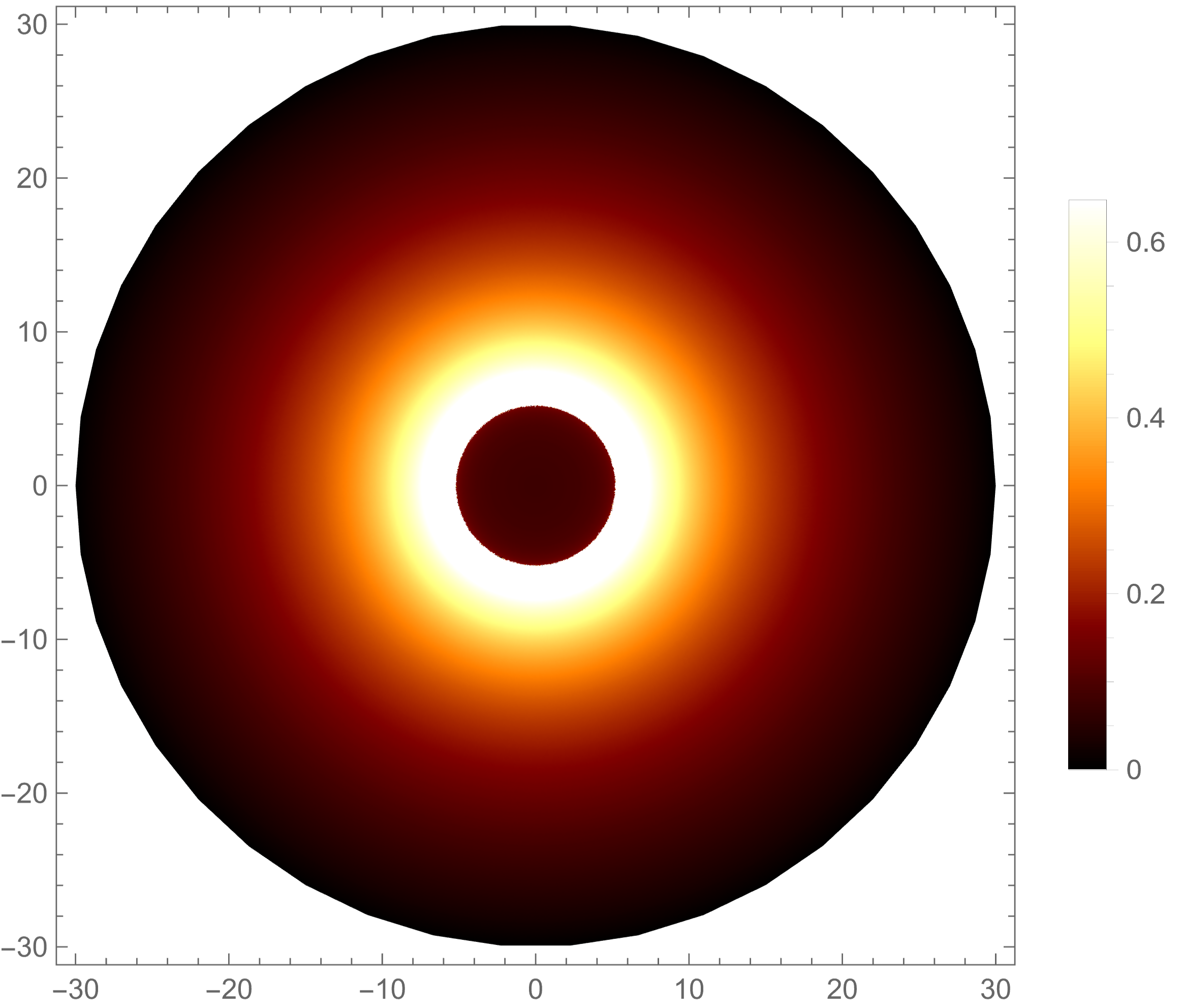}
        	\includegraphics[width = 8 cm]{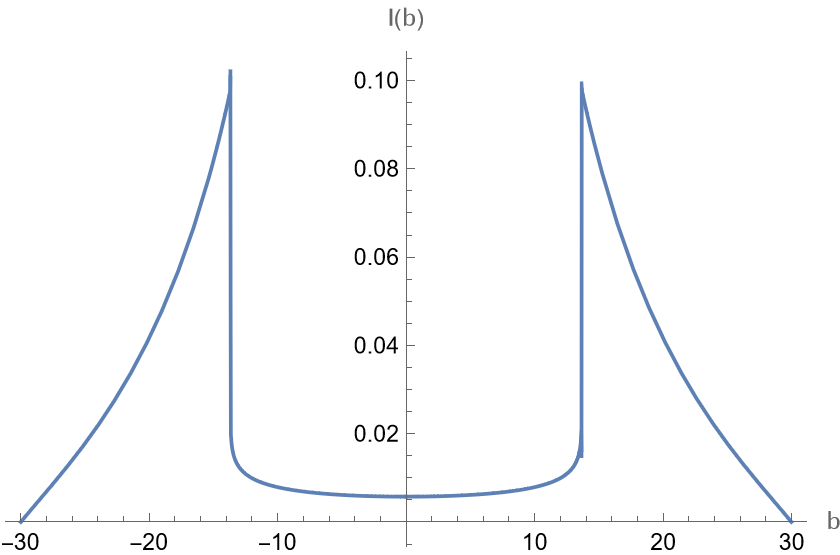}
        	\includegraphics[width = 6 cm]{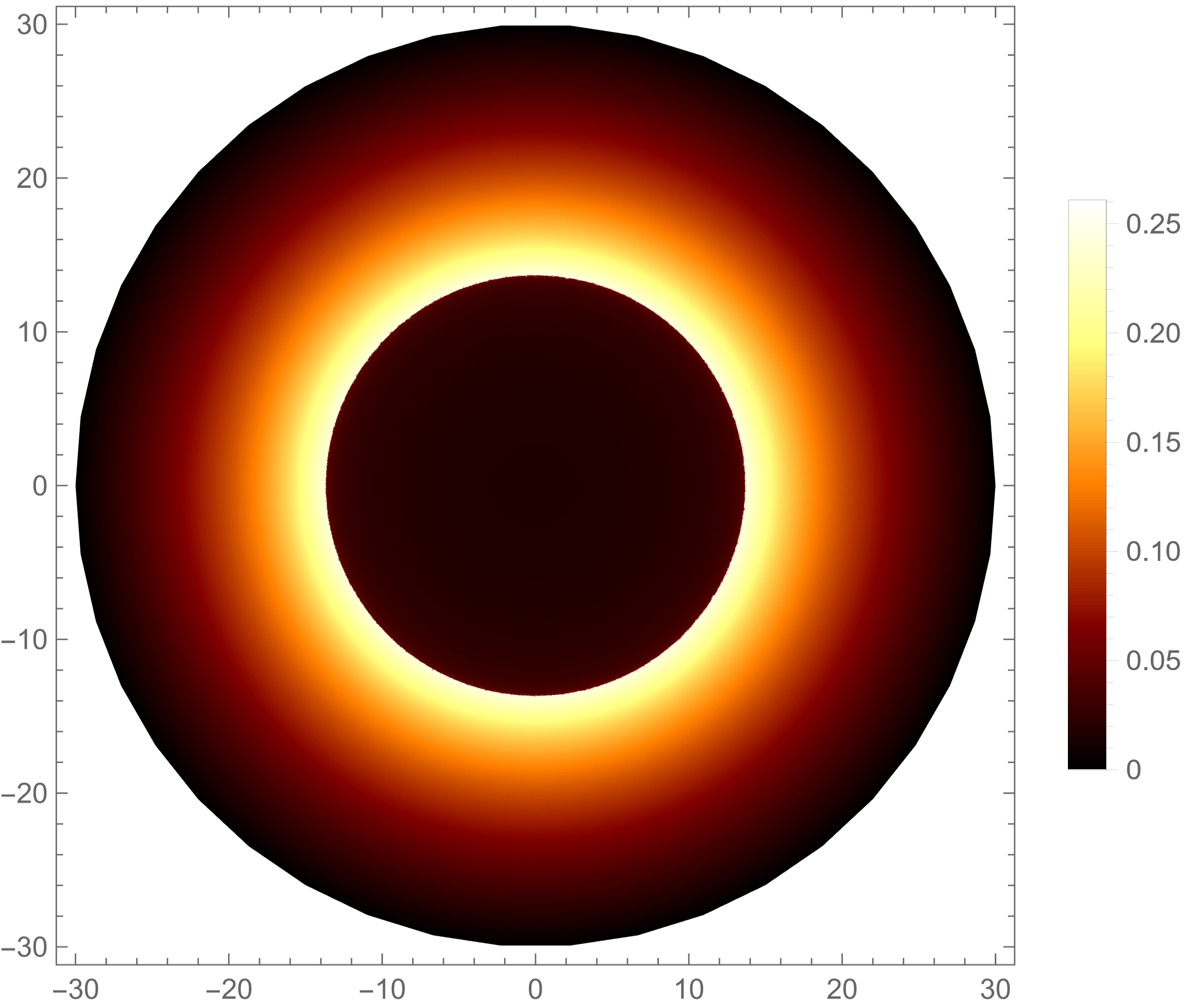}
            \caption{Observational appearance of a spherically free-falling accretion emission near a         Schwarzschild (upper) and DL-NED black hole (lower) of charge P = 0.1, $\beta = 1$. The mass for all black holes hereafter are taken to be unity. It is seen that the introduction of a charge term greatly increases the apparent size of the 'shadow', but decreases the intensity of the incoming light.
                }
        	\label{fig:sphericalSch}
        \end{figure}

\begin{figure}
        	\centering
        	\includegraphics[width = 8 cm]{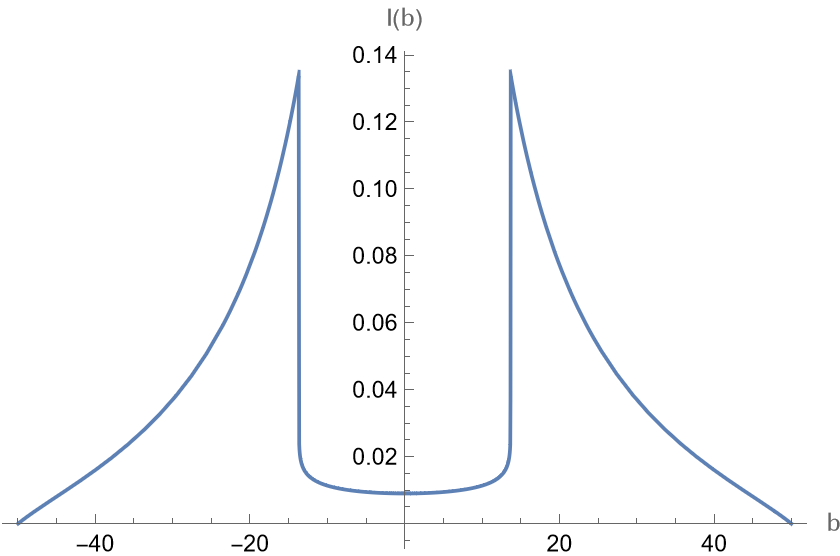}
        	\includegraphics[width = 6 cm]{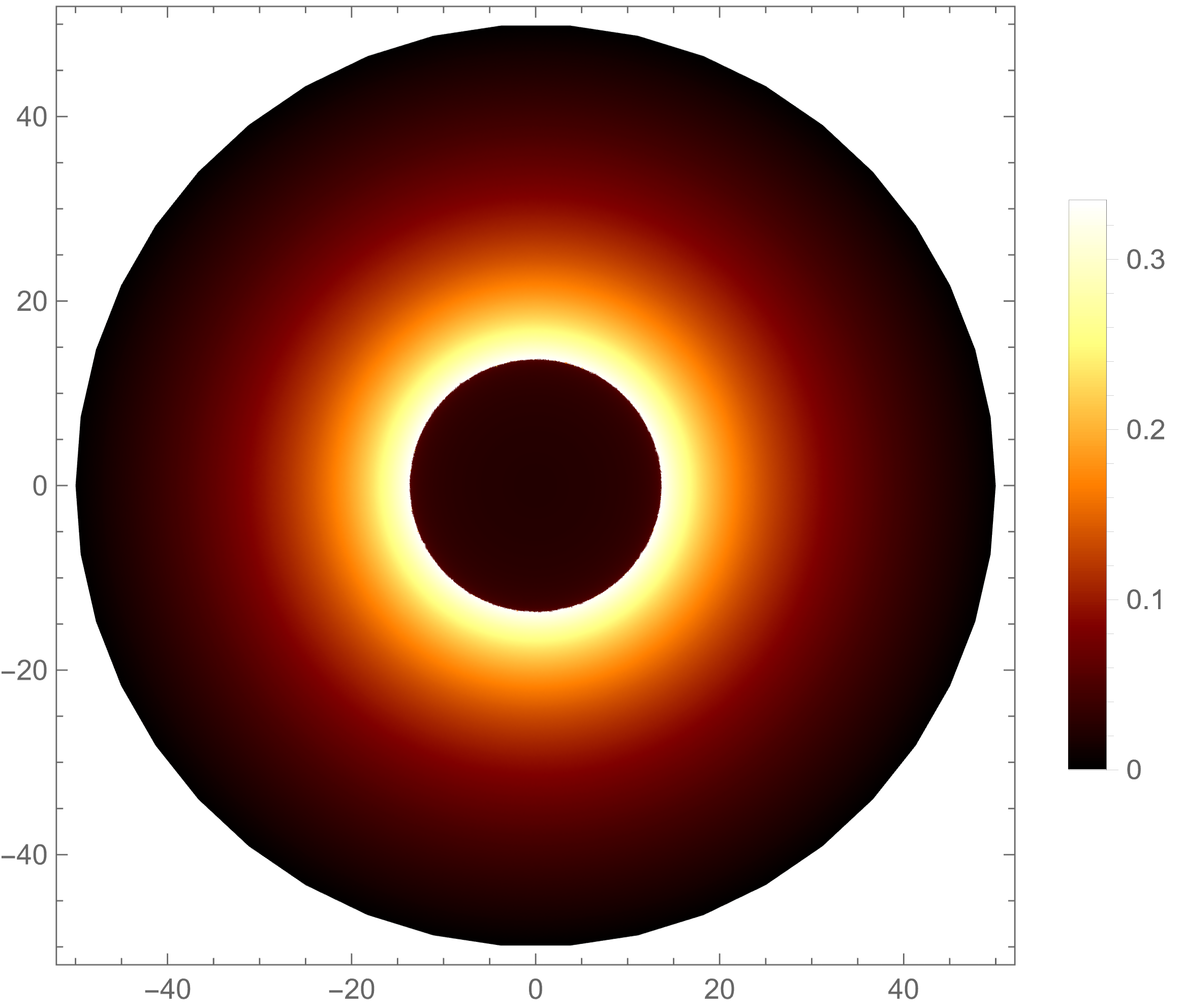}
        	\includegraphics[width = 8 cm]{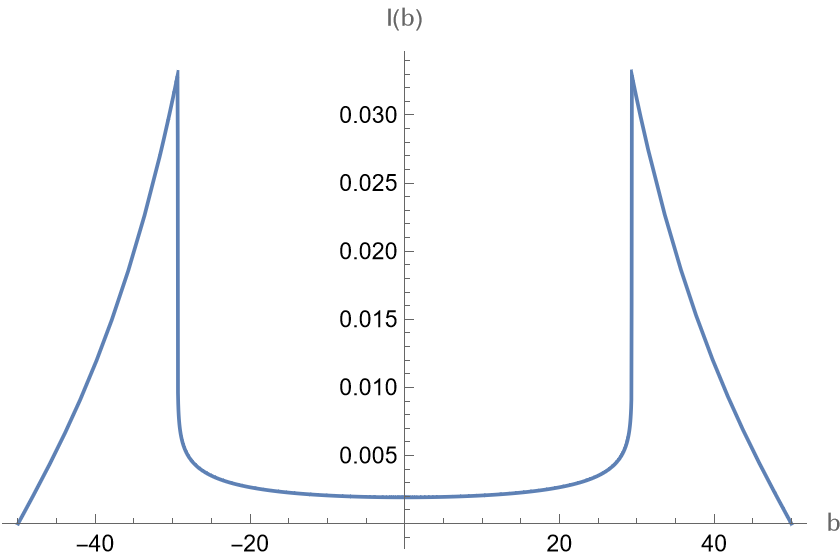}
        	\includegraphics[width = 6 cm]{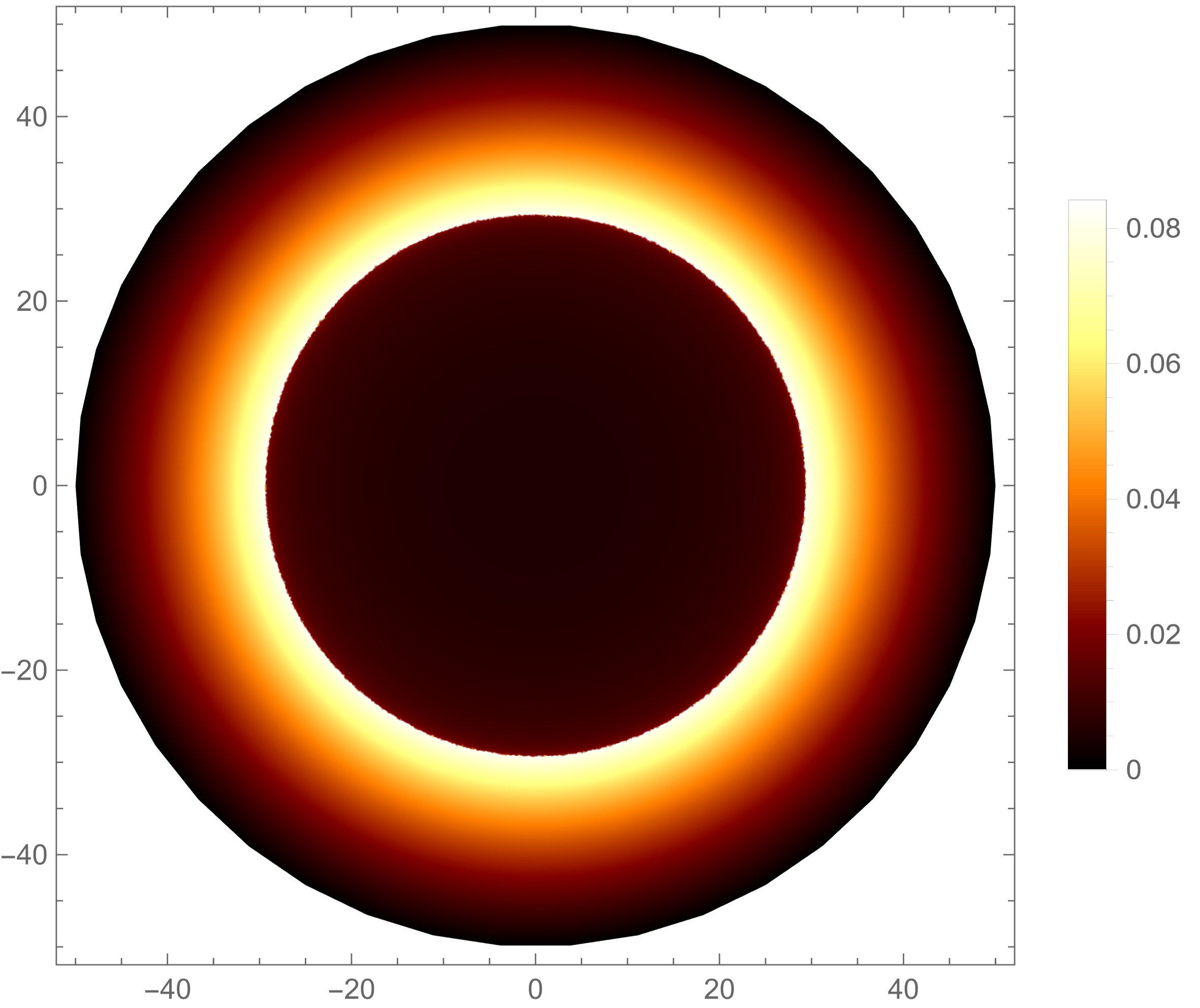}
            \caption{Observational appearance of a spherically free-falling accretion emission near a DL-NED black hole of charge P = 0.1 (upper) and P = 0.2 (lower), fixed $\beta = 1$. Similar to the comparison to the Schwarzschild black hole case, the increase of the charge term greatly increases the size of the black hole shadow. However the intensity is reduced greatly.
                }
        	\label{fig:chargecomparison}
        \end{figure}

\begin{figure}
        	\centering
        	\includegraphics[width = 8 cm]{IntensityBH0.1fov50b1}
        	\includegraphics[width = 6 cm]{SphericalBH0.1fov50b1}
        	\includegraphics[width = 8 cm]{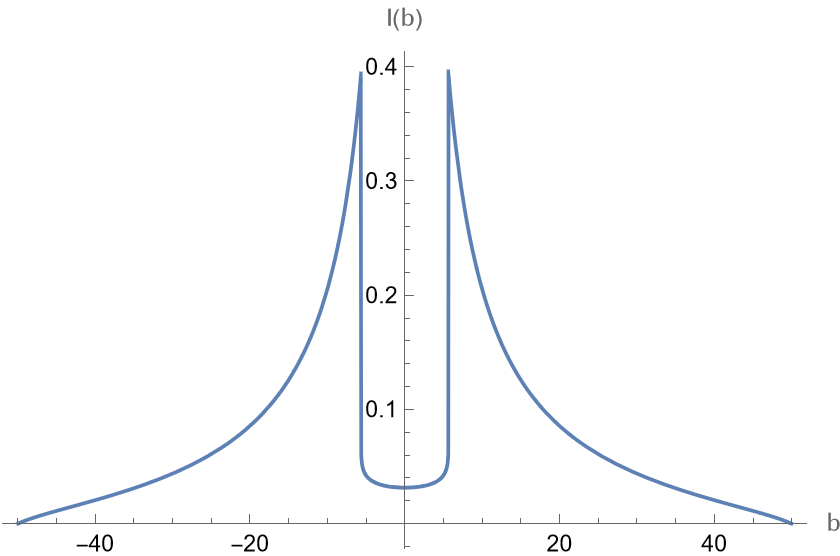}
        	\includegraphics[width = 6 cm]{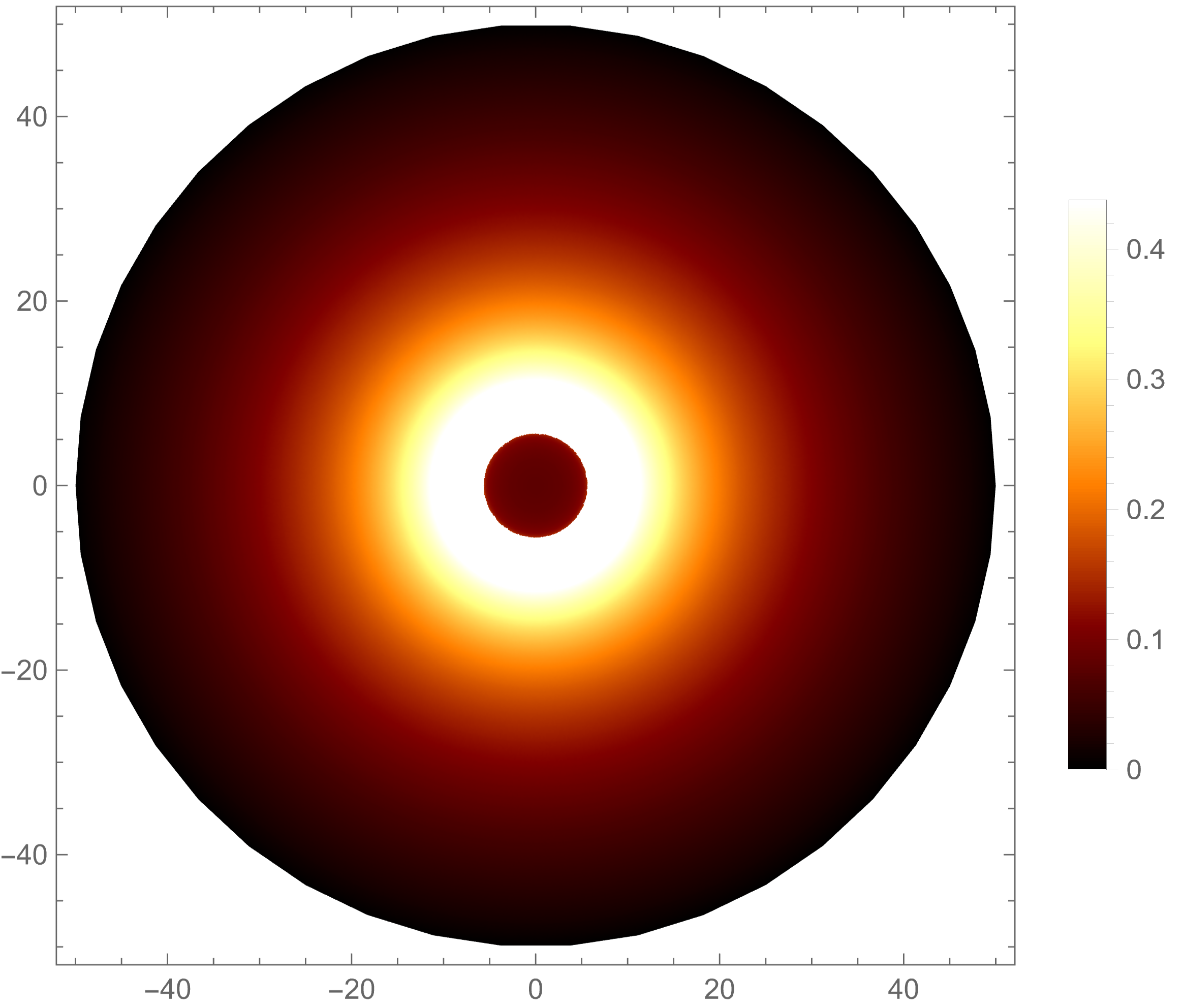}
            \caption{Observational appearance of a spherically free-falling accretion emission near a         DL-NED black hole of charge with P = 0.1 with different values of the field coupling constant $\beta$, $\beta =1 $ (upper) and $\beta = 10000$ (lower). Contrary to the charge, an increase in $\beta$ actually decreases the size of the black hole shadow, although minimally. This results in an increase in intensity.
                }
        	\label{fig:betacomparison}
        \end{figure}

        With the expressions for the flux, we created the \textit{Mathematica} code and numerically integrated the flux to see the effects of the parameters of the DL-NED theory. See figures (\ref{fig:sphericalSch}, \ref{fig:chargecomparison}, \ref{fig:betacomparison}) for examples. For an implementation in Python, see \textit{EinsteinPy} \cite{Bapat:2020xfa}.
        

\section{\label{sec:level6}Time-Domain Profile and Quasinormal modes (QNMs)}  

    In this section we wish to obtain the quasinormal modes of the DL-NED spacetime (\ref{metric_func}), by solving the wave equation in this background. An exact solution does not seem to be available, so we resort to numerical methods to approximate the frequencies. The most popular in the literature seems to be the WKB method. We further confirm the accuracy of the WKB method, by solving the wave equation itself numerically and extracting the fundamental mode using the Prony method \cite{Prony:1987}.
    
    \subsection{QNMs with the WKB method}
    
        We consider the evolution of massless scalar waves in the background, which obey the Klein-Gordon equation:
        \begin{equation}
            \frac{1}{\sqrt{-g}} \partial_{\mu}\left(\sqrt{-g} g^{\mu \nu} \partial_{v} \Phi\right)=0,
        \end{equation}
        where $m = 0$, and the covariant derivatives are expressed in terms of the metric. Due to the spherical symmtery of our metric (\ref{spacetime_metric}), we attempt to decompose the solution into spherical harmonics:
        
        \begin{equation}
            \Phi=\sum_{\ell, m} r^{-1} \Psi(t, r) Y_{\ell, m}(\theta, \phi).
        \end{equation}
        
        With this ansatz the wave equation becomes
        
        \begin{equation}
            \left(\frac{\partial^{2}}{\partial t^{2}}-\frac{\partial^{2}}{\partial r_{*}^{2}}+V(r_*)\right) \Psi=0,
            \label{eq:RWeq}
        \end{equation}
        
        where $r_*$ is the so-called tortoise coordinate given by $r_* = \int \frac{1}{f(r)} dr$. It can be analytically integrated, in terms of the horizons (\ref{eq:horizons}):
        
        \begin{equation}
            r_* = r - r_c^2\frac{\log{r-r_c}}{r_c-r_h} - r_h^2\frac{\log{r-r_h}}{r_c-r_h}.
        \end{equation}
        
        The potential $V_\ell$ is known as the Regge-Wheeler potential, given by 
        
        \begin{equation}
            V_{\ell}(r)=f(r)\left(\frac{\ell(\ell+1)}{r^{2}}+\frac{1-s^2}{r} \frac{\partial f}{\partial r}\right),
            \label{eq:RWpot}
        \end{equation}
        
        where s is the spin of the perturbation, which is s = 0 for the scalar case. If we assume a temporal dependence of the wave function of form
        
        \begin{equation}
            \Psi(t, r)=e^{-i \omega t} \psi(r),
        \end{equation}
        
        the equation (\ref{eq:RWeq}) becomes
        
        \begin{equation}
            \frac{d^{2} \psi}{d r_{*}^{2}}+\left[\omega^{2}-V(r)\right] \psi=0,
            \label{eq:timeInd}
        \end{equation}
        
        where the time dependence is eliminated and $\omega$ is the QNM frequency we want to figure out.
        To compute $\omega$, we attack (\ref{eq:timeInd}) with the WKB method. This was applied to the first order by Schutz and Will \cite{Schutz:1985km}, extended to third order by Iyer and Clifford in \cite{Iyer:1986np}, finally Konoplya derived the 6$^{th}$ order result in \cite{Konoplya:2003ii}. According to \cite{Iyer:1986np}, the WKB method essentially reduces to solving 
        
        \begin{equation}
            \frac{i\left[\omega^{2}-\left.V\left(r_{*}\right)\right|_{\bar{r}_{*}}\right]}{\sqrt{\left.2 V^{\prime \prime}\left(r_{*}\right)\right|_{\bar{r}_{*}}}}-\sum_{j=2}^{N} \Lambda_{j}(n)=n+\frac{1}{2},
        \end{equation}
        
        where $\bar{r}_{*}$ is the location of the maximum of the RW potential (\ref{eq:RWpot}), $V'(r_*) := \dv{V}{r_*}$.  $\Lambda_{j}(n)$ are the WKB correction terms and $N$ is the order. The corrections $\Lambda_{2,3}$ are given in \cite{Iyer:1986np} and $\Lambda_{4,5,6}$ can be found in \cite{Konoplya:2003ii}. Solving this equation, we obtain the modes to the 6$^{th}$ order, which we present below in Table \ref{table:qnm1}. There, we provide the QNM complex modes for different values of the overtone number $n$ and the angular momentum  $l$. As indicated in \cite{Konoplya:2003ii}, the WKB method works more accurately for lower values of $n$ and higher values of $l$. \\
        \begin{table}[]
            \centering
            \begin{tabular}{cccc}
                \multicolumn{3}{l} { Table I: $\omega$ for different values of $P$ using $6^{\text {th}}$ order WKB method \footnote{We are grateful to the R.~G.~Daghigh, M.~D.~Green and H. El Moumni for providing \textit{Mathematica} notebooks to calculate the QNMs.}} \\
                \hline  $n, l$ & $P=0$            & $P=0.1$             & $P=0.2$ \\
                \hline  0,0 & $0.22093-0.20164 i$ & $0.2197 - 0.1987 i$ & $0.2198 - 0.1991 i$ \\
                        1,0 & $0.17805-0.68910 i$ & $0.1779 - 0.6776 i$ & $0.1778 - 0.6795 i$ \\
                        
                \hline  0,1 & $0.58582-0.19552 i$ & $0.5823 - 0.1927 i$ & $0.5828 - 0.1931 i$ \\
                        1,1 & $0.52894-0.61304 i$ & $0.5272 - 0.6036 i$ & $0.5275 - 0.6048 i$ \\
                        2,1 & $0.46203-1.0843 i$  & $0.4624 - 1.0664 i$ & $0.4624 - 1.0687 i$ \\
                        
                \hline  0,2 & $0.96728-0.19353 i$ & $0.9614 - 0.1908 i$ & $0.9622 - 0.1911 i$ \\
                        1,2 & $0.92769-0.59125 i$ & $0.9231 - 0.5826 i$ & $0.9237 - 0.5837 i$ \\
                        2,2 & $0.86077-1.0174 i$  & $0.8583 - 1.0018 i$ & $0.8586 - 1.0039 i$ \\
                        3,2 & $0.78641-1.4798 i$  & $0.7864 - 1.456 i$  & $0.7864 - 1.4592 i$ \\
                        4,2 & $0.72517-1.9766 i$  & $0.7273 - 1.9431 i$ & $0.7271 - 1.9475 i$ \\
                        
                \hline  0,6 & $2.5038-0.19261 i$ & $2.4884 - 0.1899 i$ & $2.4905 - 0.1902 i$ \\
                        1,6 & $2.4875-0.57947 i$ & $2.4727 - 0.5712 i$ & $2.4747 - 0.5723 i$ \\
                        2,6 & $2.4557-0.97120 i$ & $2.4419 - 0.9572 i$ & $2.4437 - 0.9591 i$ \\
                        3,6 & $2.4098-1.3708 i$  & $2.3975 - 1.3509  i$& $2.3991 - 1.3535 i$ \\
                        4,6 & $2.3519-1.7809 i$  & $2.3414 - 1.7547 i$ & $2.3428 - 1.7581 i$ \\
                        5,6 & $2.2848-2.2036 i$  & $2.2765 - 2.1706 i$ & $2.2776 - 2.1749 i$ \\
                        6,6 & $2.2113-2.6402 i$  & $2.2054 - 2.5999 i$ & $2.2062 - 2.6052 i$ \\
                        7,6 & $2.1347-3.0914 i$  & $2.1313 - 3.0433 i$ & $2.1317 - 3.0496 i$ \\
                        8,6 & $2.0579-3.5575 i$  & $2.057 - 3.501 i$   & $2.0572 - 3.5085 i$ \\
                        9,6 & $1.9839-4.0383 i$  & $1.9855 - 3.9728 i$ & $1.9854 - 3.9815 i$ \\
                        10,6 & $1.9155-4.5334 i$ & $1.9196 - 4.4583 i$ & $1.9191 - 4.4681 i$ \\
                        11,6 & $1.8555-5.0421 i$ & $1.862 - 4.9566 i$ &  $1.8612 - 4.9678 i$
            \end{tabular}
            \label{table:qnm1}
        \end{table}
    
\begin{figure}
        	\centering
        	\includegraphics[width = 8 cm]{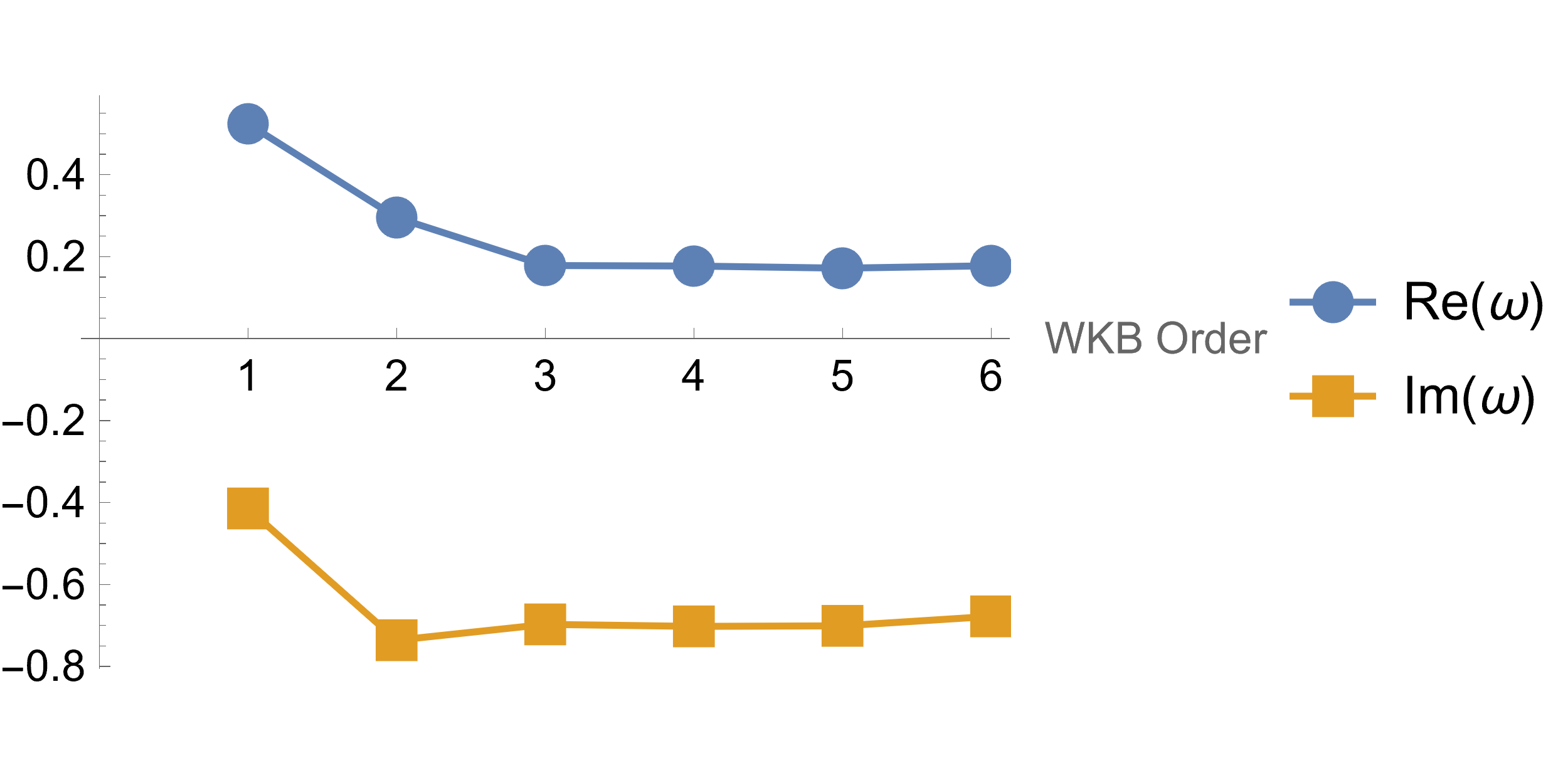}
        	\includegraphics[width = 8 cm]{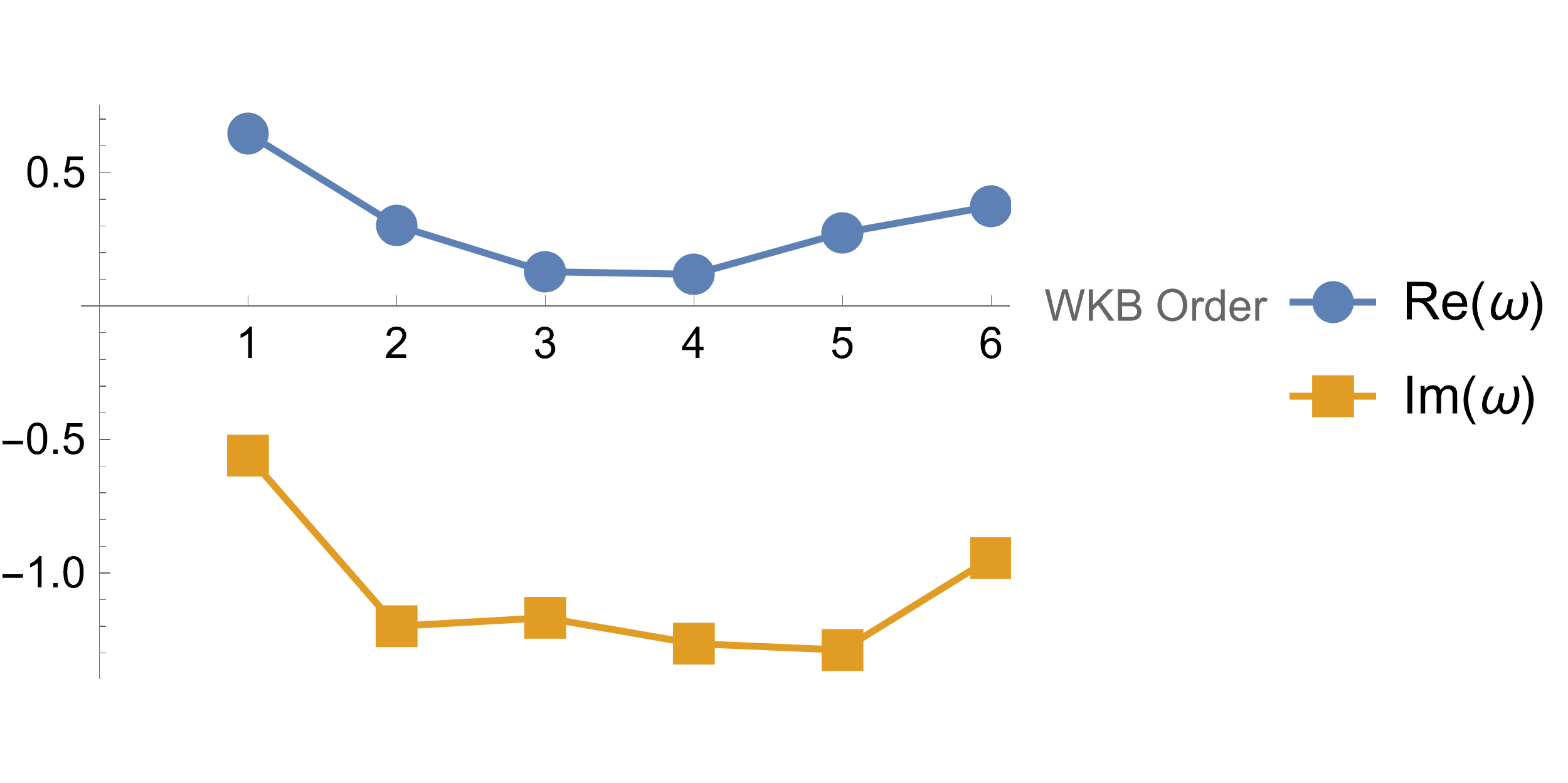}
        	\includegraphics[width = 8 cm]{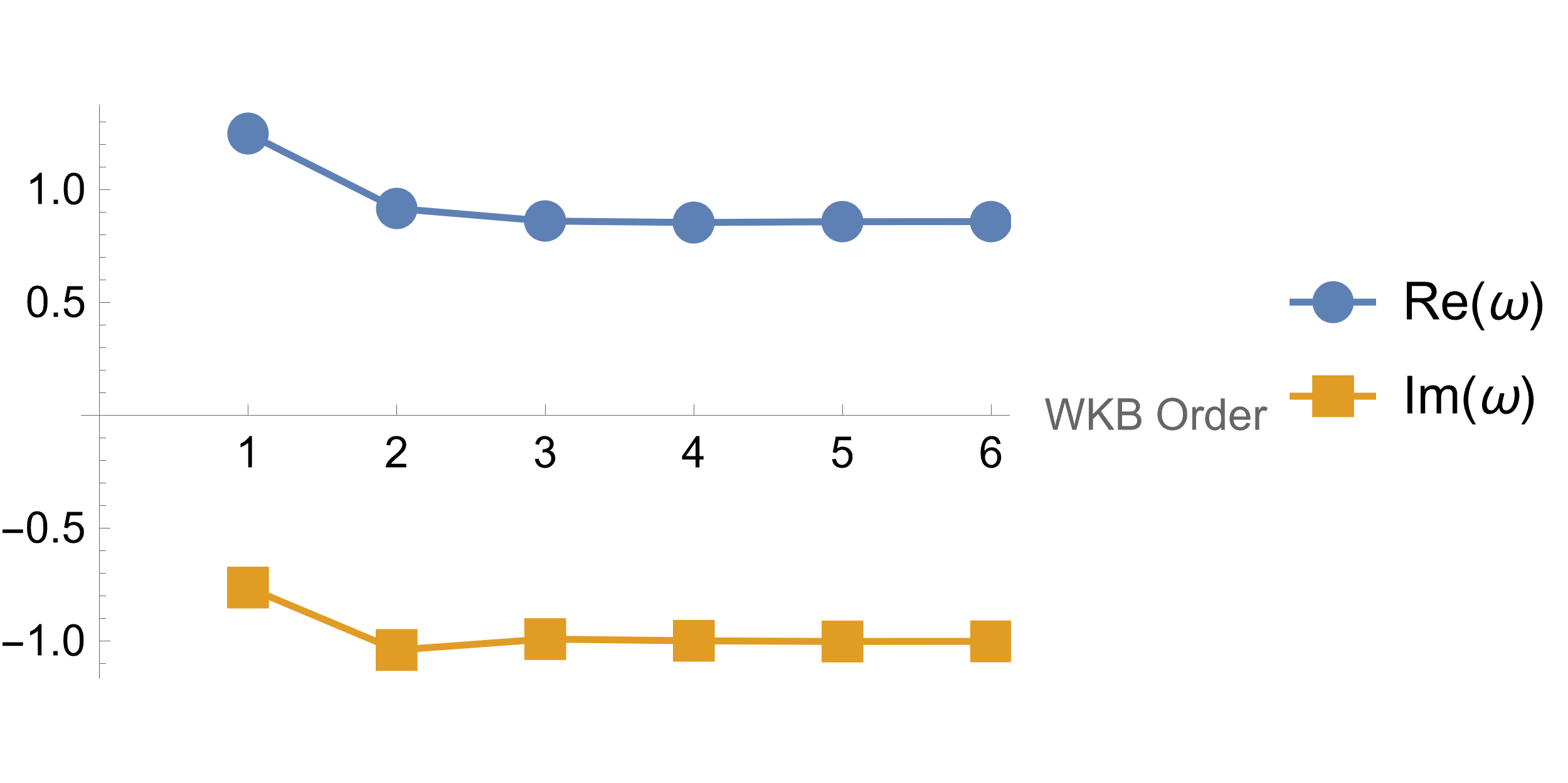}
        	\includegraphics[width = 8 cm]{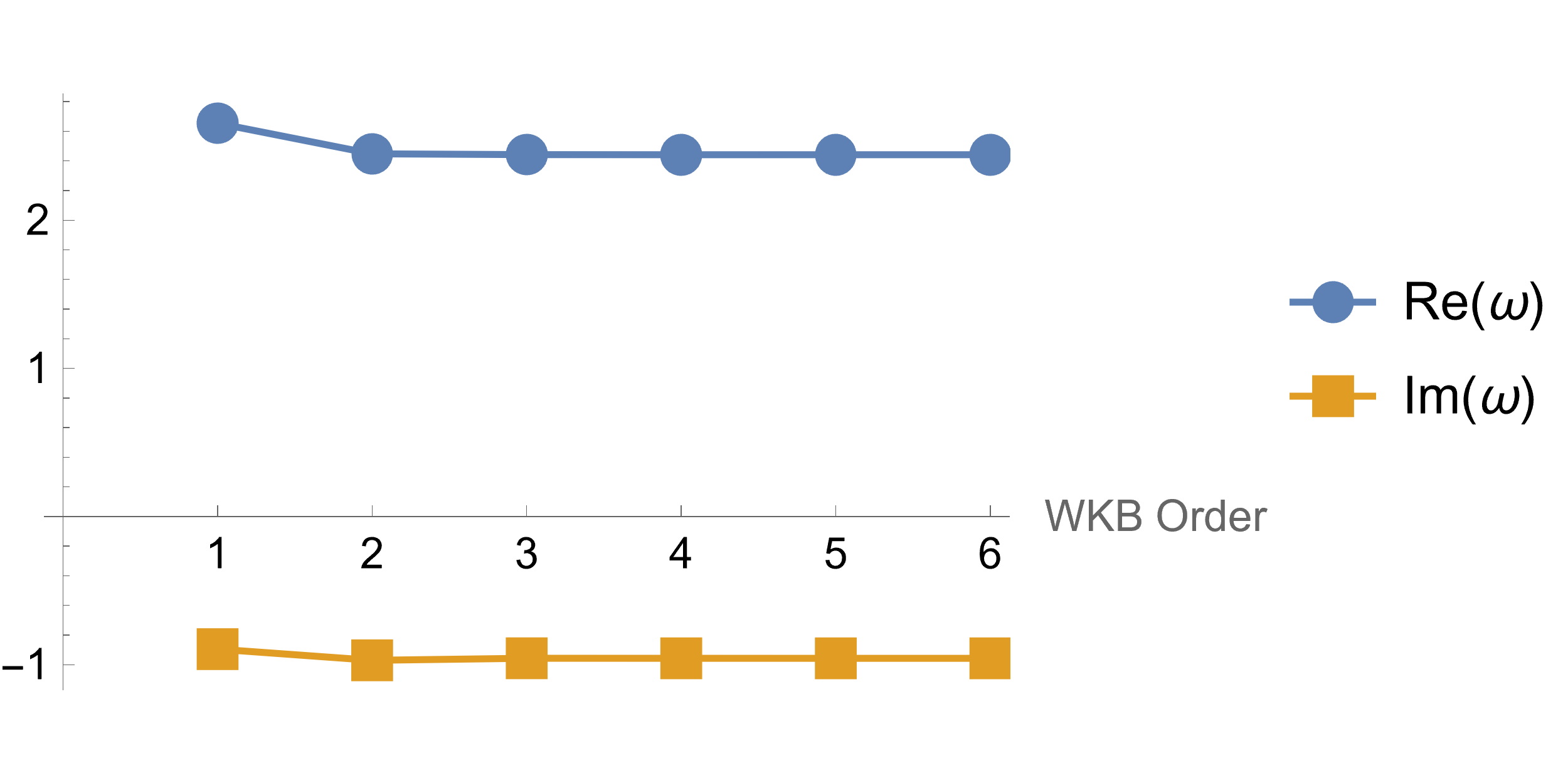}
            \caption{
            The real and imaginary parts of the QNM frequencies, versus the order of the WKB method, with different value of ($n,\,l$). Top left: $n = 1,\, l = 0$. We see that the values converge. Top right: $n = 2,\, l = 0$. the values still fluctuate, even at the highest order. Bottom left: $n = 2,\, l = 2$. Frequences seem to have converged after order 3. Bottom right: $n = 2,\, l = 6$. The values are more stable than any of the other methods.
                }
        	\label{fig:convergenceWKB}
        \end{figure}

        To test the accuracy of the WKB method for certain calculations of modes, we look for convergence of the real and imaginary components of the modes, with respect to the order of the WKB method. In Fig. \ref{fig:convergenceWKB}, we show convergence for various values of ($n,\,l$) pairs. 
        
        As mentioned in \cite{Daghigh:2020fmw}, we expect that the WKB method is reliable when the angular momentum is high and the number of overtones $n$ is low, which is confirmed by Fig. \ref{fig:convergenceWKB}. The most unstable solution is the one with $n = 2 > l = 0$ whereas the most stable is the one with $ n = 2 < l = 6$. As the WKB method is merely an approximation, we wish to compute the QNMs with another method as a way to gauge WKB's accuracy. 
            
        \subsection{\label{sec:level6B}Analyzing the QNM spectrum}
\begin{figure}
        	\centering
        	\includegraphics[width = 10 cm]{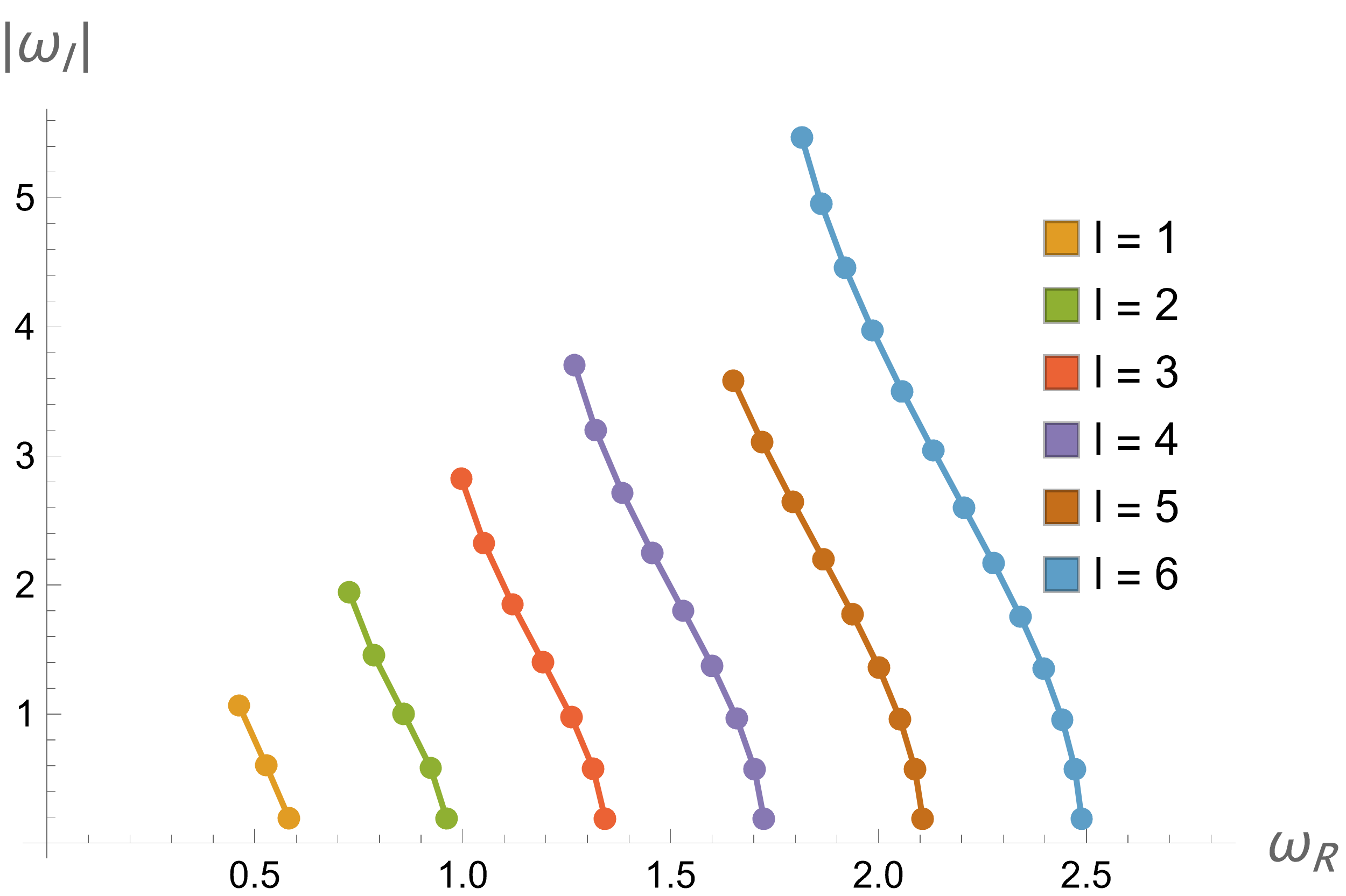}
        	\includegraphics[width = 8 cm]{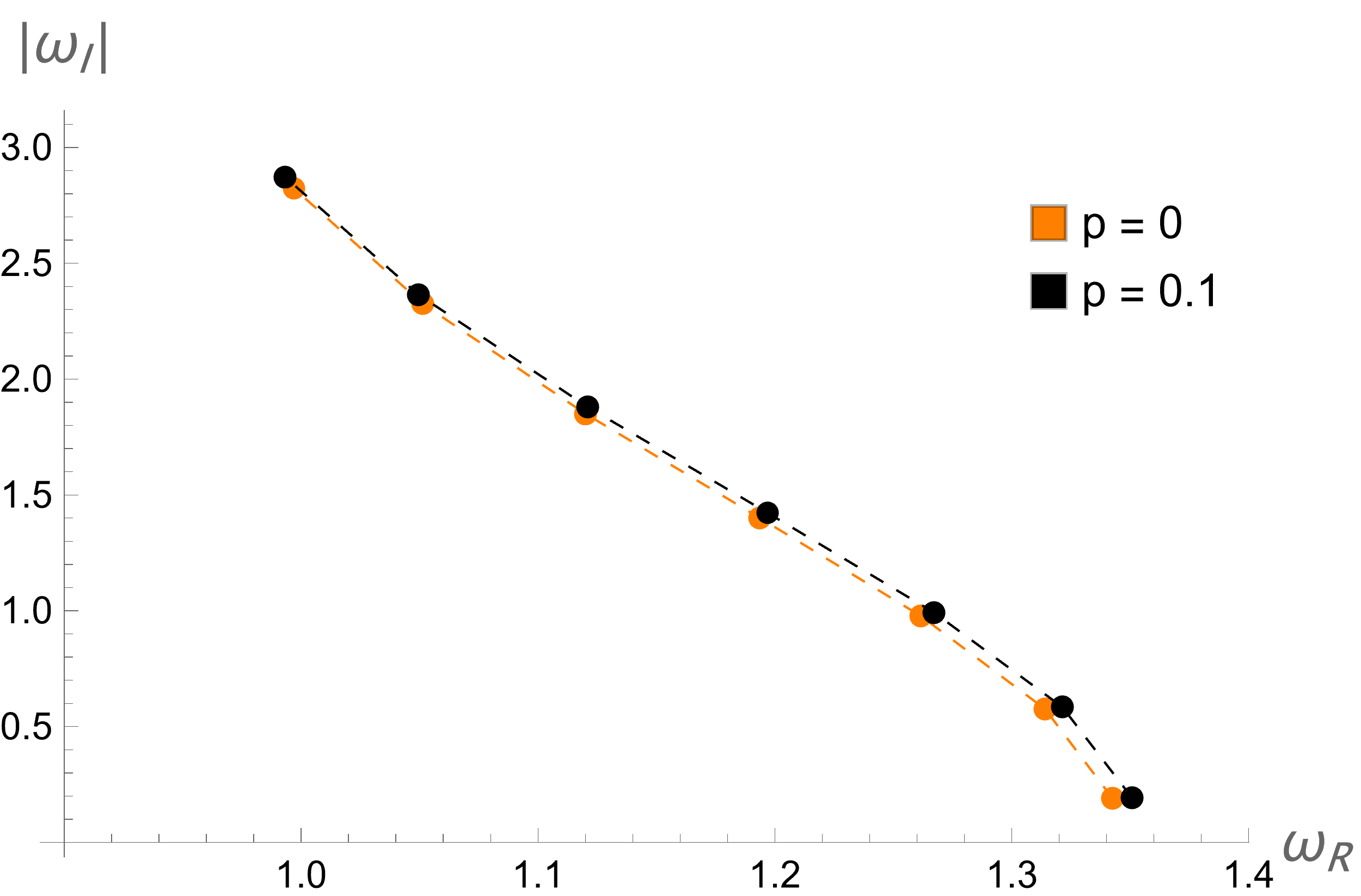}
        	\includegraphics[width = 8 cm]{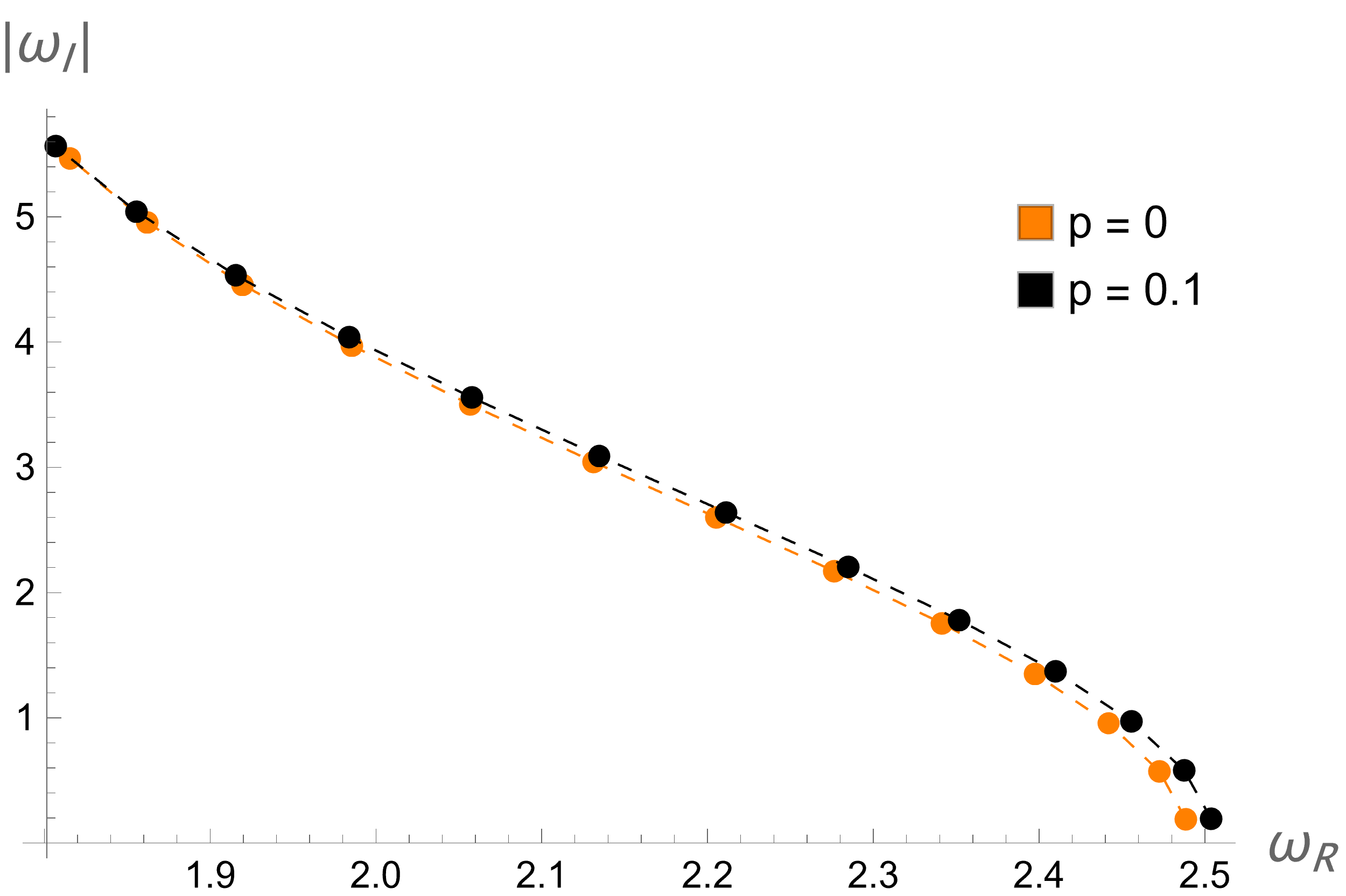}
            \caption{
            The QNM spectrum for $ P = 0.1$ (above). We show cases in which $ l \in \{1,2,3,4,5,6\}$. We also show the comparison of $ l = 3,\, 6$ cases to the Schwarzschild case. Unlike the authors in \cite{Daghigh:2020fmw}, we see that the DL-NED spectrum does not cross the Schwarzschild spectrum.}
        	\label{fig:QNMspectrum}
        \end{figure}

        We provide a visualization of the QNM spectrum for the $P = 0.1$ case in Fig. \ref{fig:QNMspectrum}. It is seen that the real part $\omega_R$ increases in discrete steps with increasing angular momentum $l$, while the imaginary part $\omega_I$ shows a minimal decrease. 
        
        The interesting point is that the cases converge in terms of their real frequencies at higher overtones, while the damping coefficient does increase faster in the Schwarzschild curve, which is lower than the DL-NED case at $n=0$ but surpasses it by the highest overtone in the plot for both $l=3$ and $l=6$. Also, the difference in the fundamental overtones suggest that there could be a detectable difference between the fundamental frequencies. Refer to section \ref{sec:level6D} for an explicit computation.
        
        \subsection{\label{sec:level6C}The time-domain solution}
        
        To calculate the QNMs in an independent way, we wish to solve equation (\ref{eq:RWeq}) in the time domain numerically. We assume that the initial disturbance is a Gaussian wave packet,
        \begin{equation}
            \Psi\left(r_{*}, t = 0\right)=\mathcal{A}\, \text{Exp} \left(-\frac{\left(r_{*}-\bar{r}_{*}\right)^{2}}{2 \sigma^{2}}\right),\left.\partial_{t} \Psi(r_*, t)\right|_{t=0}=-\partial_{r_{*}} \Psi\left(r_{*}, 0\right),
        \end{equation}
        
        where we pick $\sigma=1, \bar{r}_{*}=-40$, and $\mathcal{A}=20$, $r_{*}=90$. We impose boundary conditions such that the wavefunction vanishes at ($r_* = -200, r_* = 250$). Since our black hole is the Schwarzschild BH at the $P \rightarrow \infty$ limit, we wish to see the differences between the two spacetimes. 
        \begin{figure}
        	\centering
        	\includegraphics[width = 8 cm]{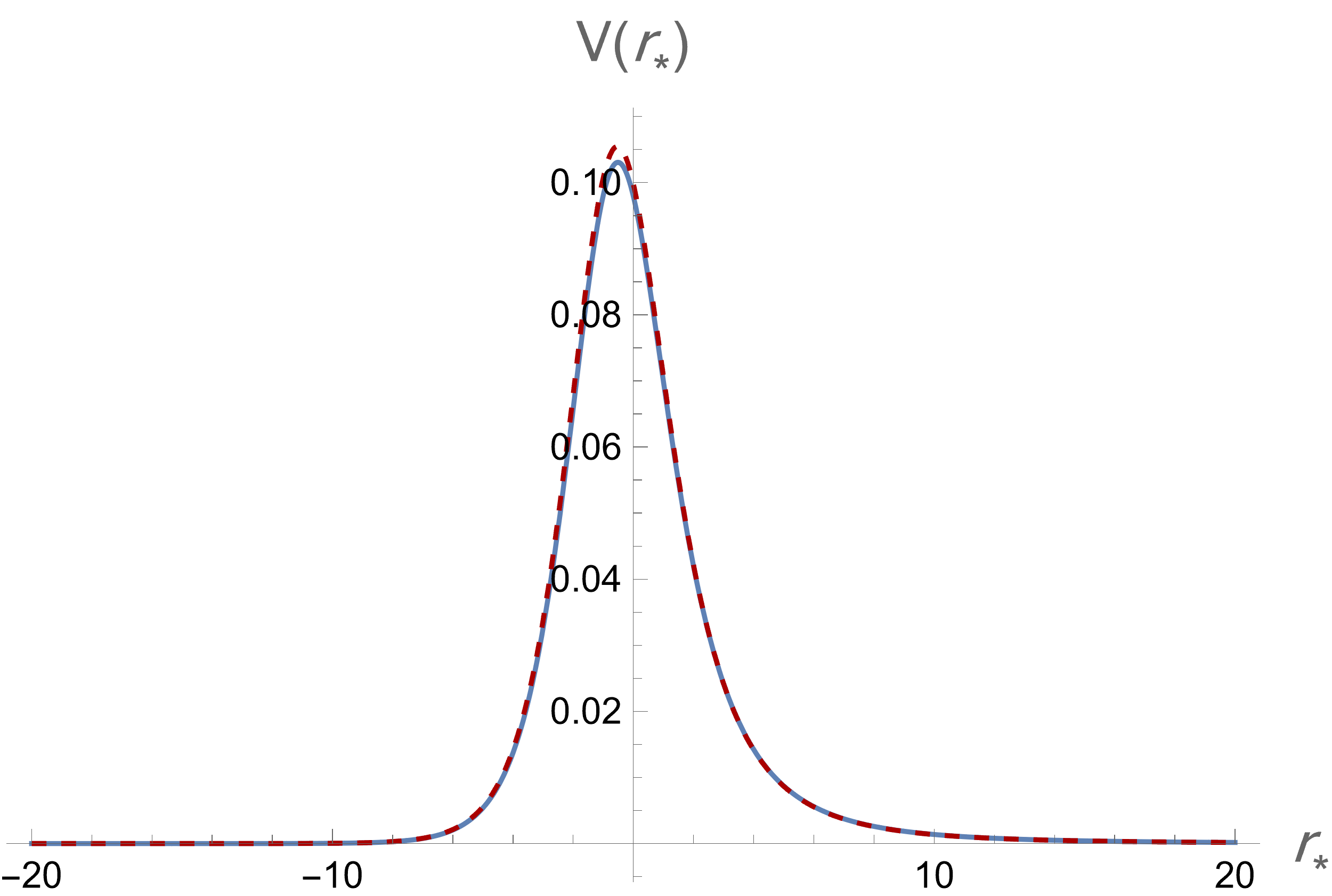}
        	\includegraphics[width = 8 cm]{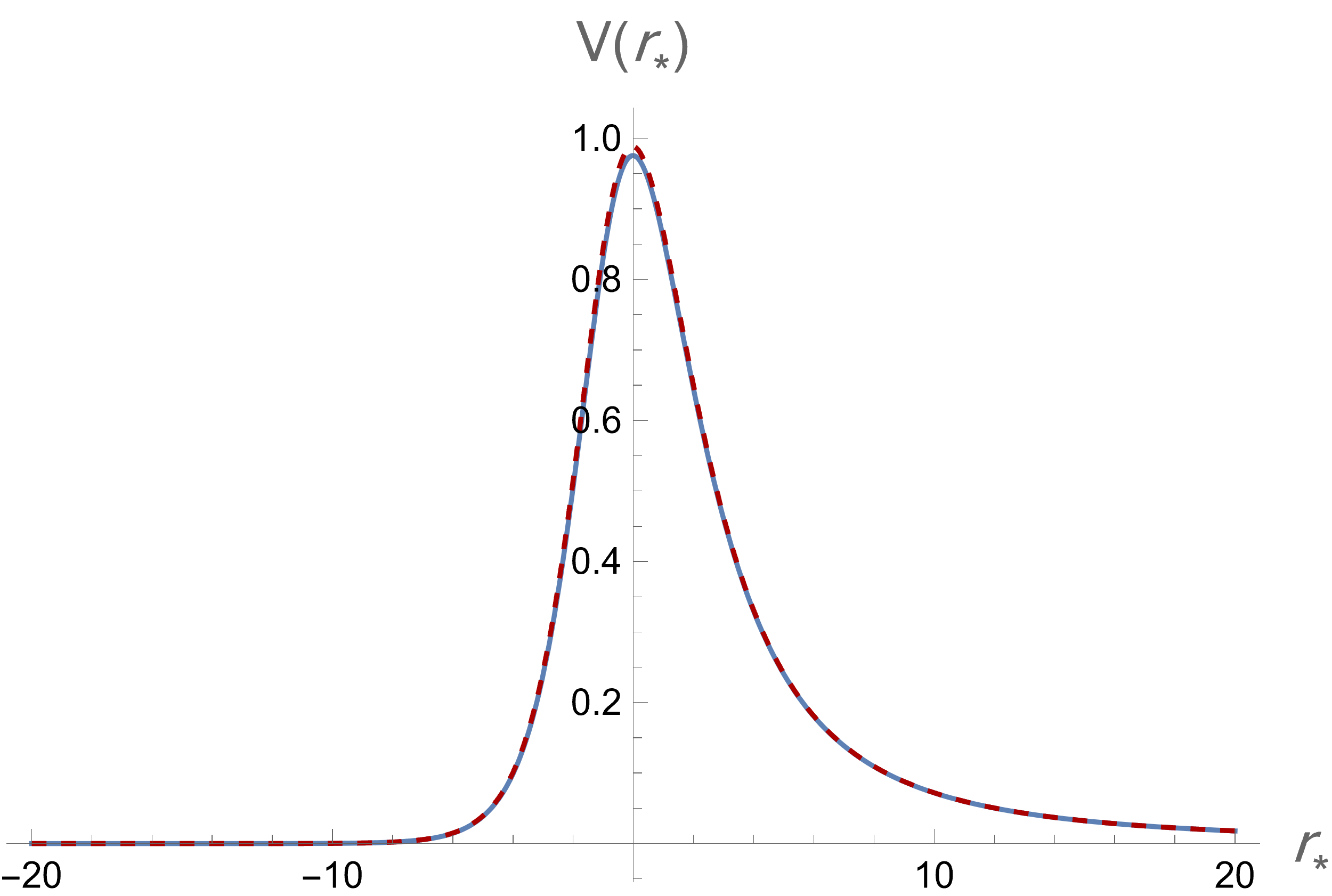}
            \caption{
            The Regge-Wheeler Potentials for $P=0$, and $P = 0.1$, with l = 0 on the right and l = 2 on the left. We show the Schwarzschild limit with dashed dark red, and the charged case in light blue. The Schwarzschild case has a slightly higher peak for both cases.
                }
        	\label{fig:RWpotl0}
        \end{figure}

\begin{figure}
        	\centering
        	\includegraphics[width = 8 cm]{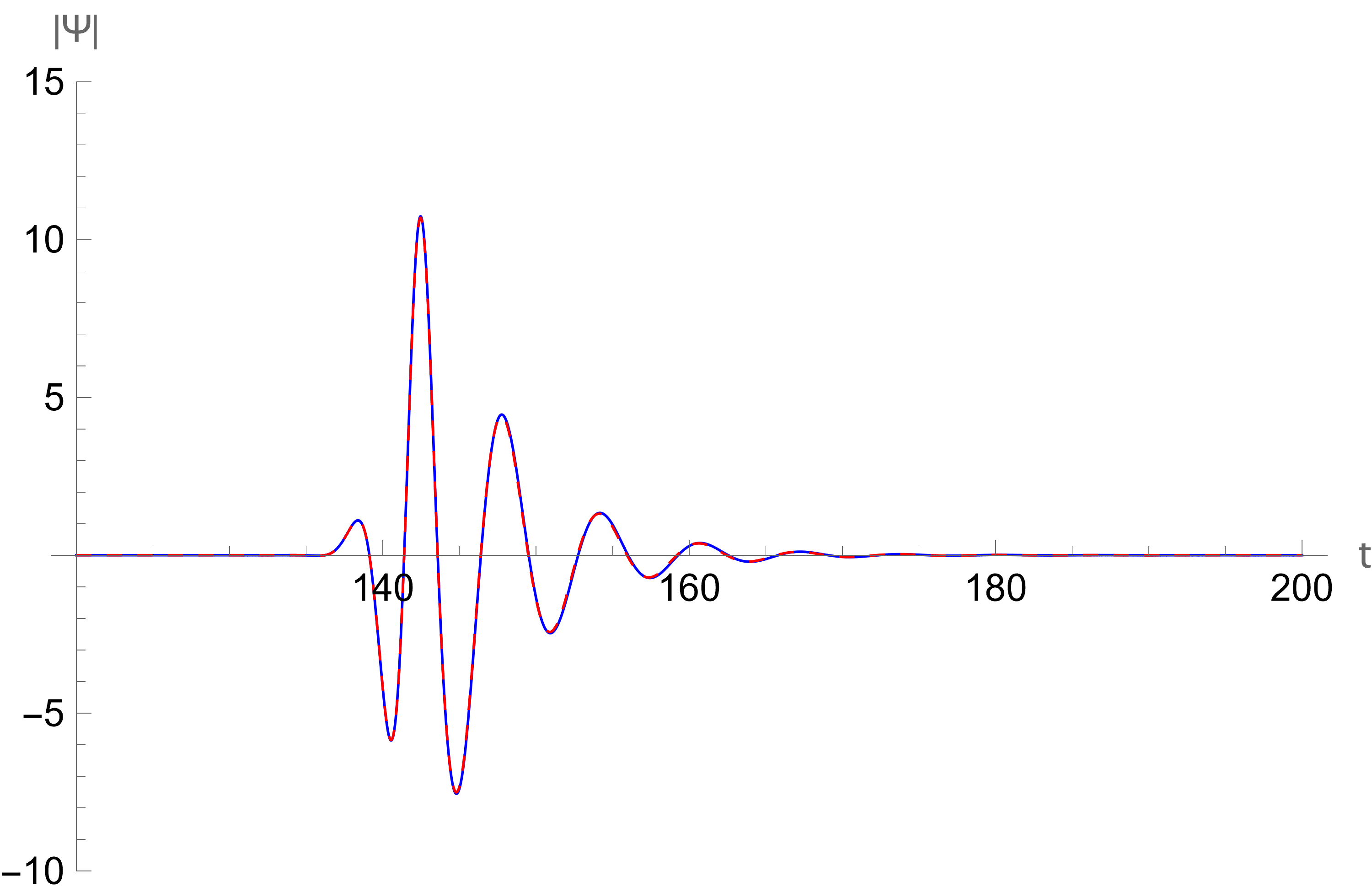}
        	\includegraphics[width = 8 cm]{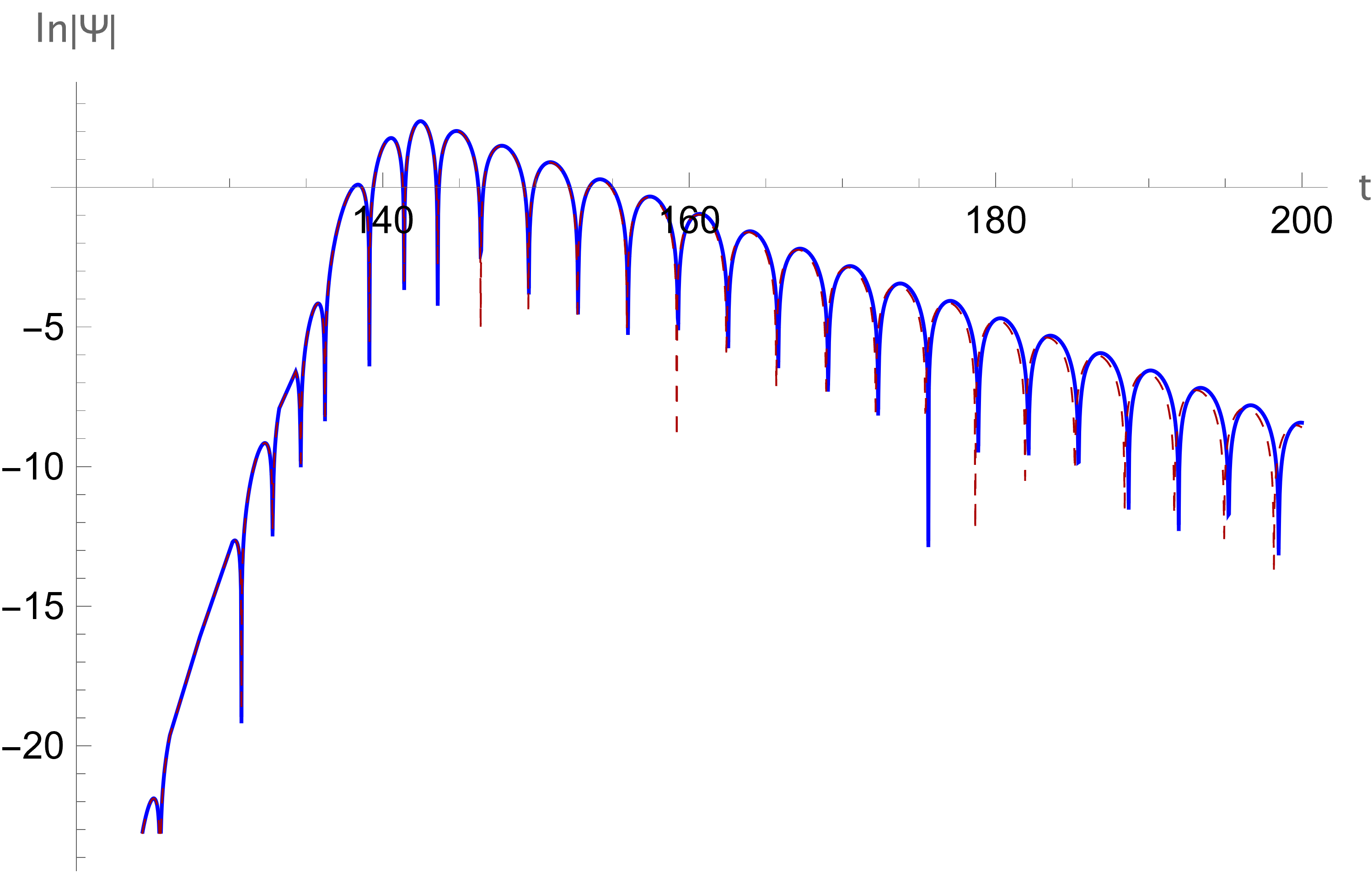}
            \caption{
            The waveforms for the l = 2 case. The DL-NED is given in blue while the Schwarzschild case is in red. In the logaritmic plot, we see that the DL-NED frequency is slightly lower, consistent with (\ref{table:qnm1}) (n=0, l=2) case fundamental overtone.
                }
        	\label{fig:waveforml2}
        \end{figure}

        In the following figures (\ref{fig:RWpotl0}, \ref{fig:waveforml0}, \ref{fig:waveforml1}, \ref{fig:waveforml2}), we show the $|\Psi|$ and $\ln{|\Psi|}$ for $l = 0, 1, 2$, for both the Schwarzschild potential and the DL-NED potential. The waveforms are essentially identical, but there are slight differences in the oscillation frequencies, which are clearer to see in the logarithmic plots. In Fig. \ref{fig:waveforml0tail}, we investigate the large-t behaviour of the l=0 case. Our expectation was that the waveforms should be essentially identical,  which is confirmed in Fig. \ref{fig:waveforml0tail}. The simplest explanation for this is that, because of the asymptotic behaviour of both the Schwarzschild and the P=0.1, 0.2 cases, both are $V_{Sch},\,V_{DL-NED} \sim \frac{1}{r^2}$.
        
        \begin{figure}
        	\centering
        	\includegraphics[width = 8 cm]{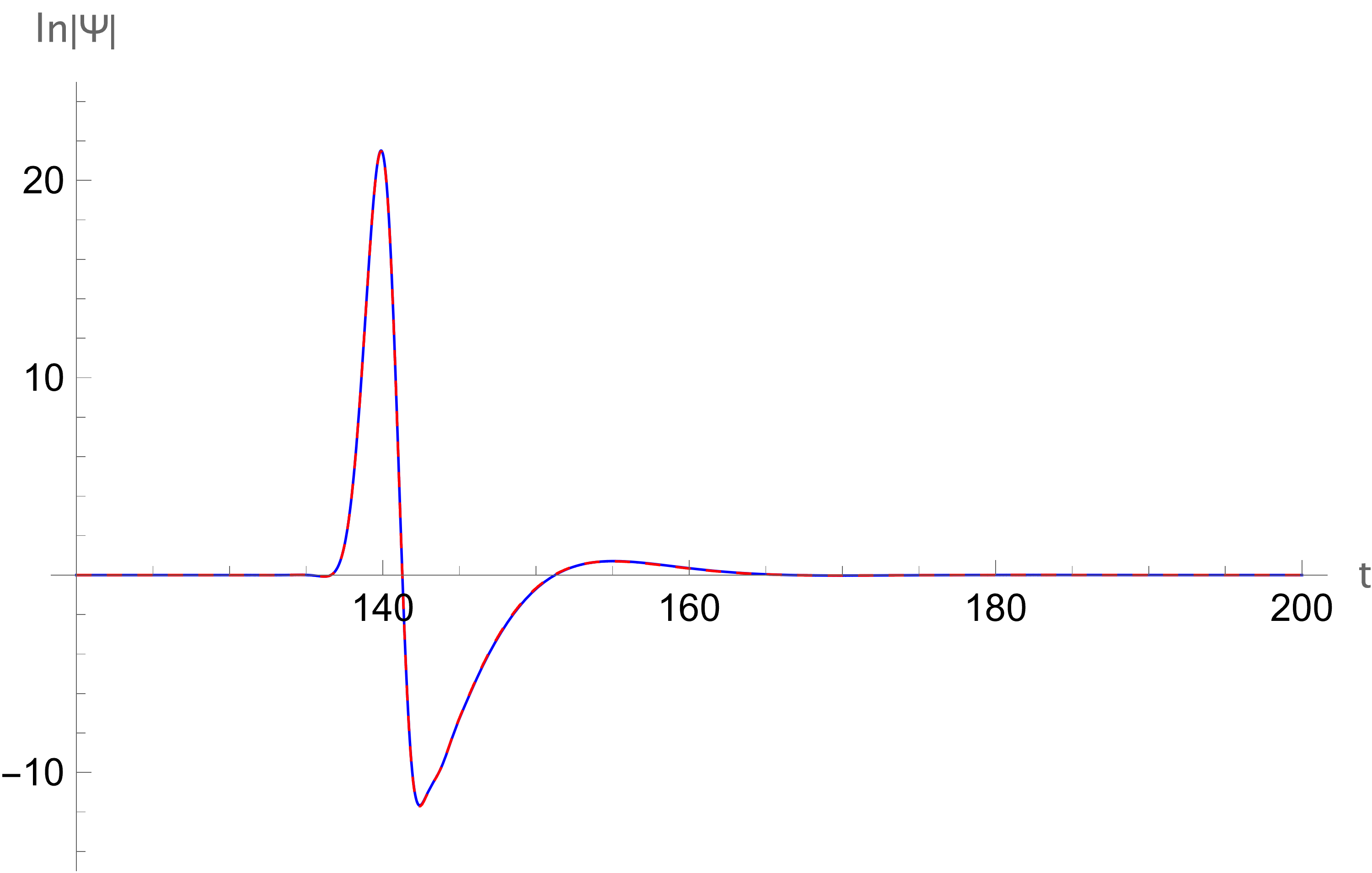}
        	\includegraphics[width = 8 cm]{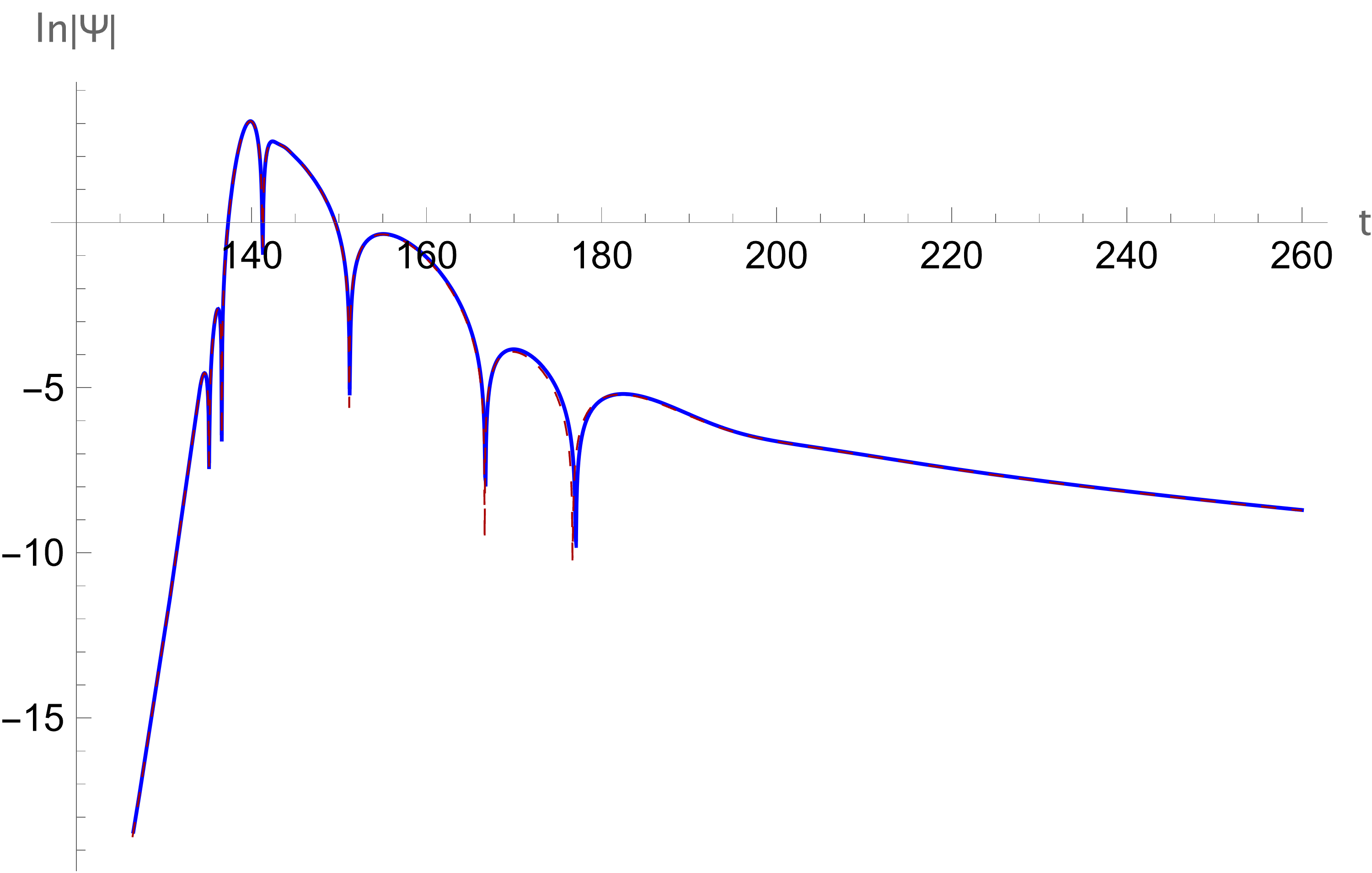}
            \caption{
            The waveforms for the l = 0 case. The DL-NED is given in blue while the Schwarzschild case is in red. In the logaritmic plot, we see that the DL-NED frequency is lower, consistent with (\ref{table:qnm1}) (n=0, l=0) case fundamental overtone. 
                }
        	\label{fig:waveforml0}
        \end{figure}
        \begin{figure}
        	\centering
        	\includegraphics[width = 8 cm]{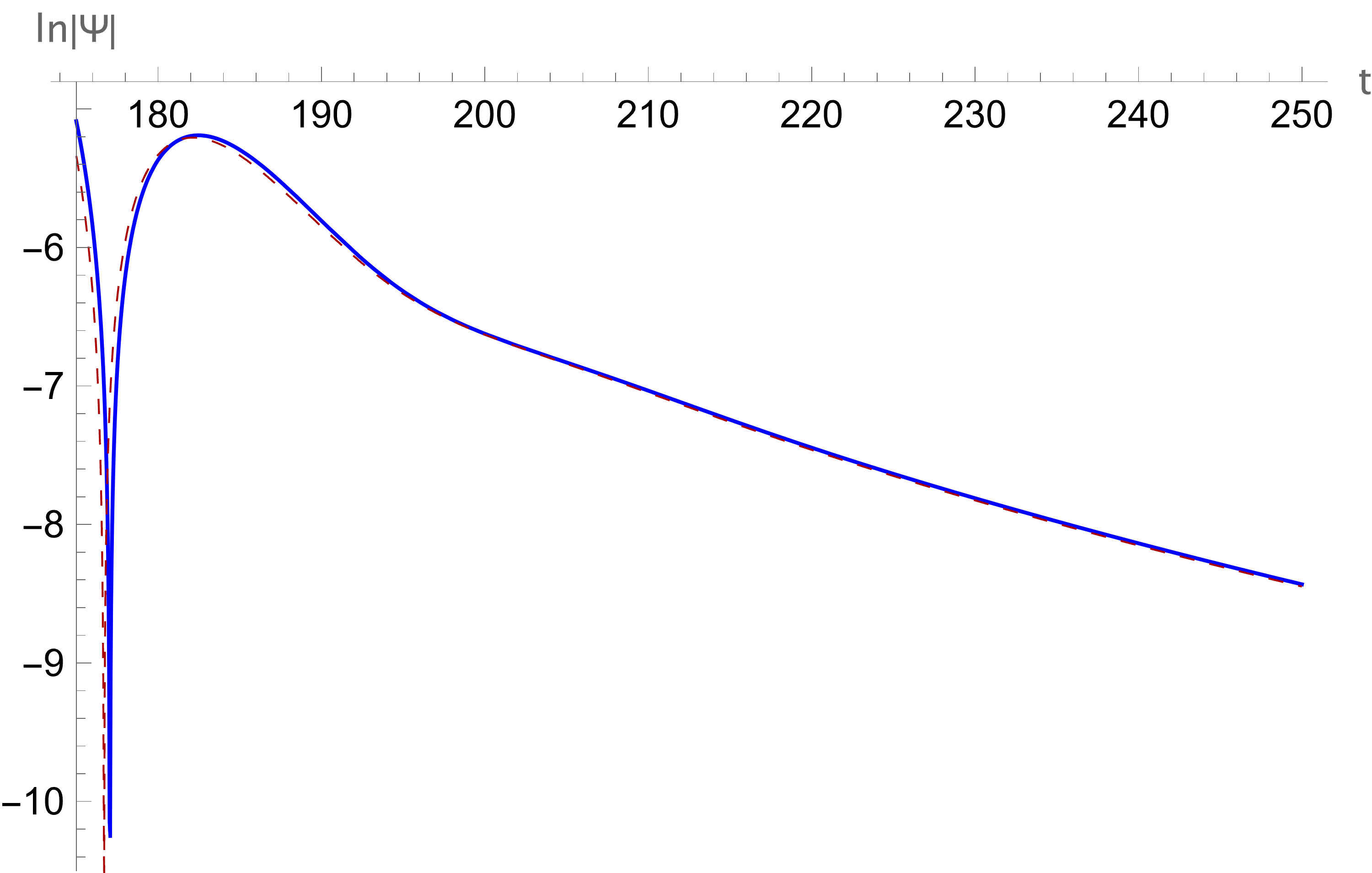}
        	\includegraphics[width = 8 cm]{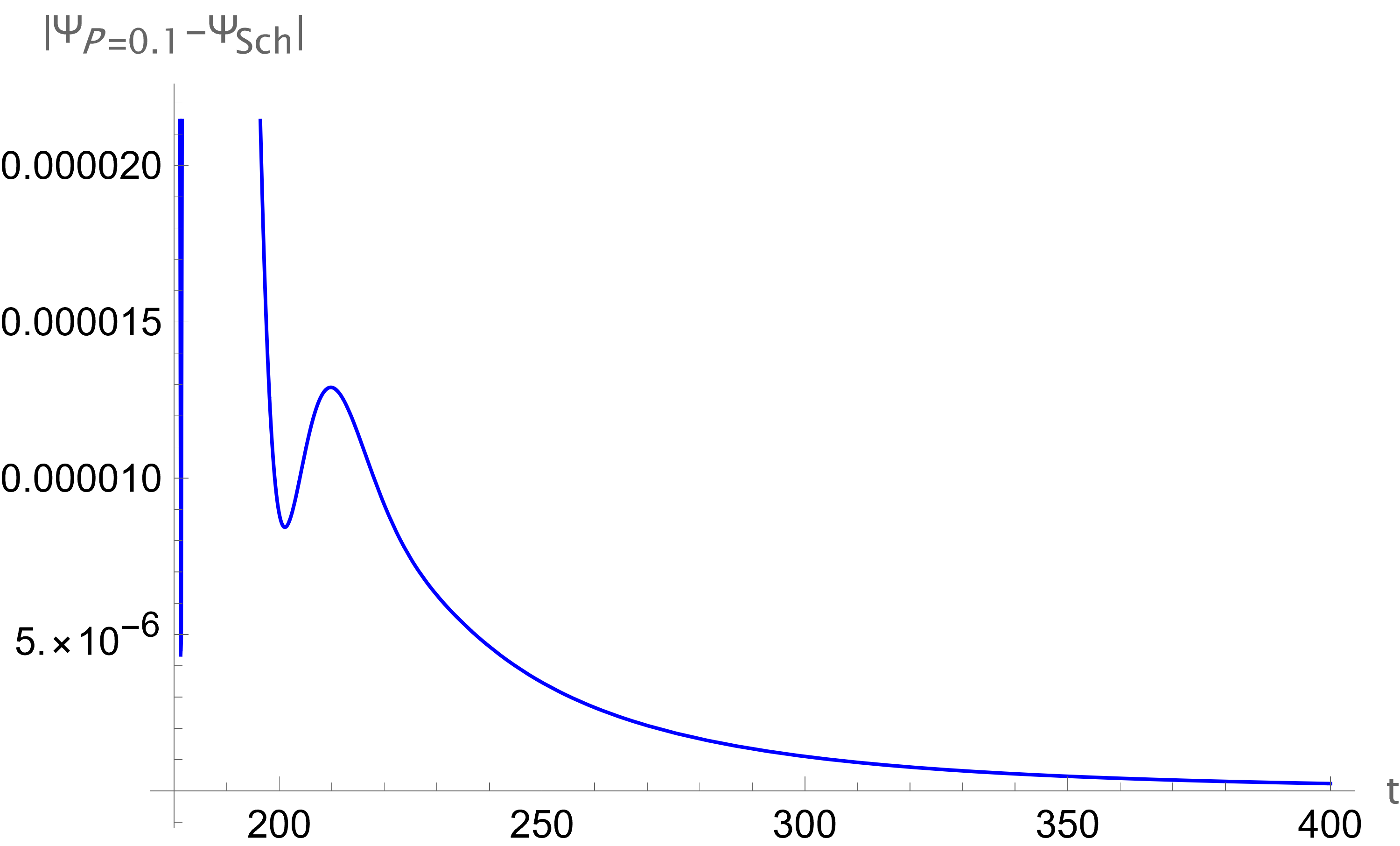}
            \caption{
            The behaviour of the tails of the l = 0 waveforms, given in (\ref{fig:waveforml0}). They follow the same paths, unlike \cite{Daghigh:2020fmw}, the behaviour of the DL-NED black hole is very similar to the Schwarzschild black hole. 
                }
        	\label{fig:waveforml0tail}
        \end{figure}
        \begin{figure}
        	\centering
        	\includegraphics[width = 8 cm]{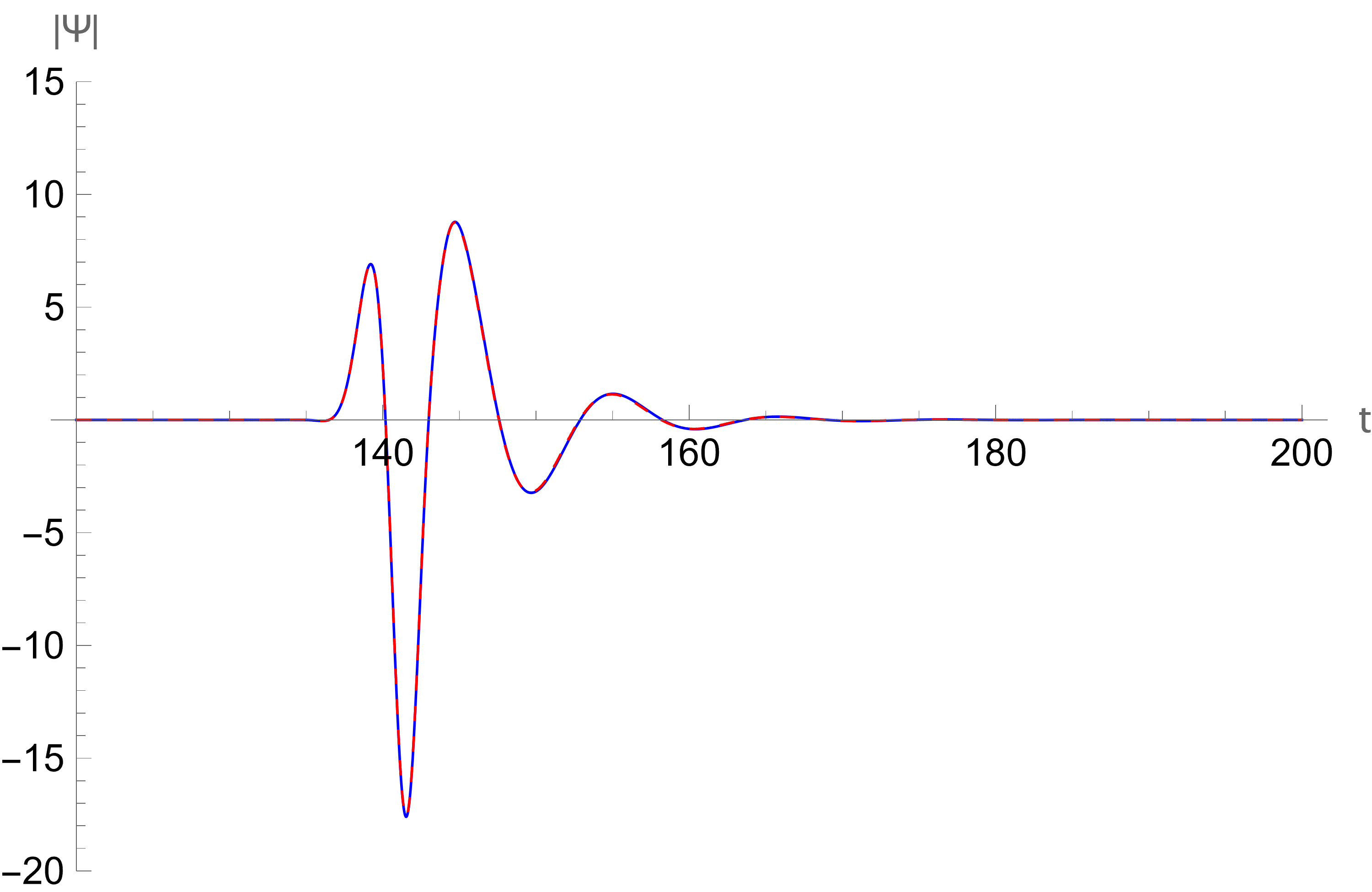}
        	\includegraphics[width = 8 cm]{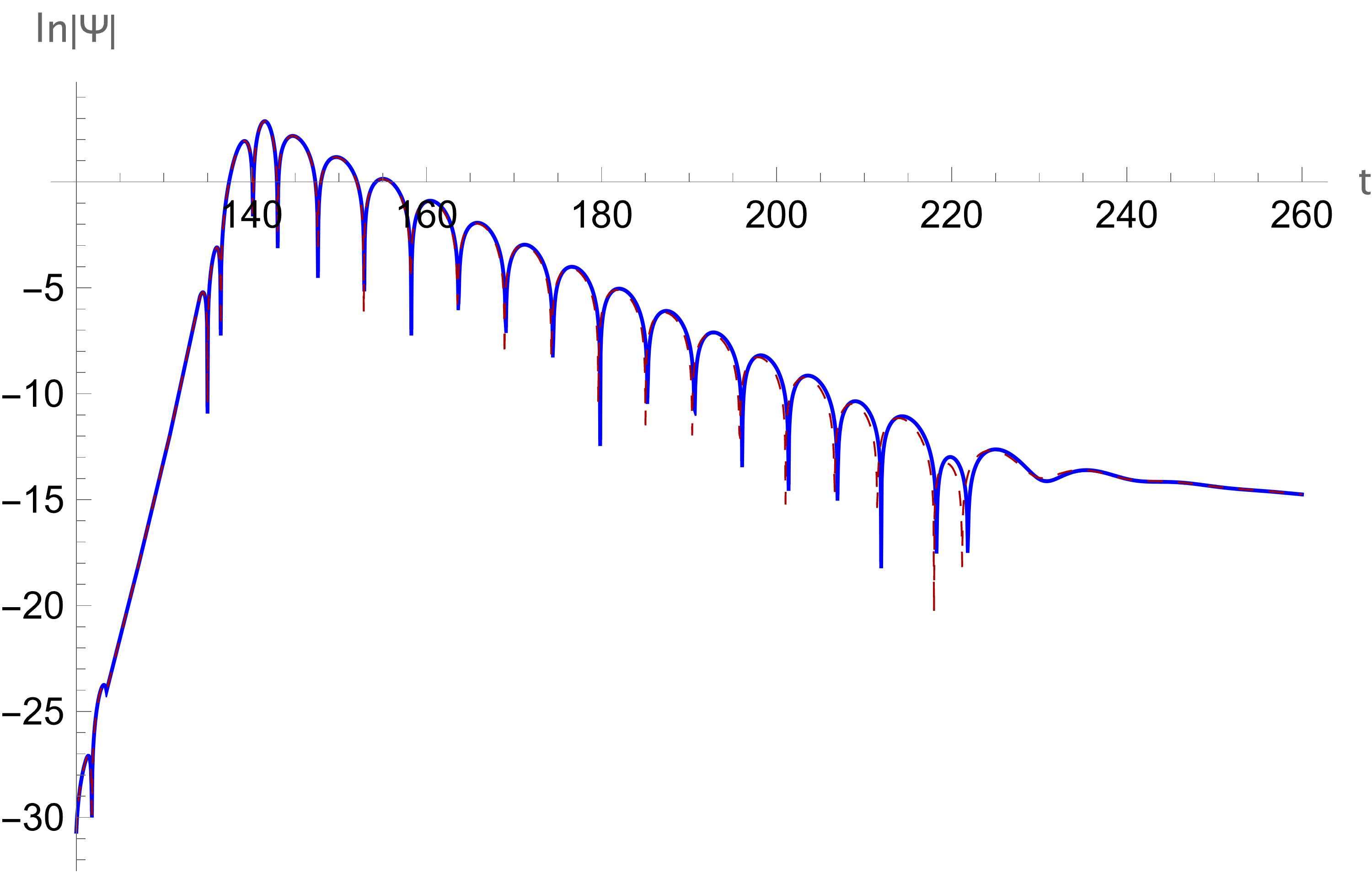}
            \caption{
            The waveforms for the l = 1 case. The DL-NED is given in blue while the Schwarzschild case is in red. In the logaritmic plot, we see that the DL-NED frequency is lower, consistent with (\ref{table:qnm1}) (n=0, l=1) case fundamental overtone. 
                }
        	\label{fig:waveforml1}
        \end{figure}
 
        Furthermore, we provide the frequencies of the fundamental modes ($n = 0$) for the numerical waveforms in Table \ref{table:qnmProny}, which we obain using the Prony method \cite{Prony:1987} . The results agree with values in Table \ref{table:qnm1}, with stronger agreement for higher values of l.
        \begin{table}[]
            \centering
            \begin{tabular}{ccc}
                \multicolumn{2}{l} { Table II: $\omega$ for the fundamental modes n=0, obtained by the Prony method \cite{Prony:1987} from the numerical simulations  \footnote{The \textit{Mathematica} notebook for the Prony method was supplied by A. Zhidenko.}} \\
                \hline  $n, l$ & $P=0$            & $P=0.1$             \\
                \hline  0,0 & $0.216889 - 0.21359 i$ & $0.210219 - 0.201568 i$\\
                        
                \hline  0,1 & $0.585765 - 0.195525 i$ & $0.58228 - 0.192675 i$\\
                        
                \hline  0,2 & $0.967308 - 0.193487 i$ & $0.961436 - 0.190716  i$ \\
                        
                \hline
            \end{tabular}
            \label{table:qnmProny}
        \end{table} 
    \subsection{\label{sec:level6D} Detectability of the QNMs}
    
        To check whether a gravitational signal emitted by an oscillating black hole in the bandwidth of the Virgo/LIGO interferometers, its fundamental frequencies have to be in the range about $10-40$ Hz or for the LISA interferometer about $10^{-4}-10^{-1}$ Hz \cite{Ferrari:2007dd}.  Assuming $M=\alpha M_{\odot}, \,\left(M_{\odot}=1.48 \cdot 10^{5}\, \mathrm{~cm}\right)$, the frequencies and damping times are calculated by \cite{Ferrari:2007dd}
    
        \begin{equation}
        \nu=\frac{c}{2 \pi \alpha \cdot M_{\odot}\left(M \omega_{0}\right)}=\frac{32.26}{\alpha}\left(M \omega_{0}\right) \mathrm{kHz}, \quad \tau=\frac{\alpha M_{\odot}}{\left(M \omega_{i}\right) c}=\frac{\alpha \cdot 0.4937 \cdot 10^{-5}}{\left(M \omega_{i}\right)} \mathrm{s}.
        \end{equation}
        
        For our fundamental overtones for angular momenta $l =0, 1,2$ of oscillating black holes with charge P=0.1, confirmed by the Prony fitting in section \ref{sec:level6C}, we compute the fundamental frequencies to determine the range of masses of oscillating DL-NED black holes LIGO and LISA would be able to detect. We find that LIGO can detect 
        \begin{equation}
            5 M_{\odot} \lesssim M \lesssim 6\cdot10^{3} M_{\odot},
        \end{equation}
        
        in the fundamental frequency range $\nu \in[12 \mathrm{~Hz}, 1.2 \mathrm{kHz}]$, and for LISA the detectable mass range is 
        
        \begin{equation}
           6.8 \cdot 10^{4} M_{\odot} \lesssim M \lesssim 8.0 \cdot 10^{8} M_{\odot},
        \end{equation}
        
        corresponding to frequencies $\nu \in\left[10^{-4}, 10^{-1}\right] \mathrm{Hz}$. Since the variations in the frequencies when the charge is increased to $P=0.2$ are on the order of $10^{-3}$, we did not expect an appreciable change in the detection mass range, which is confirmed by our calculations. \\
        Furthermore, we had pointed out that there could be a difference in the fundamental frequencies of the fundamental overtones between the Schwarzschild BH and the DL-NED BH, as seen in Fig. \ref{fig:QNMspectrum}. We find that for a black hole of mass $100\, M_\odot$, the differences in the fundamental frequencies are 
        \begin{align*}
            &\text{l=3:  } \delta\nu = 2.64 \text{Hz}\\
            &\text{l=6:  } \delta\nu = 4.97 \text{Hz}\\
        \end{align*}
        We also found that the frequency difference decreases with increasing mass of the perturbed black hole, for both cases. 
\section{\label{sec:level7}Greybody factors and High-Energy Absorption cross section via Sinc approximation}

    \subsection{The Greybody Bound}
    
        In this section we calculate the lower bound for the greybody factor of the DL-NED Black Hole. There are various methods of obtaining this bound with approximations, such as the WKB method or the matching technique. Instead, we use the rigorous lower bound, allowing us to investigate the effect of P on the bound. The rigorous bound is given by \cite{Visser:1998ke,Boonserm:2008zg}
        
        \begin{equation}
            T\geq \sech^2\Big( \int_{-\infty}^\infty \nu dr_*\Big),
        \end{equation}
        
        with
        
        \begin{equation}
            \nu = \frac{\sqrt{(h'(r_*))+(\omega^2-V(r_*)-h^2(r_*))^2}}{2h(r_*)}.
        \end{equation}
        
        Here we use choose $h(r_*) = \omega$, which trivially satisfies the boundary conditions $h(\infty) = h(-\infty) = \omega$. This yields
        
        \begin{equation}
            T_b\geq \sech^2\Big( \frac{1}{2\omega}\int_{-\infty}^\infty |V(r_*)|dr_*\Big).
        \end{equation}
        
\begin{figure}[ht]
            \centering
            \begin{minipage}[b]{0.4\linewidth}
            \includegraphics[width = 8 cm]{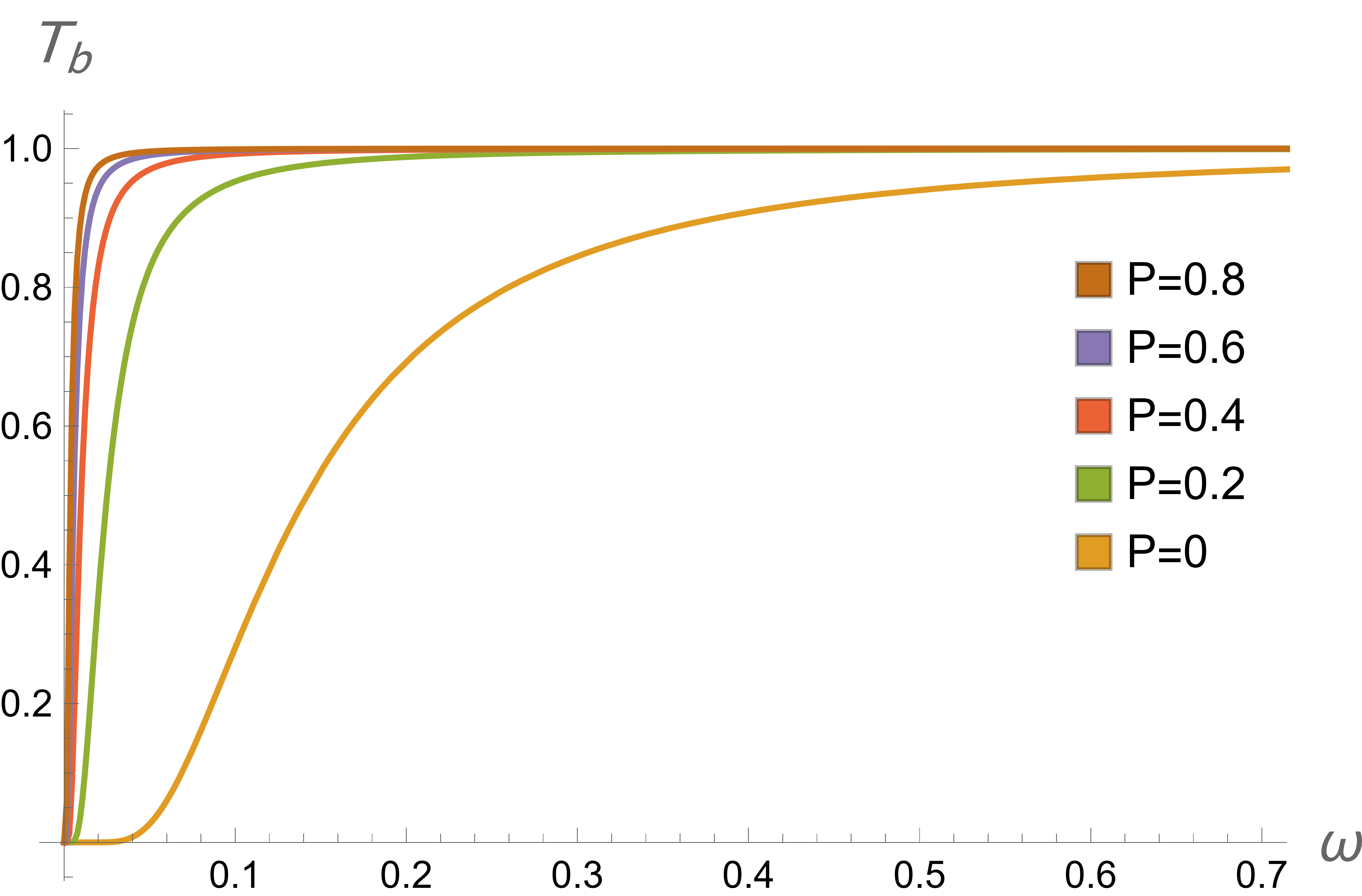}
            \caption{The Greybody Bound $T_b$ versus the magnetic charge P, with M = 1, $\beta$ = 1.}
            \label{fig:greybound}
            \end{minipage}
            \quad
            \begin{minipage}[b]{0.4\linewidth}
             	\includegraphics[width = 9 cm]{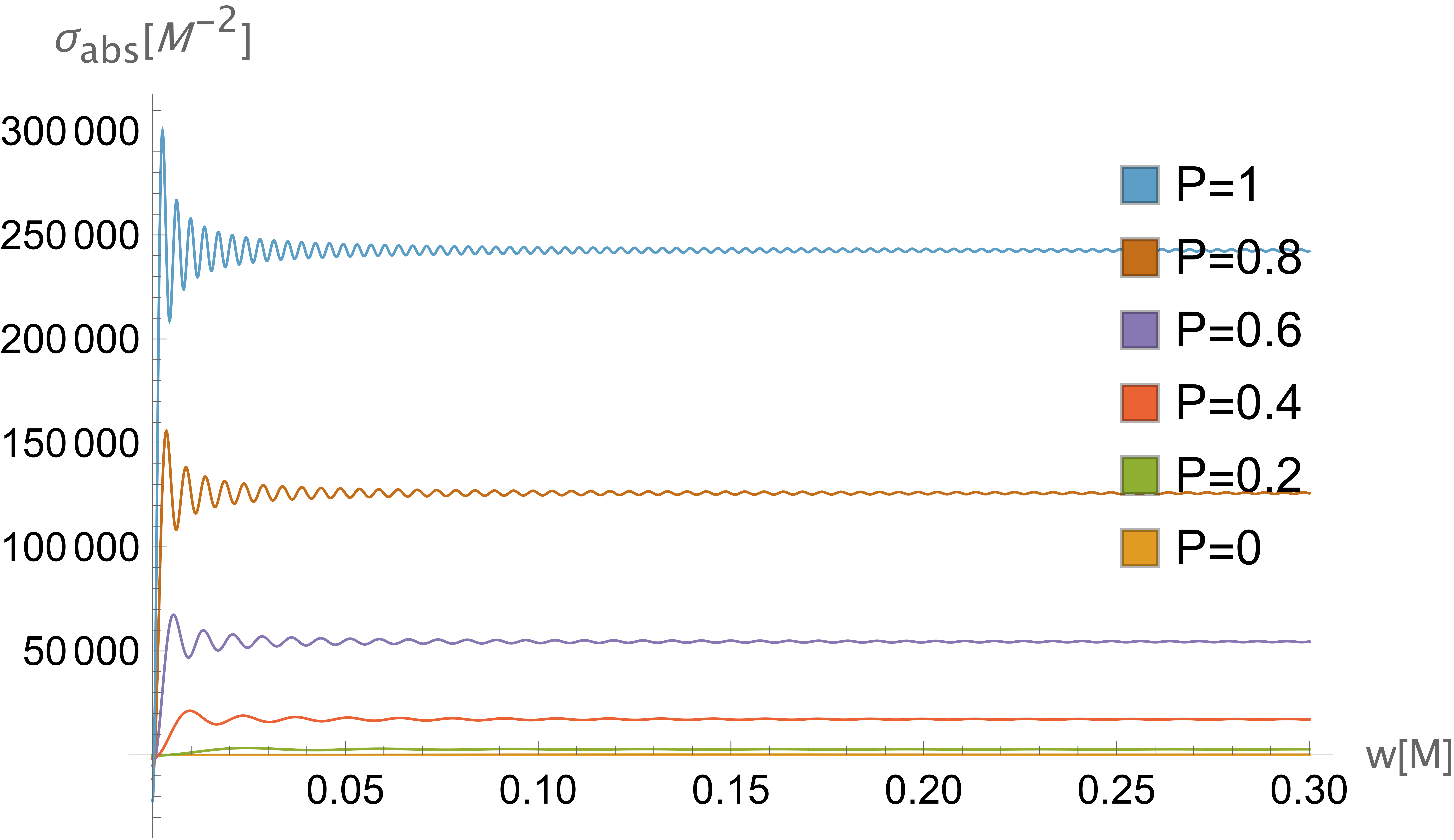}
            	\caption{The total absorption cross section for various values of the magnetic charge, $\beta$ = 1.}
               	\label{fig:absorption}
            \end{minipage}
        \end{figure}

        Using the Regge-Wheeler potential for the massless scalar field from last section, we can analytically evaluate the bound as 
        
        \begin{equation}
            T \geq T_b = \sech^2 \Bigg( \frac{1}{2\omega} \Bigg(-\frac{4 \kappa P^2}{3 r_H^3} - \frac{-3 G M - \frac{2\sqrt{2}P^2\pi \kappa}{r_0}}{3 r_H^2} + \frac{l(l+1)}{r_H}\Bigg) \Bigg),
        \end{equation}
        
        which reduces to the Schwarzschild case at $ P\rightarrow 0$ correctly, as $T_{Sch} \geq \sech^2 \big( \frac{1 + 2l(l+1)}{8 G M \omega}\big)$.
        Investigating the variation of the greybody factor with various charges, we see in Fig. \ref{fig:greybound}.

    \subsection{The High-Energy Absorption Cross Section with the Sinc approximation}
    
    The absorption cross section oscillated around the constant geometric-optics value for a black hole, which is related with the photon sphere, contrarily, it increases monotonically with increasing frequency for ordinary material sphere \cite{Sanchez:1977si}.  The cross section of the photon sphere is related with impact parameter at critical value and limiting the value of absorption cross section, the black hole's characteristic properties at low energies, where simply the cross section equals to black hole area, \cite{Das:1996we,Higuchi:2001si} and also at high energies \cite{Decanini:2011xi} using the geometrical cross section of the photon sphere can be studied. Using the Regge pole techniques, Decanini et al. \cite{Decanini:2011xi} prove that the oscillatory pattern of the high-energy absorption cross section related with a Sinc(x) function included photon sphere.
    
        The absorption cross section for high frequency waves is approximately the so-called classical capture cross section of null geodesics, given by $\sigma_{geo} = \pi b_c^2$ where $b_c$ is the critical impact parameter. 
        In the eikonal limit, the oscillatory part of the absorption cross section can be written as \cite{Decanini:2011xi}:
        \begin{equation}
            \sigma_{osc} = - 4 \pi \frac{\lambda b_c^2}{w} e^{-\pi \lambda b_c} \sin\frac{2 \pi w}{\Omega_c},
        \end{equation}
        
        where $\lambda$ is known as the Lyapunov exponent
        
        \begin{equation}
            \lambda = \sqrt{\frac{f(r_c)}{2 r_c^2} \big(2f(r_c)-r_c^2f''(r_c)\big)},
        \end{equation}
        
        and  $\Omega_c = \sqrt{\frac{f(r_c)}{r_c^2}}$ where $r_c$ is the radius of the photon orbit. Then the Sinc approximation states that the total absorption cross section at the eikonal limit is $\sigma_{abs} \approx \sigma_{osc} + \sigma_{geo}$ \cite{Sanchez:1977si,Decanini:2011xi,Magalhaes:2020sea,Paula:2020yfr,Lima:2020seq}.  In the Fig. \ref{fig:absorption} we plot the total absorption cross section for various values of $P$.
        
        The absorption cross section shows a great dependence on the magnetic charge $P$. This is supported by the direct dependence of the size of the shadow with $P$, as discussed in section \ref{sec:level5B2}. Furthermore, the magnitude of the oscillatory component of the cross section is again dependent on P with $\sigma_{abs} \propto b_c^2$, which in turn depends on P at the asymptotic limit $b_c \propto P^\frac{3}{2}$.

    Furthermore, the connection between the Sinc approximation of high-energy cross section and shadow radius of the black hole can be written in this form \cite{Decanini:2011xi,ovgun}:
        
        \begin{eqnarray} \label{sincc}
         \sigma_{\text{abs}}(\omega)=  -8\pi\eta_{\text{c}}e^{-\pi\eta_{\text{c}}}
        \operatorname{sinc}{(\omega 2\pi R_s)}\pi R_s^{2}+\pi R_s^2,
        \end{eqnarray}
        
        where
        
        \begin{equation}
            \eta_{\text{c}}=\sqrt{f(r_c)-\frac{1}{2}r_{\text{c}}^{2}f(r_c)''}.
        \end{equation}
        
\begin{figure}[ht]
        \centering
        \begin{minipage}[b]{0.4\linewidth}
        \includegraphics[width = 8 cm]{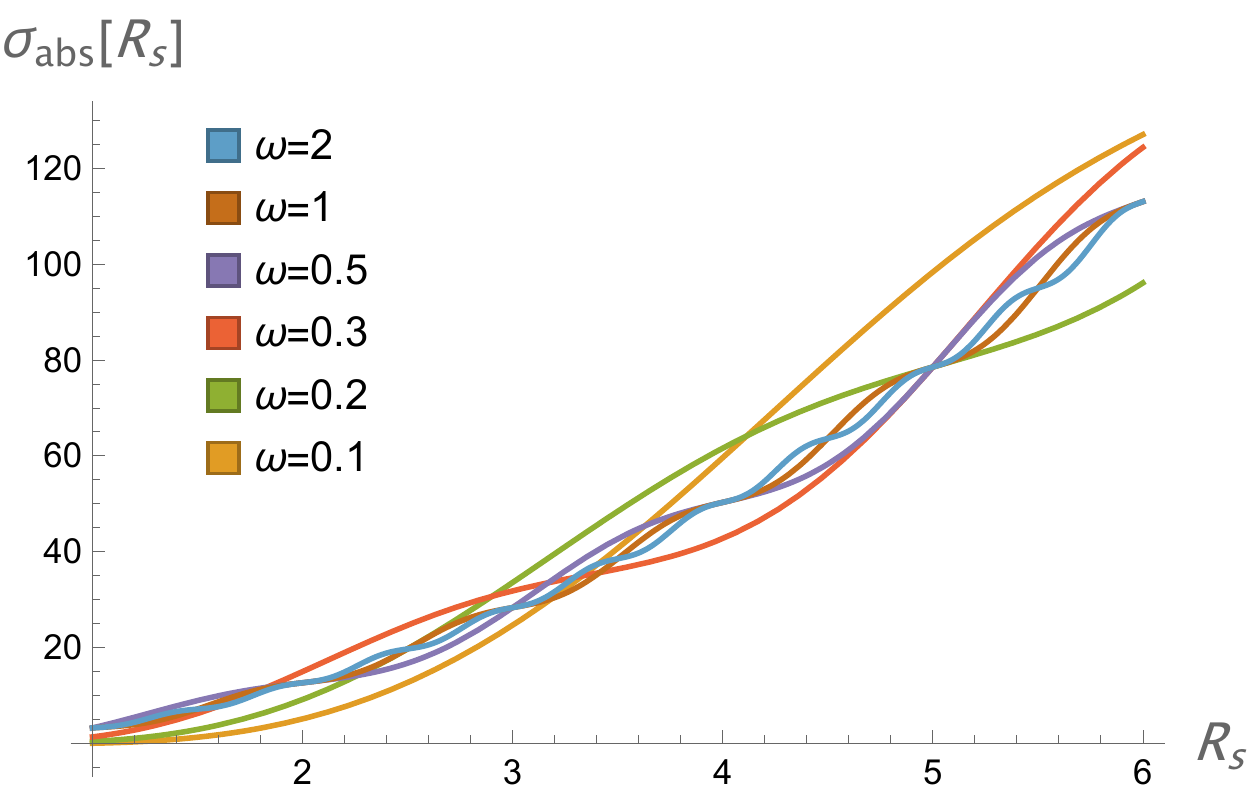}
        \caption{Figure shows the correspondence between the total absorption cross section via Sinc approximation and the shadow radius of the black hole for $R_s<4$.}
        \label{fig:corr1}
        \end{minipage}
        \quad
        \begin{minipage}[b]{0.4\linewidth}
        \includegraphics[width = 8 cm]{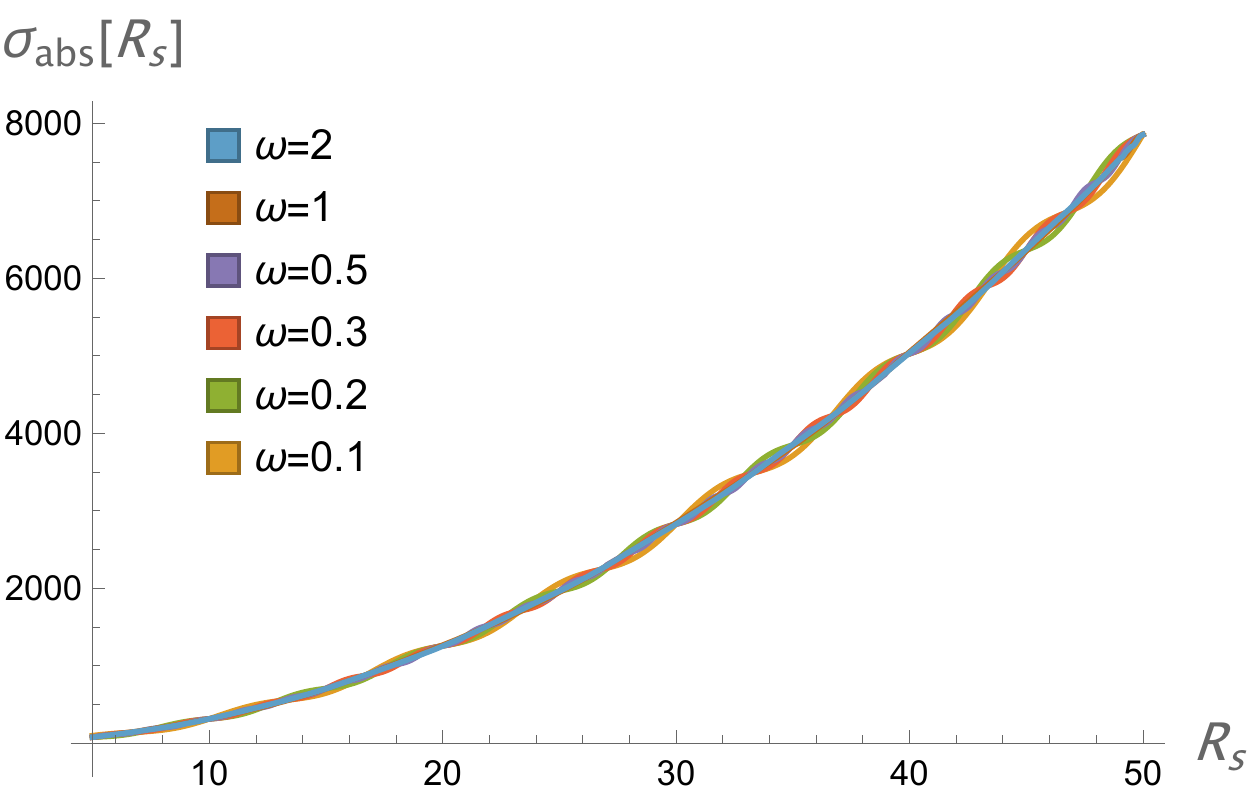}
        \caption{Figure shows the correspondence between the total absorption cross section via Sinc approximation and the shadow radius of the black hole for $R_s<50$.}
        \label{fig:corr2}
        \end{minipage}
        \end{figure}

        The Fig. \ref{fig:corr1} and Fig. \ref{fig:corr2} show that the increasing the shadow radius of the black hole, the high-energy absorption cross-section in the Sinc approximation increases slowly by fluctuated. But especially for the large shadow radius value of the black hole, the high-energy absorption cross-section in the Sinc approximation exponentially increases.
        
\section{Conclusion}

	In this paper we have investigated various properties of magnetically charged black holes in an extension of the Maxwell theory, the DL-NED theory. We initially show the outline of how the metric is derived. We calculate the horizons of the spacetime, in the asymptotically flat setting and show that, unlike the RN solution, our black hole is never extremal. We then look into the Hawking thermodynamics of the black hole solution, looking into how the magnetic charge affects the thermodynamic properties. We further investigate the effects of a generalized uncertainty principle correction to the thermodynamics. We find that in both cases, the Hawking temperature is inversely related to the charge whereas the Bekenstein entropy is proportional to the charge by $P^4$. By computing the heat capacity in both cases, we find that the black hole can be stable, which the GUP corrections amplify. After, we investigate the angle that null geodesics are deflected in the weak-field limit through the Gauss-Bonnet theory and the geodesic equation. The deflection angle peaks right above the critical impact parameter $P= 0.1$ case, dying off with increasing charge. The geodesic equation is then used to visualize the observational appearance of the black hole, for a static accretion disk and a free-falling spherical shell of gas. It is found that with increasing magnetic charge, the size of the black hole is greatly increased and with increasing field coupling parameter $\beta$, the size of the shadow is slightly diminished. We then compute the QNM frequencies of small oscillations of the black hole, for the cases $P = 0, 0.1, 0.2$, using the WKB approximation. We solve the Regge-Wheeler equation numerically and confirm the accuracy of the WKB method for the fundamental mode ($n =0$) for varying angular momenta. At the end, we investigate the absorption cross-section of our black hole, first computing the greybody bound and then using the sinc approximation. We find that the greybody bound is gets steeper with increasing charge. We finally find that the oscillatory part of the cross section, as well as the total absorption, get increasingly larger with increasing charge, confirming the calculations of the section on the observational appearance of the black hole shadow. This similarity is further confirmed by the final computation of the relation between the absorption cross section and the shadow radius.

	While we provide a survey of theoretical computations about the properties of the black hole, we also provide possible avenues of experimental observation of the DL-NED magnetic black hole. We show that neglecting the rotational properties of the M87* black hole, the image of the shadow bounds its possible magnetic charge between $P=0$ and $P=0.024$ in units of $M_{BH}$. This result agrees with the general idea that any sort of electromagnetic charge on an astrophysical black hole is either zero of minimally positive. We also provide observational possibilities in the gravitational wave avenue, by showing that the Virgo/LIGO ground detectors and the LISA space detector could find DL-NED black holes with masses ranging from $5M_{\odot}$ up to $8.0 \cdot 10^8 M_{\odot}$, with a small gap between $6\cdot 10^3 M_\odot \;\text{and}\; 6\cdot 10^4 M_\odot$, which corresponding frequencies not covered by these detectors. These values are for the $P = 0.1$ case, and as for increasing values of the magnetic charge, the metric approaches the Schwarzschild case, the detection range therefore also converges to the Schwarzschild case given in \cite{Ferrari:2007dd}.

	In the future, the most natural extension of our study is the generalization of the DL-NED metric to the rotating case. In the real universe, we do expect that most black holes have nonzero angular momentum. In our study of the EHT image as a DL-NED black hole, what remains to be done is a more accurate investigation of the nontrivial rotational effects that are present, allowing for a stronger constraint on the possibility of a net magnetic charge. Furthermore, we also wish to investigate the quasinormal modes from other approaches. Particular topic of interests are the pseudospectrum of the QNMs, from which we can investigate the stability. A further consideration is utilizing and improving neural network methods to compute quasinormal modes. 
	
\acknowledgments
The \textit{Mathematica} codes of shadow calculations will be shared on reasonable request to the authors by email. The authors are grateful to the anonymous referees for their valuable comments and suggestions to
improve the paper.

\end{document}